\documentclass[traditabstract]{aa}
\usepackage{aas_macros}
\usepackage{hyperref}
\hypersetup{colorlinks=true,citecolor=blue}

\usepackage{bm}
\usepackage{booktabs,adjustbox}
\usepackage{longtable}
\usepackage{array}
\usepackage{url}
\usepackage{algorithmicx}
\usepackage{algorithm}
\usepackage{algpseudocode}
\usepackage{scrextend}
\usepackage{multirow}
\usepackage{xcolor}
\usepackage{graphicx}
\usepackage{rotating}
\usepackage{caption}

\usepackage{amsmath}
\usepackage{appendix}
\usepackage{rotating}
\usepackage{float}
\usepackage{txfonts}
\usepackage{natbib} 
\usepackage{xspace}
\usepackage{url}
\bibpunct{(}{)}{;}{a}{}{,} 
\usepackage{subfigure}

\newcolumntype{x}[1]{>{\centering\arraybackslash\hspace{0pt}}p{#1}}

\begin{document}
\title{Sparse Lens Inversion Technique (SLIT): \\
lens and source separability from linear inversion of the source reconstruction problem}
\titlerunning{Lensed source reconstruction and deblending with MCA}

\author{R. Joseph\inst{1} \and F. Courbin\inst{1}  \and J.-L. Starck\inst{2} \and S. Birrer\inst{3}
}
\authorrunning{R. Joseph et al.}

\institute{
Institute of Physics, Laboratory of Astrophysics, Ecole Polytechnique F\'ed\'erale de Lausanne (EPFL), Observatoire de Sauverny, CH-1290 Versoix, Switzerland \and  
Laboratoire AIM, CEA/DSM-CNRS-Universite Paris Diderot, Irfu, Service d'Astrophysique, CEA Saclay, Orme des Merisiers, 91191 Gif-sur-Yvette, France \and 
Department of Physics and Astronomy, University of California, Los Angeles, 475 Portola Plaza, Los Angeles, CA 90095-1547, USA
}

\date{Received ; accepted }

\abstract{Strong gravitational lensing offers a wealth of astrophysical information on the background source it affects, provided the lensed source can be reconstructed as if it was seen in the absence of lensing. In the present work, we illustrate how sparse optimisation can address the problem. As a first step towards a full free-form lens modelling technique, we consider linear inversion of the lensed source under sparse regularisation and joint deblending from the lens light profile. The method is based on morphological component analysis, assuming a known mass model. We show with numerical experiments that representing the lens and source light using an undecimated wavelet basis allows us to reconstruct the source and to separate it from the foreground lens at the same time. Both the source and lens light have a non-analytic form, allowing for the flexibility needed in the inversion to represent arbitrarily small  and complex luminous structures in the lens and source. in addition, sparse regularisation avoids over-fitting the data and does not require the use of any adaptive mesh or pixel grid. As a consequence, our reconstructed sources can be represented on a grid of very small pixels. Sparse regularisation in the wavelet domain also allows for automated computation of the regularisation parameter, thus minimising the impact of arbitrary choice of initial parameters. Our inversion technique for a fixed mass distribution can be incorporated in future lens modelling technique iterating over the lens mass parameters. 
The python package corresponding to the algorithms described in this article can be downloaded via the github platform at \url{https://github.com/herjy/SLIT}
}



\maketitle

\section{Introduction}
\label{sec:Intro}

Strong gravitational lensing is a powerful tool to study astrophysics and cosmology, from structures at a wide range of scales to probing dark matter and dark energy. At cluster scale, strong gravitational lensing allows us to map the distribution of dark matter and to compare it with the visible mass \cite[e.g. see recent work in the Hubble Frontier Fields][]{Priewe2017, Harvey2016, Sebesta2016, Diego2016}. In merging or interacting galaxy clusters, it is used to infer a limit on the dark matter cross-section \cite[e.g.][]{Harvey2015}. At galaxy scale, strong lensing is a powerful way of studying the interplay between visible and dark matter, i.e. to study galaxy formation and evolution. The growing sample of known galaxy-scale strong lenses includes early-type galaxies \citep[SLACS;][]{Bolton2008}, spiral galaxies \citep[SWELLS;][]{Treu2011}, emission lines galaxies \cite[BELLS;][]{Bolton2012}, groups of galaxies \citep{More2012} and even lensing by AGNs \citep{Courbin2012}. Strong lensing is very sensitive to small substructures in the lensing galaxy and consists in a unique way of detecting such small structures, hinting at the nature of dark matter and the cosmological model \cite[e.g.][]{Vegetti2010}. Finally, the measurement of the so-called time delay between the images of strongly lensed quasars \citep{Refsdal1964} allows to constrain cosmology independently of any other cosmological probes \citep[for a review see][]{TM2016}, with high sensitivity to $H_0$ and little dependence on the other cosmological parameters \cite[e.g.][]{Suyu2010, Suyu2014, Tewes2013, Bonvin2017}.   

Most of these applications require reliable mass models for the lens, which, in turn, requires robust methods to unlens the original source, whatever the level of complexity in the shape may be. They also require to deblend the light of the foreground lens from the background source either prior to the mass modelling or together with it, in a single-step process. Different methods exist to carry out these tasks, all with their own assumptions, advantages and drawbacks. In the case of cluster-scale lensing, the positions of multiply imaged background sources give sufficient constraints on the mass macro-models. However, Additional information about the distortion of extended images might allow to break certain degeneracies in the cluster lens models, while, allowing us to recover the morphologies of highly magnified sources to probe galaxy evolution at high redshift, as recently illustrated in \cite{Cava2018}.

With the quality of modern lens observations, taken with the HST or ground-based Adaptive Optics, it becomes increasingly important to design modelling techniques, both for galaxies and clusters, that are both flexible and robust: they must simultaneously capture all possible observational constraints and remain robust with respect to noise. This is particularly important to explore the degeneracies inherent to the lensing phenomenon \citep[e.g.][]{SS2013, SS2014} without applying strong priors both on the source and lens shape. 

Some of the modelling techniques currently in use consider a full analytical lens mass and light distribution \citep{Lenstool, Bellagamba_2017,Oguri2010}. Others use a semi-analytic approach, where the source is pixelated and regularised but where the lens has an analytical representation \citep{Dye2005, Warren2003,Suyu2006,Vegetti2009} or where the lens is represented on a pixelated grid with regularisation or assumptions on its symmetry \citep[e.g.][]{Coles2014, Nightingale2018}. Further work in this direction involve adaptive pixel grids to represent the source \citep{Nightingale2018,Nightingale2014} or an analytic decomposition of the source on a predefined dictionary  as  was done in \cite{Birrer2015}, where the authors used shapelets \citep{Refregier2003}. In the following, a dictionary is  a collection of atoms (vectors) that, together, form a generative set of $\mathbb{R}^{n}$, with $n$ the number of samples in the data. 

In the present work we address the problems of source reconstruction and deblending as a single linear inverse problem. By using a family of functions called starlet \citep[alt. undecimated isotropic wavelet transform][]{Starck2005}, we are able to use sparse regularisation over the lens and source light profiles. Sparsity with starlets has the advantage of performing model-independent reconstructions of smooth profiles and allows for deterministic expression of the regularisation parameter. Because the lensed source can be represented using only a limited number of starlet coefficients, the pixel grid can be almost as fine as desired and the reconstructed source is denoised and deconvolved from the instrumental Point Spread Function (PSF). 

The choice of wavelets is motivated by the successful use of this decomposition in recent years to model astronomical objects \citep{Ngole2015, Garsden2015, Lanusse2016, Farrens2017, joseph2016, Livermore2017,Pratley2016}. In the present application, the separation between lens and source light is performed through morphological component analysis \citep{Starck2005,starckbook2015}. The method relies on the sparsity of the lens and source light profiles in their respective dictionaries to perform the separation.We apply sparse regularisation to the source and lens light distributions and illustrate with numerical simulations the effectiveness of the method at reconstructing the source and deblending lens and source.

This paper is intended as a proof of concept to show how convex optimisation under a sparsity prior for the source light profile can be used as an adequate minimisation technique for lens modelling. The scope of this paper remains limited to the modelling of light distribution alone and to the potential of using morphological component analysis to provide a new framework for lens modelling. Our ultimate goal is to come up with a full lens modelling technique that would perform joint free-form modelling of the light and mass density profiles. The latter being a non-linear problem and the problem of free-form lens modelling as a whole being largely underconstrained and degenerate, this will be the subject of another paper.

The paper is organised as follows: Sec.~\ref{sec:lensing} introduces the basics of strong gravitational lensing; in Sec.~\ref{sec:problem} we describe the problem of lens/source light profile estimation as a linear inverse problem; in Sec.~\ref{sec:method} we present the detail of our method; Sec.~\ref{sec:numexp} describes the testing of our method on simulated images and the comparison with a state-of-the-art technique: {\tt lenstronomy} \citep{Birrer2018}. Sec.~\ref{sec:Repro} details the content of the public package we produced to release our code. 

\section{Source reconstruction given a known lens mass}
\label{sec:lensing}

In this section, we give the basics of the gravitational lensing formalism we use to back-project lensed images to the source plane.

We note $\pmb{\theta}$ the angular position on the sky of an object seen through a gravitational lens (image plane coordinates), with intrinsic angular position $\pmb{\beta}$ (source plane coordinates). The mapping from source to image plane is described by the lens equation:

\begin{equation}
\pmb{\beta} = \pmb{\theta}-\pmb{\alpha}(\pmb{\theta}),  \label{eq:mapping}
\end{equation} 
where:
\begin{equation}
\pmb{\alpha}(\pmb{\theta}) = \frac{1}{\pi}\int_{\mathbb{R}^2}\kappa(\pmb{\theta}^\prime)\frac{\pmb{\theta} - \pmb{\theta}^\prime}{|\pmb{\theta}- \pmb{\theta}^\prime |^2}{d}^{2}{\pmb{\theta}}^\prime,
\end{equation}
is the deflection angle in the lens plane.The value $\kappa(\theta)$ is the dimensionless convergence of the gravitational lens at position $\theta$. Convergence $\kappa$ is defined as:
\begin{equation}
\kappa = \frac{\Sigma(D_L \pmb{\theta})}{\Sigma_c},    
\end{equation}
where $\Sigma_c$ is the critical surface mass density defined as:
\begin{equation}
    \Sigma_c = \frac{c^2}{4\pi G}\frac{D_S}{D_L D_{LS}},
\end{equation}
and $\Sigma(D_L \pmb{\theta})$ is the surface mass density of the lens defined as the projected volume density of the lens on a plane perpendicular to the line of sight. Parameters $D_S$, $D_L$ and $D_LS$ are the angular diameter distances between the observer and the source, the observer and the lens, and between the lens and the source respectively.

The problem of inverting Eq.~\ref{eq:mapping} from photometric observations only (meaning only $\pmb{\theta}$ is known) is a non-linear and highly under-constrained problem with two unknowns: The source's position, $\pmb{\beta}$, and the convergence map of the lens. In the case of extended sources, the goal of lens inversion is to recover the light profile of a lensed galaxy as seen in the source plane, which implies being able to calculate the flux at each position $\pmb{\beta}$ knowing the flux at position $\pmb{\theta}$. 

In practice, current techniques for lens inversion rely on an iterative process that consists in successively reconstructing the source profile brightness and the $\kappa$ map. In an effort of developing an automated, model independent method for lens inversion, we choose to decompose the problem. In this paper, we address the problem of source light profile reconstruction given a known convergence map, in which case, the problem is linear. In \cite{Warren2003}, the authors express the mapping between source and image surface brightness using an operator $F_{\kappa}$, such that the observed surface brightness of a lensed galaxy can be written:

\begin{equation}
Y = F_{\kappa}S+Z \label{eq:lensing},
\end{equation}
where $Y$ is the observed surface brightness, flattened as a vector with length $N_{pix}$. Vector $S$ is the unknown source surface brightness vector in the source plane with length $N_{ps}$. $F_{\kappa}$ is a $N_{ps}\times N_{pix}$ matrix where, following \cite{Warren2003}'s formalism, element $f_{i,j}$ is the $j$th pixel in image plane of the mapping of a source that has only its $i$th pixel set to one. In other words, $F_{\kappa}$ indicates which pixels from the source plane have to be combined to predict the value of a pixel in the image plane. The elements of $F_{\kappa}$ are entirely determined by the mass density distribution $\kappa$. Vector $Z$ is an additive noise map. In this work, we consider $Z$ as a white Gaussian noise with standard deviation $\sigma$, but the method can easily be extended to Poisson statistics, or more generally to noise with known root mean square.

\subsection{Pixel-to-pixel mapping}

As illustrated in Fig~1 of ~\cite{Wayth2006}, a square pixel in the image plane is a diamond shaped pixel in source plane, with a total area depending on the magnification at the pixel location. Although this phenomenon should strictly be accounted for, we choose to make the approximation that each photon hitting a pixel whether in source or image plane hits at the centre of the pixel. This way, we lose part of the information provided by the distribution of photons over the whole surface of a pixel, but this approximation has the advantage of avoiding correlating the noise when back projecting from the image plane to the source plane. Furthermore, using a small pixel size limits this imperfect modelling and allows to compensate for the variation in the light profile. This inverse problem being ill-posed, it admits no unique and stable solution, hence calling for a regularisation to solve it. Increasing the pixel size, and therefore losing resolution can be seen as a naive regularisation. We will show how advanced regularisation techniques can be used efficiently.

\subsection{Projection and back-projection between source and lens planes}
\label{projection}

To compute the elements of $F_{\kappa}$ for each of the $N_{ps}$ pixels in the source plane, we associate a pixel in the image plane by shooting a photon from the centre, $\pmb{\beta}$ of a source plane pixel and recording the position(s) $\pmb{\theta}$, given by Eq.~\ref{eq:mapping}, of the pixel(s) where the deflected photon will hit the image plane. This boils down to recording the positions $\pmb{\theta}$ where $\pmb{\beta}+\pmb{\alpha}(\pmb{\theta})-\pmb{\theta} = 0$. The element of $F_{\kappa}$ at position(s) $(\beta,\theta)$ is(-are in the case of a multiply imaged pixel) set to one to indicate the mapping. Since $F_{\kappa}$ is a sparse matrix with only very few non-zero coefficients, we choose to only store the positions $(\beta,\theta)$ that map into one another in order to save memory and hence, computation time in the following steps.

The projection of a source light profile into a lensed light profile in the image plane is then performed by allocating to each image pixel the sum of the intensities of the corresponding source pixels according to $F_{\kappa}$. Conversely, back-projection is performed by allocating to each source pixel the average value of all its lensed counterparts according to $F_{\kappa}$. This ensures conservation of surface brightness between the source and lens planes.

\section{Linear development of strong gravitational lens imaging}
\label{sec:problem}

In real imaging data of strong gravitational lenses, the problem of finding the delensed light profile of a lensed galaxy is harder than solving Eq.~\ref{eq:lensing}, which is already non-trivial. First, one has to include the impulse response of the instrument that acquired the image. This effect corresponds to a convolution of the images described by $F_{\kappa}S$, by the point spread function (PSF). Let the linear operator $H$ account for the convolution by a known PSF, Eq.~\ref{eq:lensing} becomes:

\begin{equation}
Y = HF_{\kappa}S+Z, \label{eq:Lens_PSF}
\end{equation}
which is the problem one has to solve when dealing only with the lensed light profile of a source, assuming that the light profile from the foreground lens galaxy has been perfectly removed prior to the analysis.

In practice, images of strongly lensed galaxies are contaminated by light from a foreground lens galaxy, $G$. Taking this into account, Eq.~\ref{eq:Lens_PSF} then writes:

\begin{equation}
Y = H(F_{\kappa}S+G)+Z. \label{eq:Lens_G_PSF}
\end{equation}
When $Z$ is a white Gaussian noise, solving Eq.~\ref{eq:Lens_G_PSF} reduces to finding $S$ and $G$ such that:
\begin{equation}
||Y-H(F_{\kappa}S+G)||_2^2 < \epsilon, \label{eq:min_epsi}
\end{equation}
Where $\epsilon$ accounts for the precision of the reconstruction and depends on the noise level. Given that we have an $N_{pix}$-sized image and aim at finding an $N_{pix}$-sized galaxy light profile and an $N_{ps}$-sized source light profile, we need to impose further constraints on these unknowns. Classically, the light distribution of the lens is approximated by an analytic profile such as a S\'{e}rsic or deVaucouleur profile. 

Reconstructing the source light profile being an ill-posed problem , where unknowns largely outnumber the number of observables, several strategies have been investigated in the literature, for instance: adaptive pixel grids \citep{Dye2005}, negentropy minimisation \citep{Warren2003, Wayth2006}, Bayesian inference over the regularisation parameters of the source \citep{Suyu2006},  perturbative theory \citep{Alard2009} or model profile fitting \citep{Bellagamba_2017}. Although these methods have their own advantages and disadvantages, only few of them are able to reconstruct complex sources without degrading the resolution of the output. In \cite{Birrer2015}, the authors used a family of functions to reconstruct the source light profile with promising results. Here we propose to push this idea further by exploiting a family of functions that is well suited to represent galaxies, and that possesses properties of redundancy allowing for the use of sparse regularisation.

\subsection{Source reconstruction in absence of light from the lens}
\label{sec:FS}
 We propose a new approach to solve Eq.~\ref{eq:Lens_PSF}. Given that galaxies are compact and smooth objects, their decomposition over the starlet dictionary \citep{Starck2007} will be sparse, meaning that only a small number of non-zero starlet coefficients will contain all the information in a galaxy image. This property allows to constrain the number of coefficients used in starlet space to reconstruct galaxy profiles, therefore offering a powerful regularisation to our problem. Starlets are a class of discrete undecimated, isotropic wavelets, formally introduced in \citep{Starck2007} and extensively described with regard to computation algorithms in \cite{starckbook2015}.

\subsubsection{Sparse regularisation}

Assuming a signal is sparse in a dictionary $\Phi$,  the solution to an inverse problem like in Eq.~\ref{eq:Lens_PSF} is the solution that uses the least number of coefficients in the $\Phi$ dictionary while minimising the square error between the observables and the reconstruction. In a more formal way, sparsity is enforced by minimising the $\ell_1$-norm of the decomposition over $\Phi$ of a signal known to be sparse in this dictionary. In addition, because the mapping of an image from lens plane to source plane does not conserve shapes, the edges of the image in lens plane does not match the borders of the image in the source plane, leaving parts of the source image unconstrained as they map into pixels outside the field of view of the lens plane. Let us call $\mathbb{S}$ the set of pixels in source plane that have an image in lens plane. We impose that the coefficients of the solution outside set $\mathbb{S}$ be set to zero. This allows us to write the problem of finding $S$ as an optimisation problem of the form:
 
 \begin{equation}
 \underset{\alpha_S}{argmin} ||Y-HF_{\kappa}\Phi \alpha_S||_2^2 + \lambda||W \odot \alpha_S||_1, \label{eq:L1}
 \end{equation}
where $\Phi$ is the starlet dictionary, $\alpha_S$ are the starlet coefficients of $S$ such that $\alpha_S = \Phi^T S$. The operator $\odot$ is the term by term multiplication operator, and $W$ is a vector of weights that serves the purpose of setting to zero all coefficients outside $\mathbb{S}$, while keeping the $\ell_1$-norm constraint from biasing the results (more on that in the following paragraphs). In practice, minimising the $\ell_1$-norm of a vector is done by soft-thresholding the vector. this consists in decreasing by a positive factor $\lambda$ the absolute value of all its coefficients and by setting to zero the coefficients smaller than $\lambda$, as shown in the following equation: 

\begin{equation}
 ST_{\lambda}(x) =
  \begin{cases}
    sign(x)\times(|x|-\lambda) & \quad \text{if } |x|>\lambda\\
    0  & \quad \text{otherwise}\\
  \end{cases}
\label{eq:ST}
\end{equation}
The regularisation parameter $\lambda$ controls the trade-off between fitting the observed data and enforcing sparse solutions. From the definition of Eq.~\ref{eq:ST}, it appears that solutions derived with soft-thresholding will present a bias due to the subtraction by $\lambda$. In order to mitigate this effect, we use the reweighting scheme from \cite{Candes2008}. In order to prevent the most significant coefficients from being truncated, we multiply the regularisation parameter $\lambda$ by:
\begin{equation}
W = \frac{2}{1+exp(-10(\lambda-\alpha_0))}, \label{eq:weight}
\end{equation}
where $\alpha_0$ is the solution of Eq.~\ref{eq:L1} with $W=1$. With this definition for W, the coefficients that are much larger than $\lambda$ are less affected by soft thresholding than others. Values of $W$ for coefficients outside $\mathbb{S}$ are set to infinity, hence naturally ensuring that the corresponding wavelet coefficients are set to zero. 

This approach can be used to recover the source light profile in systems involving a faint foreground lens galaxy, a large Einstein radius, or when a reliable deblending of the lens and source light profiles is available prior to the source reconstruction scheme presented here.

\subsection{Source reconstruction and deblending of the foreground lens light profile}
\label{sec:separability}

In a more general case, one has to deal with the separation between the lens and source light profiles. Although several techniques allow for their separation prior to the analysis of the lens system \citep[e.g.][]{Joseph2014,joseph2016,Brault2015}, each of these methods have limitations, in the sense that they require specific inputs (field of view, or multiband images) or do not take into account the lensed source profile when fitting the lens, resulting in potential biases. Another approach consists in fitting an analytic lens light profile while reconstructing the lens mass density profile and the source \citep{Birrer2015, Tessore2016}. Here, we propose a solution to reconstruct and separate the lens and source light profiles using the fully linear framework provided by morphological component analysis \citep[henceforth, MCA,][]{Starck2005}. 

Very importantly, a galaxy is sparse in starlets in its own plane (source or image), meaning that, given a mapping $F_{\kappa}$, with $\kappa(\theta)>1$ at several positions $\theta$, between source and lens plane, a galaxy in the source plane and a galaxy in lens plane are both sparse in starlets. We justify and illustrate this statement with simulations in Sec.~\ref{sec:numexp} using simulated lenses. 

Morphological component analysis allows for separation of two mixed components in a signal, based on the fact that each component can be sparsely represented in its own dictionary but not in others. In the context of lens source separation, the explicit dictionaries are the starlet transform of a back-projection in source plane on one hand and the starlet transform for the lens, in lens plane, on the other hand.

We can therefore iteratively project the mixed signals in their own respective dictionaries, impose a sparsity constraint on each projection and therefore reconstruct the corresponding components separately. As seen in Sec.~\ref{sec:FS}, sparsity is imposed by minimising the $\ell_1$-norm of both decompositions. Due to the problem of reconstructing $S$ being ill-posed, a sparse version of $S$ has to be computed at each iteration, which requires subiterations.

\subsection{Optimisation problem}

In mathematical terms, the aforementioned MCA problem boils down to finding the model $\{\hat{S},\hat{G}\}$ that provides the best approximation of the data set $Y$ according to Eq.~\ref{eq:Lens_G_PSF}, while minimising the $\ell_1$-norms of the starlet coefficients $\alpha_S$, and $\alpha_G$, with $\alpha_G= \Phi^T G$. This writes:
    
\begin{eqnarray}
\hat{\alpha_S},\hat{\alpha_G} = \underset{\alpha_S, \alpha_G}{argmin} && ||Y-H(F\Phi\alpha_S-\Phi\alpha_G)||_2^2 \label{eq:MCA} \\
&& +\lambda_S||W_S\odot\alpha_S||_1  +\lambda_G||W_G\odot\alpha_G||_1, \nonumber
\end{eqnarray}
Here, $\lambda_S$ and $\lambda_G$ account for the sparsity of $\alpha_S$ and $\alpha_G$ respectively. Similarly to Eq.~\ref{eq:L1}, $W_S$ and $W_G$ are weights that play the same role as in Eq.~\ref{eq:weight}.We describe the calculation of these values in Sec.~\ref{sec:SLIT}.

Since, our main interest is to fully reconstruct the source $S$, it is not necessary to compute the fully deconvolved vector $G$. Instead, we limit ourselves to estimating the convolved vector $G_H = HG$ so that we extract the convolved foreground lens galaxy $G_H$ and decrease the computational time by avoiding several convolution steps of over $G$ when solving Eq.~~\ref{eq:MCA}, which becomes:

 \begin{eqnarray}
\hat{\alpha}_{S},\hat{\alpha}_{G_H} = \underset{\alpha_S, \alpha_{G_H}}{argmin} && ||Y-HF\Phi\alpha_S-\Phi\alpha_{G_H}||_2^2 \\ &&+\lambda_S|| \alpha_S||_1 \label{eq:SLIT_MCA} 
 +{\lambda_G}_H||\alpha_{G_H}||_1. \nonumber 
\end{eqnarray}


\begin{figure*}[t!]
\begin{tabular}{ccccc}
\includegraphics[trim = 5cm 1cm 4cm 1cm, clip = true, scale = 0.25]{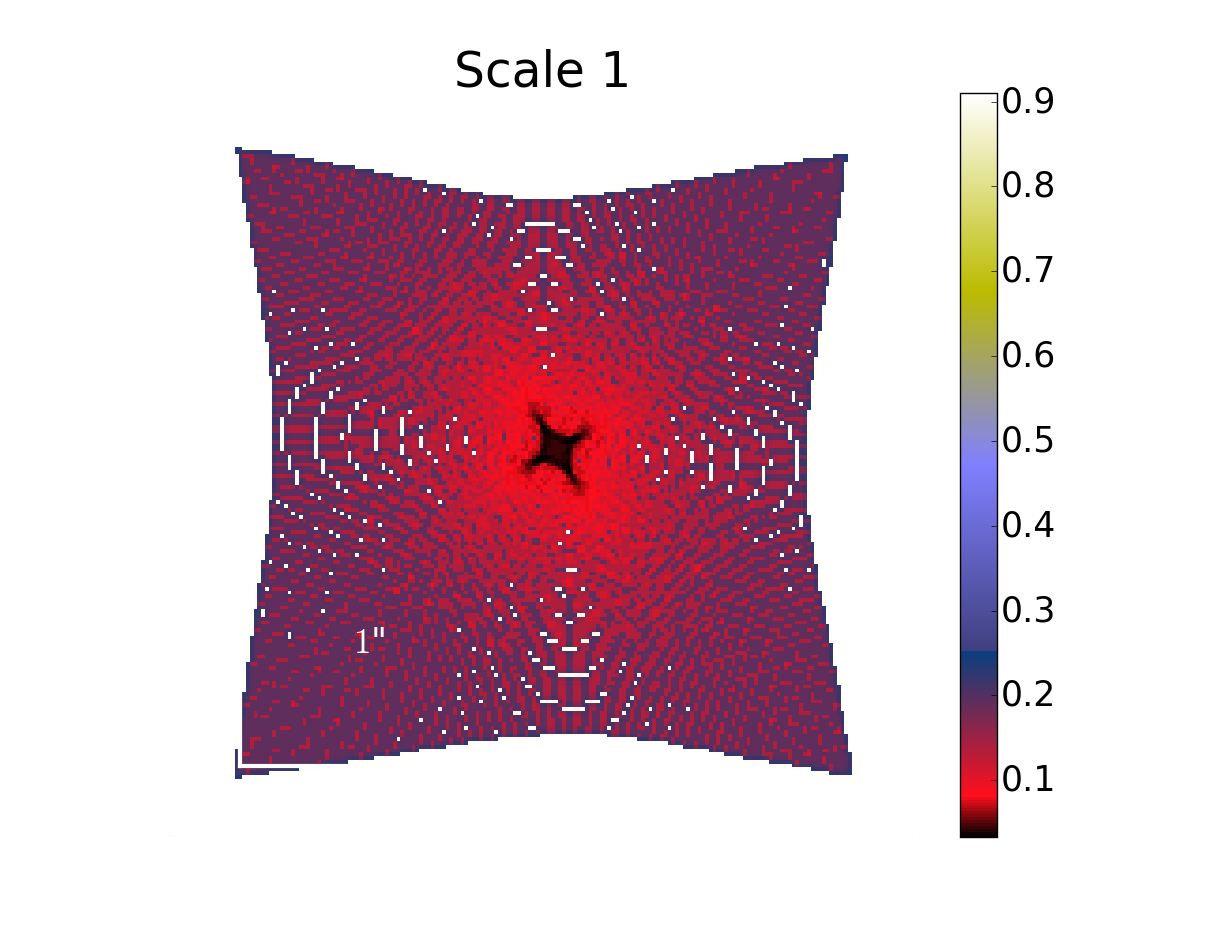}&
\includegraphics[trim = 5cm 1cm 3cm 1cm, clip = true, scale = 0.25]{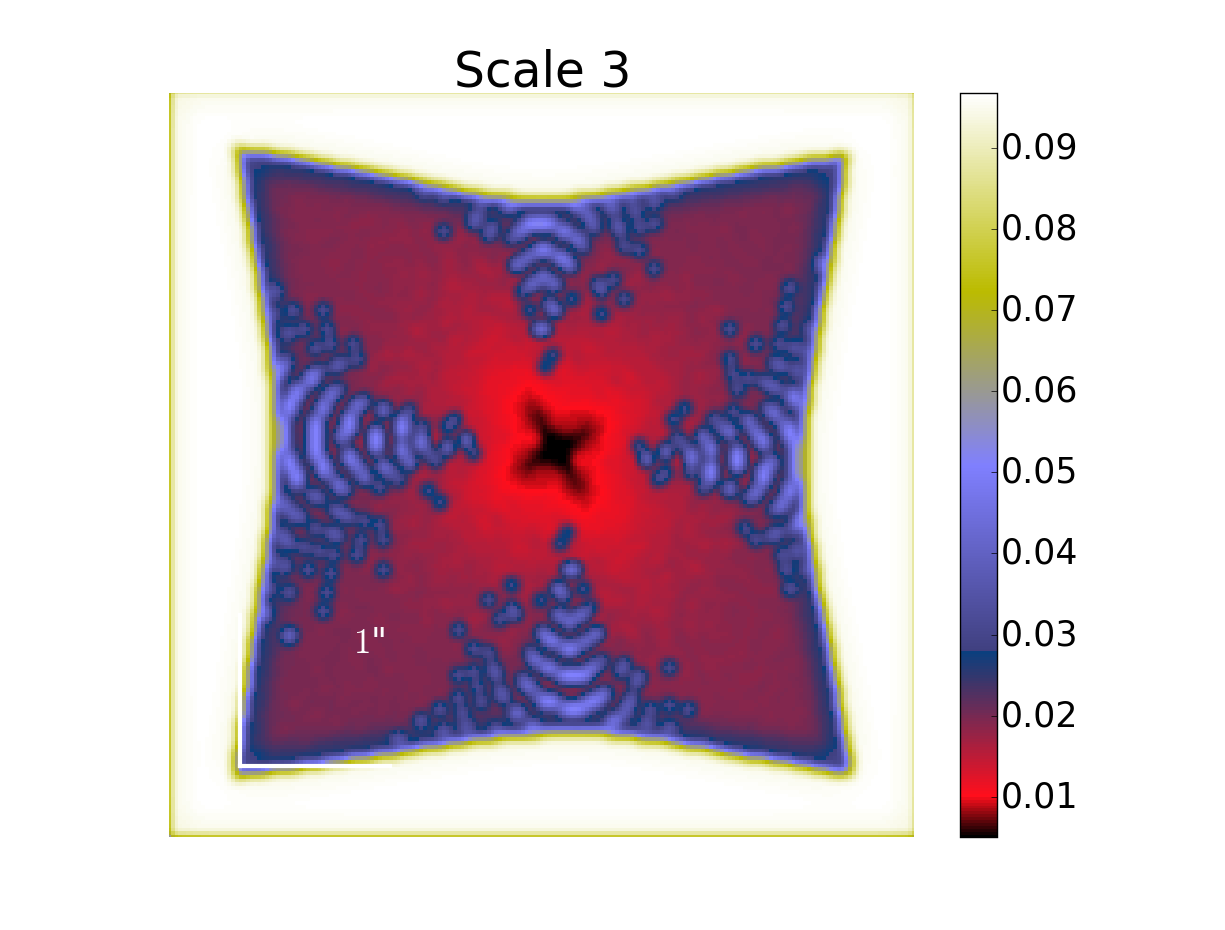}&
\includegraphics[trim = 5cm 1cm 2cm 1cm, clip = true, scale = 0.25]{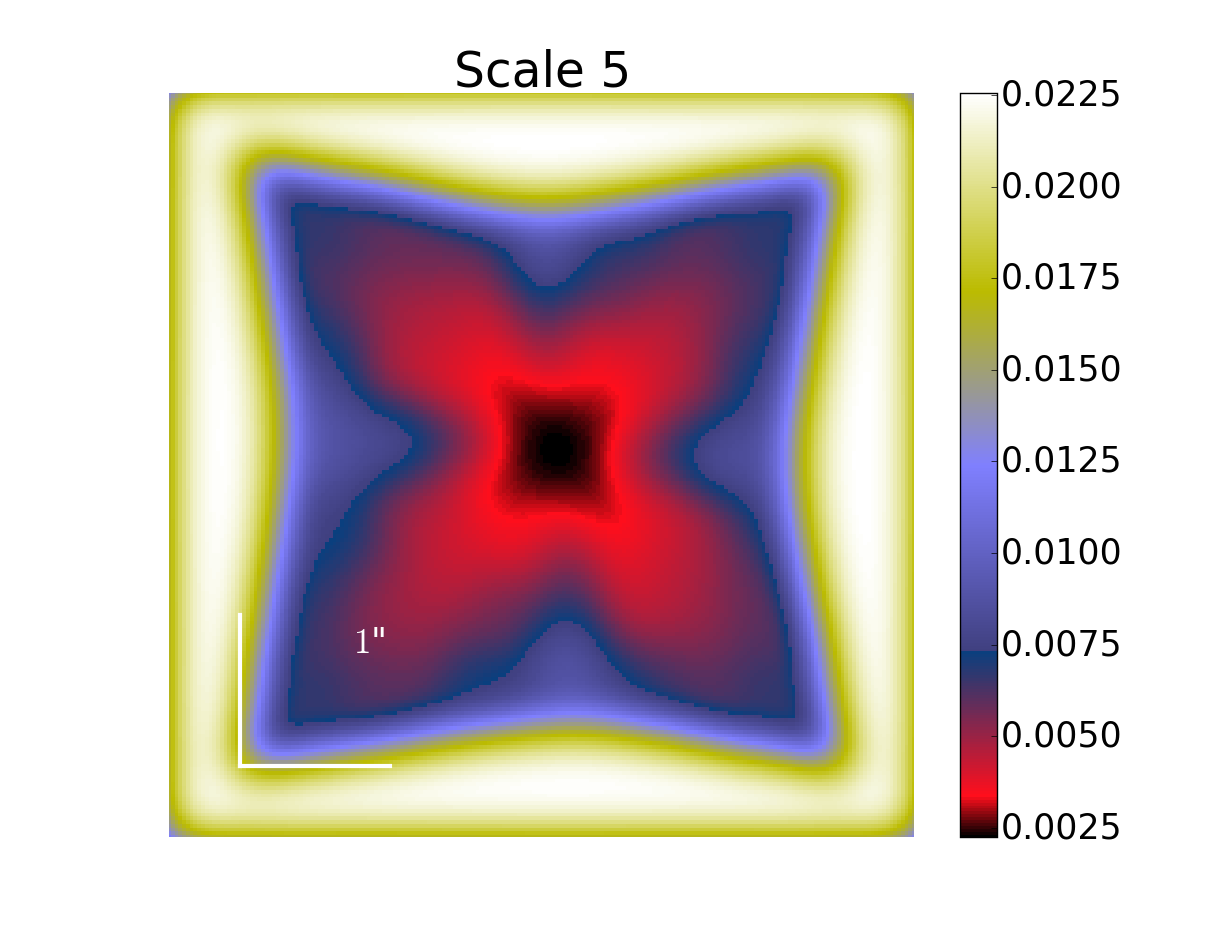}
\end{tabular}
\caption{Noise levels in the source plane ($\lambda_S$)for three starlet scales (scale 1, 3 and 5) out the five computed for $100\times 100$ pixels images with noise standard deviation $\sigma = 1$.   \label{fig:scales}}
\end{figure*}

\section{Method: the {\tt SLIT} algorithms}
\label{sec:method}

In this section, we describe the two algorithms, {\tt SLIT} and {\tt SLIT\_MCA} that we implemented to solve Eq.~\ref{eq:L1} (no lens light) and Eq.~\ref{eq:SLIT_MCA} (full light reconstruction problem) respectively.

\subsection{Source delensing: {\tt SLIT} algorithm}
\label{sec:SLIT}

Starting with the simpler case of solving Eq.~\ref{eq:Lens_PSF}, we made use of the fast iterative soft thresholding algorithm \mbox{\citep[FISTA][]{beck2009}}. This iterative algorithm is similar to a forward backward \citep{Gabay1983} algorithm with an inertial step that accelerates the convergence. We show the pseudo-algorithm for one iteration in Algo.~\ref{algo:FISTA}. It consists in a step of gradient descent (steps \ref{FISTA:state1} and \ref{FISTA:state2}) followed by a soft-thresholding of the starlet coefficients of the source (step \ref{FISTA:state3}), which acts as a sparse regularisation. Step \ref{FISTA:state5} aims at pushing forward the solution in the direction of smaller error, which accelerates the convergence. The process is repeated until convergence, as shown in Algo.~\ref{algo:SLIT}.

 \begin{algorithm}
 \begin{algorithmic}[1]
 \Procedure{FISTA}{$Y,\alpha_{i},H,F,\xi_i, t_{i}, \lambda, W$}
 
 \State $R \gets  F^{T}H^{T}(Y-HF\Phi \alpha_{i})$ \label{FISTA:state1}
 \State $\gamma \gets \xi_{i}+\mu\Phi^{T} R$ \label{FISTA:state2}
 \State $\alpha_{i+1} \gets ST_{\lambda\odot W}(\gamma)$ \label{FISTA:state3}
 \State $t_{i+1} \gets \frac{1+\sqrt{1+4{t_{S}}_{i-1}^2}}{2}$ \label{FISTA:state4}
 \State $\xi_{i+1} \gets \alpha_{i}+\frac{t_{i}-1}{t_{i+1}}(\alpha_{i+1}-\alpha_{i})$ \label{FISTA:state5}

 \State \textbf{return} $\xi_{i+1}, \alpha_{i+1}, t_{i+1}$ 
 \EndProcedure
 \end{algorithmic}
 \caption{FISTA iteration}
 \label{algo:FISTA}
\end{algorithm}

In this algorithm, $Y$ is the original image of a lensed galaxy, $\xi_i$ and $\gamma$ are local variables used to carry local estimates from one iteration to another, $\alpha_{i}$ is the starlet decomposition of the estimated source at iteration $i$ and $t_{i}$ gives the size of the inertial step. This sequence has been chosen to ensure that the cost function convergence is bounded by the Euclidean distance between the starting point for $S$ and a minimum of the cost function. This is explained in more details in   \cite{Chambolle2015}  (thm. 1) and \cite{beck2009} (thm 4.1). The gradient step $\mu$ is chosen to be $\mu = (||HF\Phi||_s^2)^{-1}$, with $||.||_s$ being the spectral norm of a matrix, defined by: 
\begin{equation}
||M||_s = \underset{x\neq 0}{max} \frac{||Mx||_2}{||x||_2}.
\end{equation}
Function $ST_\lambda$ is the soft-thresholding operator described by Eq.~\ref{eq:ST}.

 \begin{algorithm}
 \begin{algorithmic}[1]
 \Procedure{SLIT}{$Y,H,F,\lambda, Niter, W$}
 \State $\xi_{0}, \alpha_{0}, t_{0} \gets 0,0,1$
 \For{$0 < i \leq N_{iter}$}
 	\State $\xi_{i+1}, \alpha_{i+1}, t_{i+1} \gets FISTA(Y,\alpha_{i},H,F,\xi_i, t_{i}, \lambda, W)$
\EndFor
\State $S \gets \Phi \alpha_{N_{iter}}$
\State $W \gets \frac{2}{1+exp(-10(\lambda-\alpha_0))}, \label{eq:weight_algo}$
 \State \textbf{return} $S, W$ 
 \EndProcedure
 \end{algorithmic}
 \caption{SLIT}
 \label{algo:SLIT}
\end{algorithm}

Parameter $\lambda$ has to be chosen with care as it accounts for the sparsity of the solution. In practice, $\lambda$ is a threshold that is applied to each starlet coefficient of the solution in order to reduce its $\ell_1$-norm in starlet space. In the present case, given the presence of noise in the input data $Y$, it is important to choose a threshold above the noise levels. This is done by propagating the noise levels in image $Y$ to the starlet coefficients  $\alpha_{i}$. The starlet transform being an undecimated multi-scale transform, coefficients $\alpha_{i}$ can be ordered as a set of images, each image representing the variations in the data at different scales. Therefore, we have to estimate how noise levels translate from the data to each scale of the starlet transform. In the current implementation, the noise from image $Y$ also has to be propagated through the $H^T$ and $F^T$  operators as shown in step \ref{FISTA:state1} of Algo.~\ref{algo:FISTA}. Because the convolution $H^{T}$ correlates the noise in the data and the back-projection to source plane induces varying multiplicity of the delensed pixels across the field of view, it is necessary to estimate a different threshold $\lambda$ at each pixel location in each scale of the starlet transform of the source. 
In practice, for measurements affected by noise with known covariance $\Sigma$, noise standard deviation in the starlet domain of the source plane are given by the square root of the diagonal elements of 
\begin{equation}
\Sigma_S = \Phi^T F_{\kappa}^TH^T\Sigma H F_{\kappa} \Phi \label{eq:cov}
\end{equation}
In the case of the "a trou" \cite[french for "with whole",][]{holschneider1989,Shensa1992} algorithm, which relies on filter bank convolution to perform the starlet transform, the elements of $\Phi$ are never explicitly calculated. Instead, the noise standard deviation at scale $s$ and pixel $p$ in the starlet domain of the source plane is given by:
\begin{equation}
\Xi_{s,p}^2 = \Delta_{s}^2*\Gamma_{i,j, (i=j)}^2, \label{eq:cov2}
\end{equation}
Where $\Delta_s$ is the starlet transform of a dirac function at scale $s$, $*$ is the convolution operator, and $\Gamma_{i,j, \{i=j\}}$ is the vector containing the diagonal elements of $F_{\kappa}^TH^T\Sigma H F_{\kappa}$.

The result is the noise level in source space at each location and each scale of the starlet transform. By construction, the last scale contains the coarse details in the image and is left untouched in the thresholding process. Fig.~\ref{fig:scales} shows the noise levels in the source plane calculated from a simulation where the surface mass density is a singular isothermal ellipsoid (SIE). The PSF is a simulated Tiny Tim PSF \citep{Krist2011}, for the F814W filter of the  ACS/WFC instrument on the HST and the noise standard deviation is set to $1$. The original image is a $100\times 100$ pixels image which is decomposed in $6$ starlet scales, i.e. the maximum number of scales that we can possibly compute, given the size of the image.

\subsection{Deblending and source delensing: {\tt SLIT\_MCA} algorithm}

In real data, the source and lens light profiles are blended, i.e. the light of the lens impacts the quality of the source reconstruction. Handling the deblending and delensing simultaneously can be done using MCA.

In classical source separation problems where two components are to be separated, solutions are obtained through MCA, by performing a gradient step and by alternatively regularising over each component in its own dictionary. In the present case, solving Eq.~\ref{eq:Lens_G_PSF} requires that we solve an inverse problem each time we aim at reconstructing the quantity $HF_{\kappa}S$. This inverse problem corresponds to solving Eq.~\ref{eq:Lens_PSF} for which we already presented a solution in Algo.~\ref{algo:SLIT}.

Our MCA algorithm is therefore an iterative process that consists in alternatively subtracting a previous estimation of $G_H$ and $S$ to the data: 
\begin{equation}
D_S = Y-G_H \label{eq:DS}
\end{equation}
and 
\begin{equation}
D_G = Y-HF_{\kappa}S, \label{eq:DG}
\end{equation}
 as detailed in Algo.~\ref{algo:MCA_SLIT}. At each iteration, the previous subtractions $D_S$ and $D_G$ are used to estimate $S$ and $G_H$ respectively. Estimating $S$ requires running the full {\tt SLIT} algorithm on $D_S$ until convergence. Estimating $G_H$ at a given iteration is simply done by running one single iteration of the FISTA algorithm on $D_G$ with inputs $F_{\kappa}$ and $H$ being identity matrices. We found empirically that using the projections of $S$ and $G_H$ on the subsets of vectors with positive values (meaning that we set to zero all negative coefficients), we could achieve faster convergence towards more realistic solutions. Although this is not a formal positivity constraint since we do not apply positivity on the solutions themselves, we found that in practice this leads to galaxy profiles with less negative structures which is not a physical feature we find in galaxy light profiles.

Estimating $\lambda_G$ is as crucial as estimating $\lambda_S$ but is much simpler given that there is no inverse problem to solve in this case. The threshold $\lambda_G$ only depends on the noise level in the image. Given that we impose sparsity in starlet space, we still have to evaluate noise levels at each scale of the starlet transform. To do so, we simply compute how a unitary signal in direct space translates into starlet space and multiply it by the noise standard deviation. In other words, we take the starlet transform of a Dirac function and compute the $2$-norm of each scale of the starlet transform. This tells us how energy is distributed into starlet space. For a decomposition over 6 starlet scales, the values we obtain for the first five scales in order of increasing scale are: $\lambda_{G} = [0.891,$ $0.200,$ $0.086,$ $0.041,$ $0.020]$. As in the previous section, the last scale is left untouched. The obtained values are then multiplied by a scalar that accounts for the desired detection level in units of noise. The scalar is often chosen to be between $3$ and $5\sigma$ as seen previously. A detection at $3\sigma$ will produce very complete but noisy reconstruction of the signal, when $5\sigma$ will lead to a more conservative reconstruction of the highest signal-to-noise ratio features only. The obtained thresholds are applied uniformly across each scale.

 \begin{algorithm}
 \begin{algorithmic}[1]
 \Procedure{SLIT\_MCA}{$Y,H,F_{\kappa},N_{iter}, N_{subiter}, \lambda_S, \lambda_G$}
 \State $\tilde{S} \gets 0$
 \State $\tilde{G}_H \gets 0$
 \State $[{\xi_S}_0, {\xi_G}_0] = [0,0]$
 \State $[{\alpha_S}_0, {\alpha_G}_0] = [0,0]$
 \State $[{t_S}_0, {t_G}_0] = [1,1]$
 \For{$0 < i \leq N_{iter}$}
	\State $D_S \gets Y-G_H$
	\State $S \gets SLIT(D_S, H, F_{\kappa}, \lambda_S, N_{subiter})$
    \State $D_G \gets Y-HF_{\kappa}S$
    \State ${\xi_G}_{i}, {\alpha_G}_{i}, {t_G}_{i}  \gets FISTA(D_G,{\alpha_G}_{i-1},I_d,I_d, {\xi_G}_{i-1}, {t_G}_{i-1}, \lambda_G, 1)$
    \State $G_H \gets \Phi^T {\alpha_G}_{i}$
  \EndFor
 \State \textbf{return} $\tilde{S}, G$ 
 \EndProcedure
 \end{algorithmic}
 \caption{{\tt SLIT} MCA algorithm}
 \label{algo:MCA_SLIT}
\end{algorithm}

\section{Numerical experiments with simulations}
\label{sec:numexp}

In the following we illustrate the performances of our algorithms with numerical experiments that mimic different observational situations. We also apply our algorithms to a set of simulated images and show comparisons of reconstructions with a state-of-the-art method: {\tt lenstronomy} \citep{Birrer2018}.

\subsection{Creating realistic simulated lenses}



In order to make the simulations as realistic as possible, we use galaxy light profiles extracted from deep HST/ACS images taken in the F814W filter. The images are part of the Hubble Frontier Fields program and the specific data we used\footnote{the frames were recovered from the HFF site at \url{http://www.stsci.edu/hst/campaigns/frontier-fields/FF-Data}} were taken from the galaxy cluster Abell 2744 \citep{Lotz2016}. We selected various patches, each containing a galaxy that we use to represent a lens or a source. The HFF images were cleaned using starlet filtering with a 5-sigma threshold. Source galaxies have been chosen to display visually apparent substructures with several modes, that we aim at recovering with our lens inversion methods. The lens light profile was chosen to present a smooth monomodal distribution, as expected for a typical massive early-type galaxy, e.g. like in the SLACS samples \citep{Bolton2008}.






To generate our simulations we then lens the sources following the recipe in Sec.~\ref{projection}, using various lens mass profiles. We then add the lens light and convolve it with a PSF created with the Tiny Tim software \citep{Krist2011} for the ACS/WFC and the F814W filter. 

The images shown in this section were created from images taken with the ACS/WFC instrument on HST. Flux units are showed in $e^-$ and pixels in image plane are 0.05 arc-seconds on-a-side.

\subsection{Plane-wise sparsity of galaxy light profiles}
\label{sect:NLA}
The {\tt MCA-SLIT} algorithm consists in projecting the mixture image of the lens galaxy and the lensed galaxy, back and forth between the source and lens plane, and thresholding the starlet coefficients of each projection. Our hypothesis is that the starlet thresholding favours, in each plane, the corresponding galaxy: source galaxy in source plane and lens galaxy in lens plane. This assumes that a lens galaxy in lens plane is well reconstructed with only a few starlets coefficients, while a source galaxy projected to lens plane is not. Conversely, it implies that a source galaxy in source plane is sparser in starlet domain than a lens galaxy projected to source plane. 

\begin{figure*}[t!]
\centering
\includegraphics[trim = 3cm 1cm 2.5cm 1cm, clip = true,scale = 0.27]{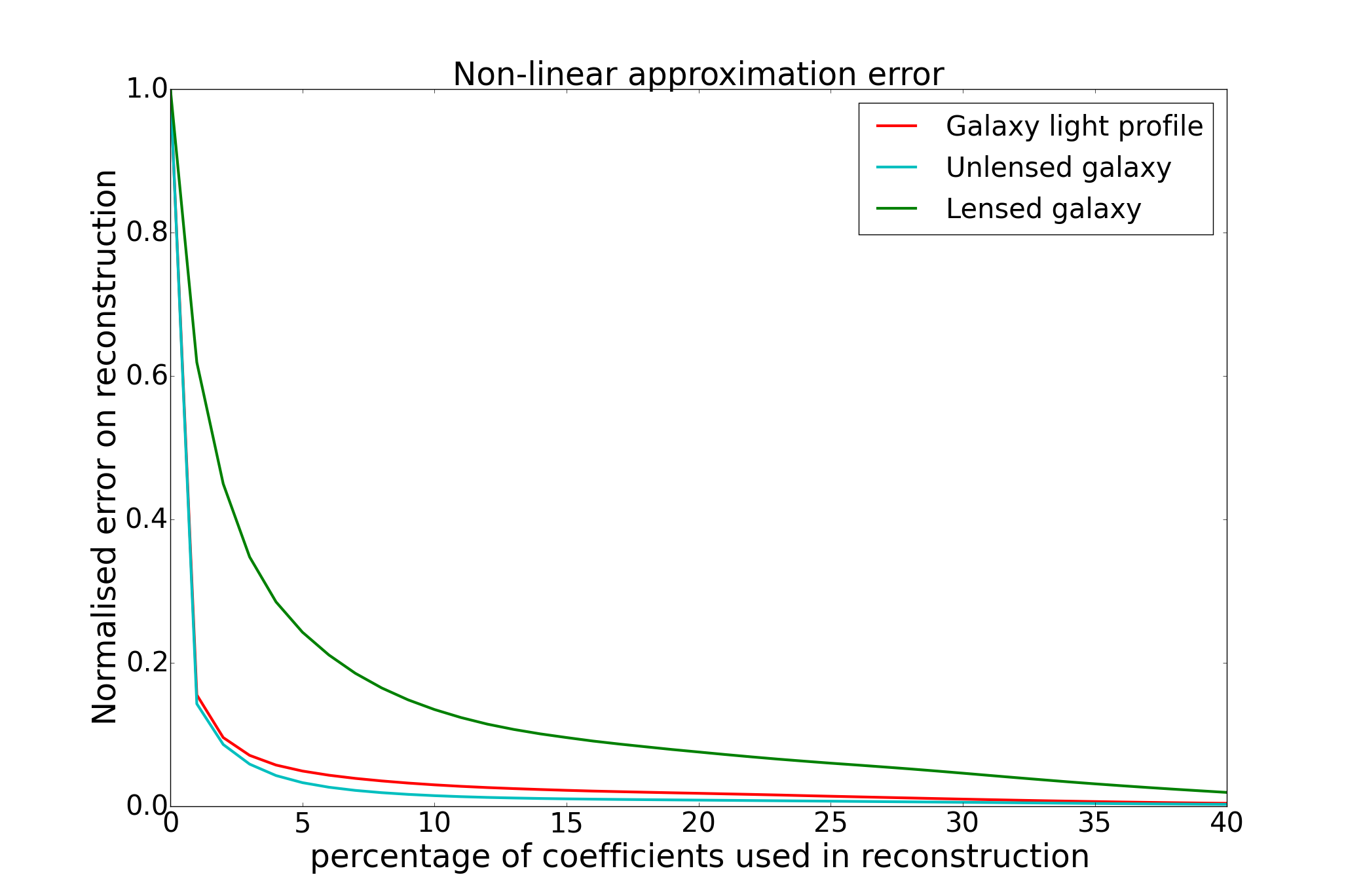} 
\caption{Normalised non-linear approximation (NLA; \ref{sect:NLA}) of galaxies projected in source and lens plane. The red curve stands for the average NLA of galaxy images that can be seen as source or lens galaxies. The cyan curve stands for the average NLA of the same galaxies once projected from lens to source plane. The green curve stands for the NLA of the same galaxies projected to lens plane.} \label{fig:NLE}
\end{figure*}

To verify this hypothesis, we selected 167 images of galaxies from cluster MACS J0717 from the HFF survey, in the F814W filter of the ACS/WFC camera of HST. We used the HFF-DeepSpace catalogue by \cite{Shipley2018} to select galaxies with a semi-major axis of at least 5 pixels and a flag at 0 in the F814W filter to insure that the galaxies are isolated in their stamp. We performed the starlet decomposition of these images along with their projections to source and lens plane using three different mass profiles (SIE, SIS and elliptical power law) with realistic draws of the lens parameters. For each of these starlet decompositions we set to zero the $p\%$ smallest coefficients and reconstruct the image in pixel space. We then compute the error on the reconstruction as a function of $p$. The resulting curve is the non-linear approximation error \citep[NLA; see][]{starckbook2015}, shown in Fig. \ref{fig:NLE}. The NLA can be used as a metric for the sparsity of a galaxy profile, in the sense that a sparse galaxy sees its NLA decreasing rapidly with $p$.

In Fig.~\ref{fig:NLE}, we see in particular that the NLA of galaxies (in red) decreases faster than the one of the lensed source (in green). This means that when keeping only a small percentage of the highest coefficients in starlets (e.g. 10\%) of the decomposition of an image $Y$, will reconstruct well the lens galaxy but not the lensed source. When comparing the NLA of a galaxy (red curve in Fig.~\ref{fig:NLE}) to that of its projection to source plane (cyan curve in Fig.~\ref{fig:NLE}), we see that the NLAs of both profiles are very similar, making it difficult to disentangle between them. In practice, the reconstructed source images can be contaminated with features belonging to projections of the lens galaxy if not converged properly. On the bright side, since lens galaxy light profiles are being reconstructed in lens plane very efficiently, the signal from lens galaxy in source plane decreases very rapidly with iterations. 

\begin{figure*}[h!]
\centering
    \includegraphics[trim = 5cm 2cm 3cm 0.5cm, clip = true,scale = 0.15]{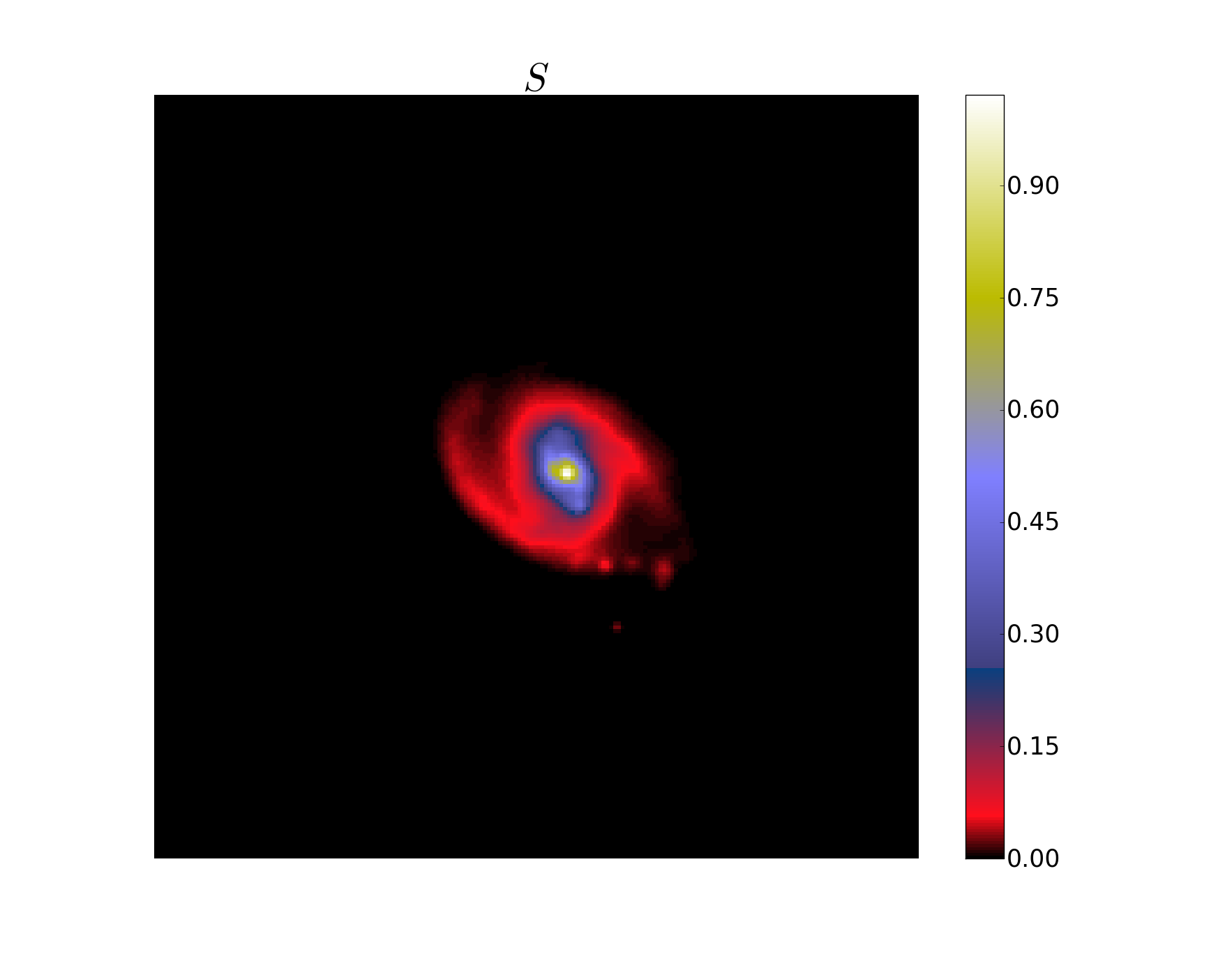}
    \includegraphics[trim = 5cm 2cm 3cm 0.5cm, clip = true,scale = 0.15]{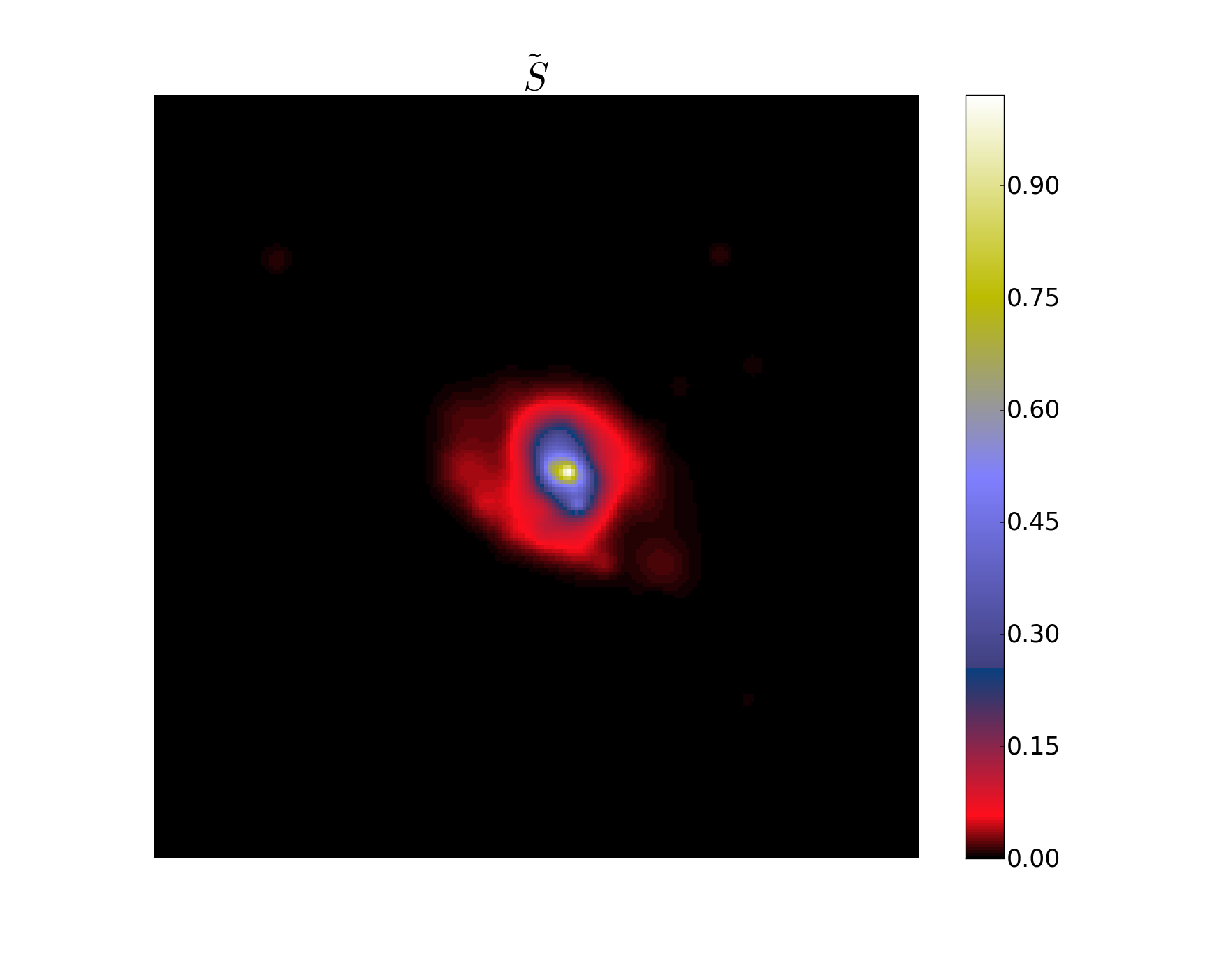}
    \includegraphics[trim = 5cm 2cm 3cm 0.5cm, clip = true,scale = 0.15]{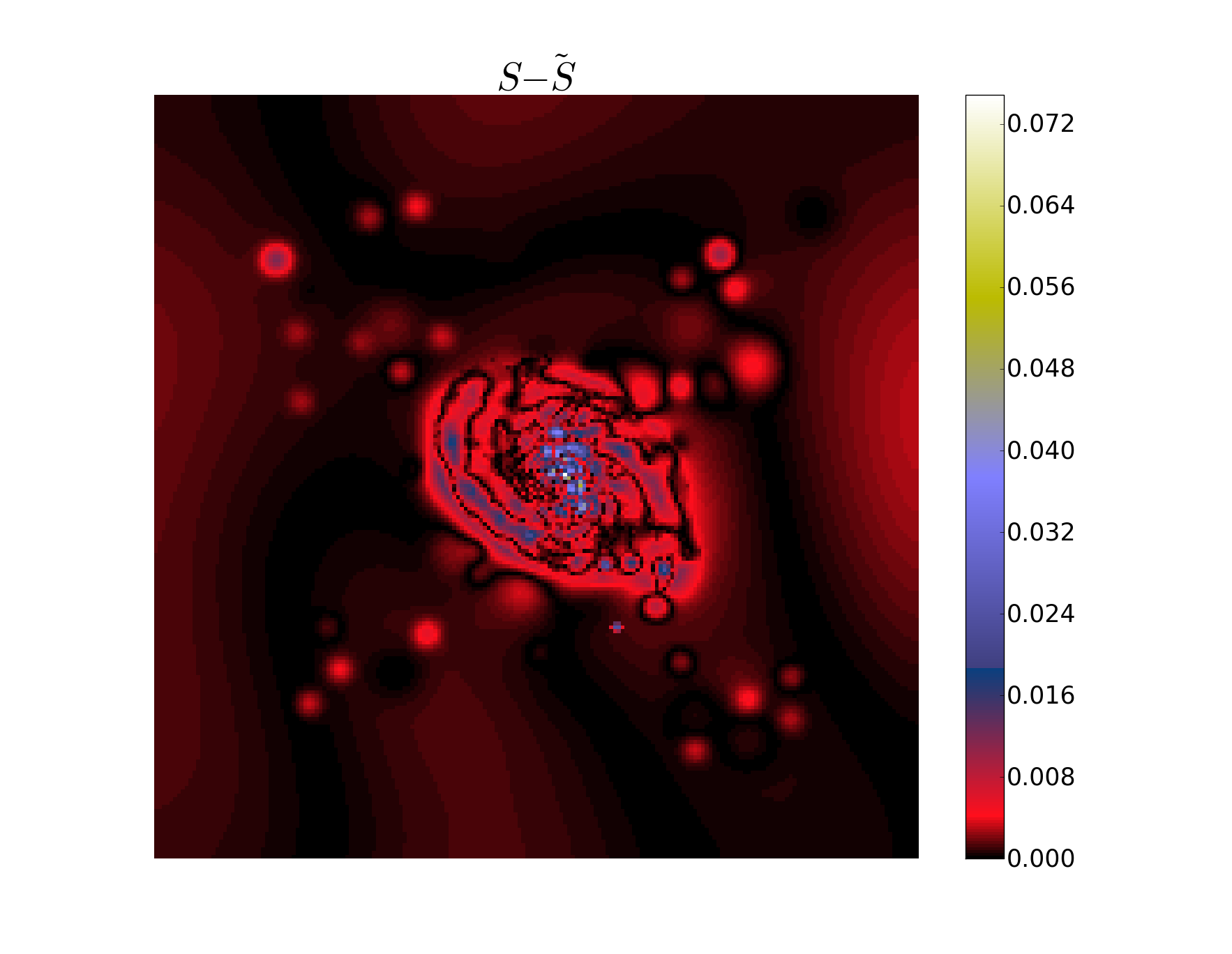} \\
    \includegraphics[trim = 5cm 2cm 3cm 0.5cm, clip = true,scale = 0.15]{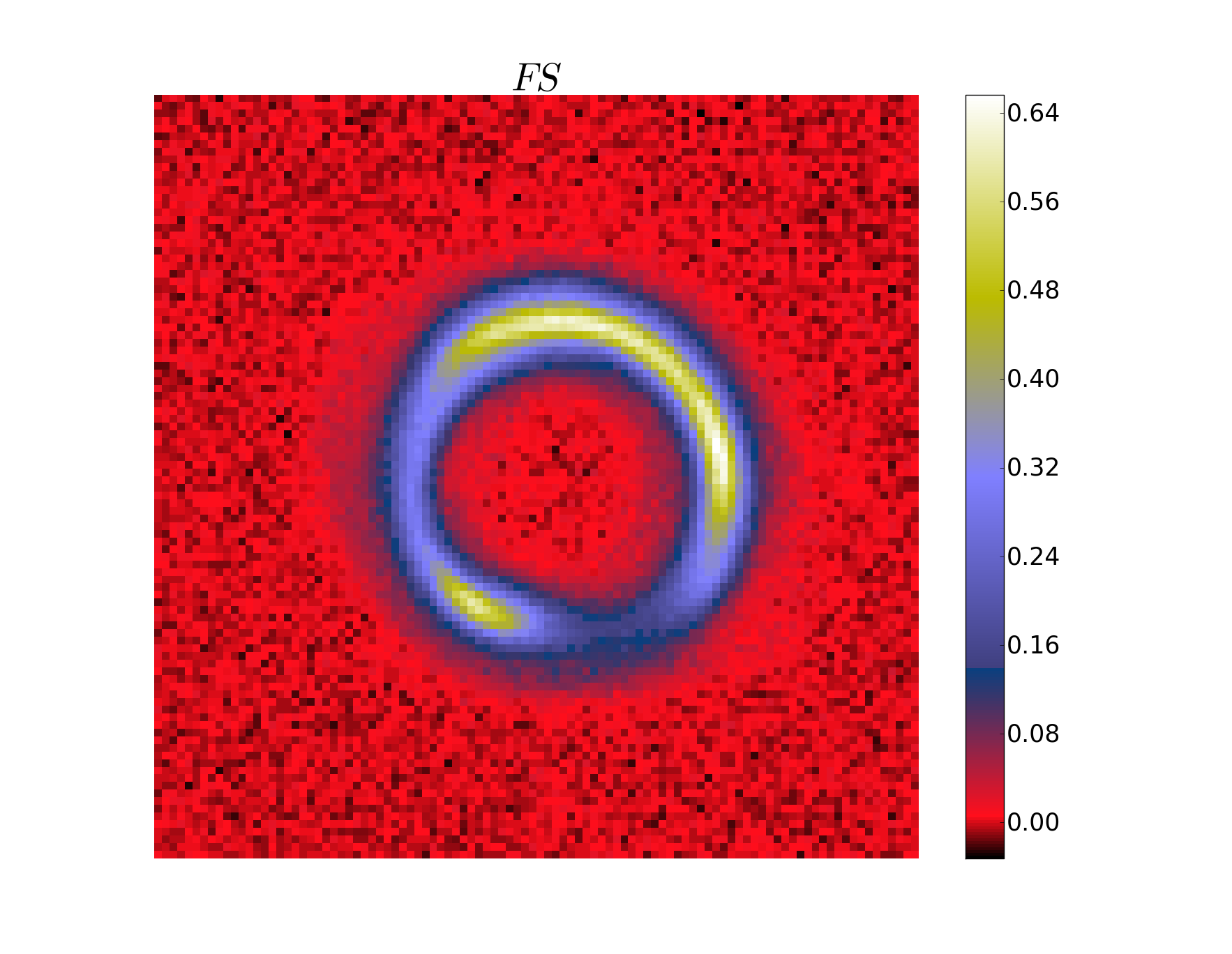}
    \includegraphics[trim = 5cm 2cm 3cm 0.5cm, clip = true,scale = 0.15]{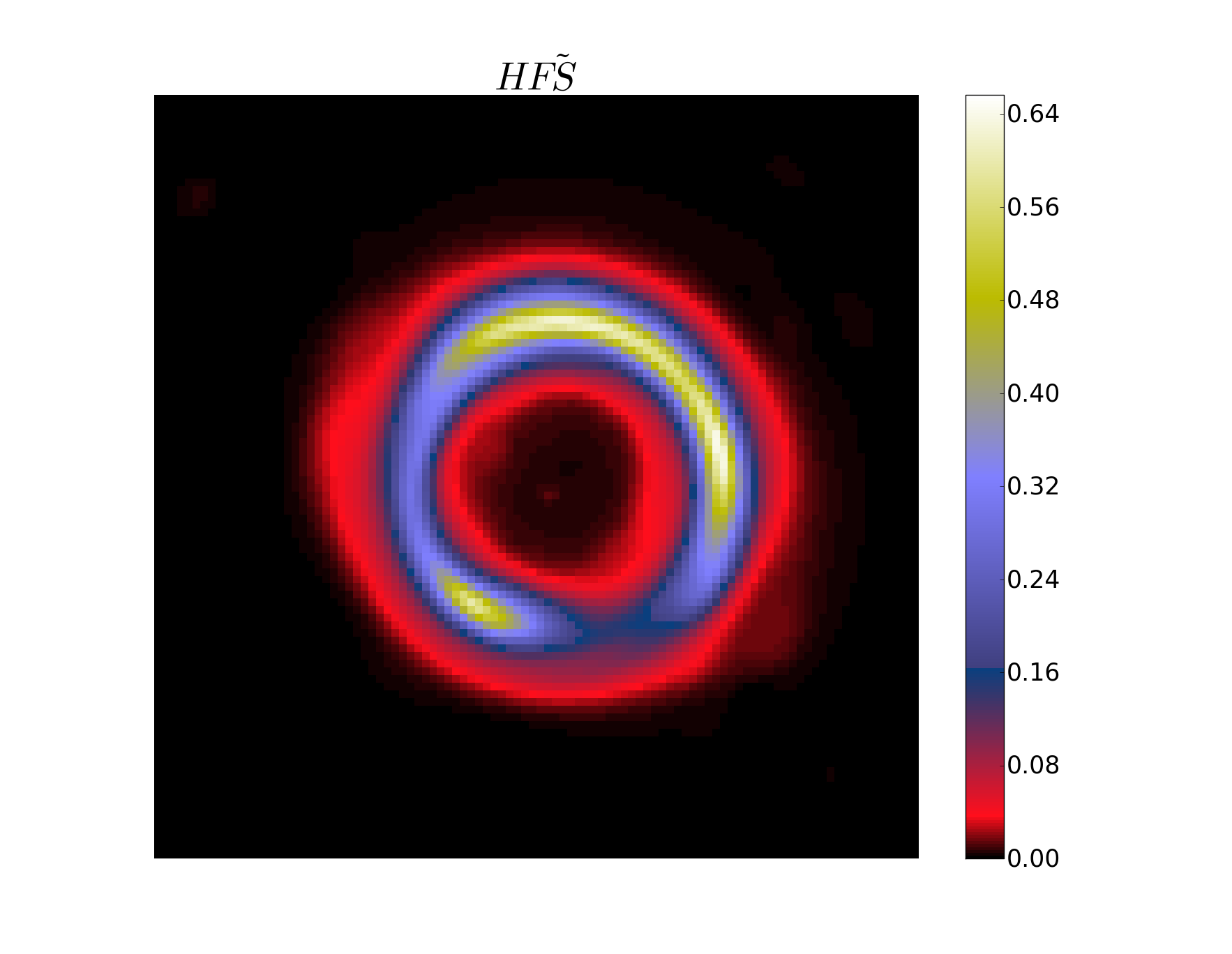}
    \includegraphics[trim = 5cm 2cm 3cm 0.5cm, clip = true,scale = 0.15]{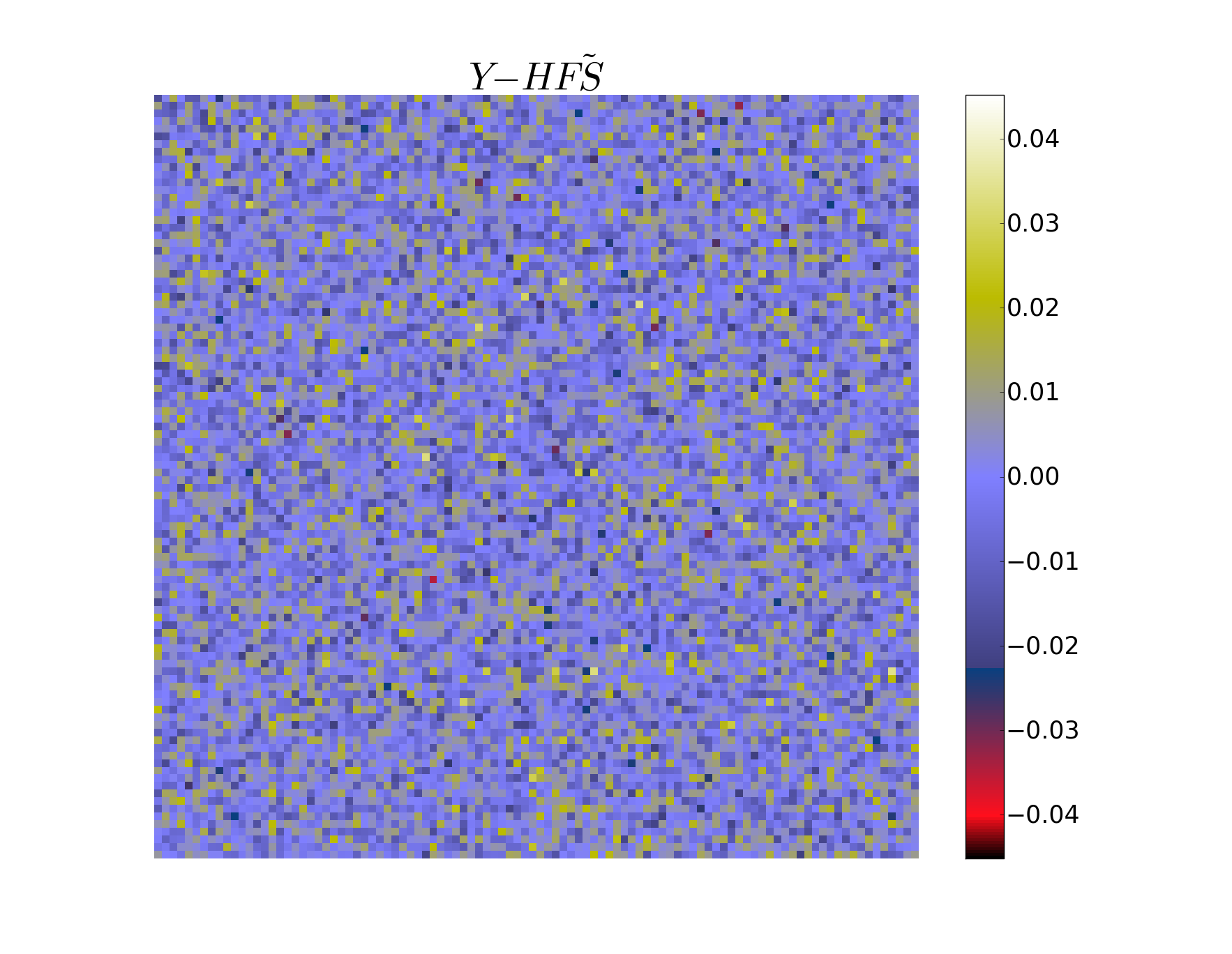}
    \caption{Application of the {\tt SLIT} algorithm to a simulated lensing system in the simple case where there is no light from the lensing galaxy. {\it Left:} the  simulated source is shown on the top while its lensed and noisy version is shown on the bottom. Both include a PSF convolution. {\it Middle:} source recovered with the {\tt SLIT} algorithm and the lensed version of it. Note that both are still convolved with the PSF. {\it Right:} the difference with the true source (top) and the residuals in the lens plane (bottom). The original and reconstructed images are displayed with the same colour cuts. The residuals in the bottom right panel are shown with $\pm 5\sigma$ cuts.}\label{fig:Results_SLIT}
\end{figure*}
%
\subsection{Testing {\tt SLIT} and {\tt SLIT\_MCA} with simulations}
\label{sec:Test}

 In the present work, our goal is to show the potential of MCA-based algorithms as a simultaneous source reconstruction and source-lens deblending technique. All our tests therefore assume that the mass density profile of the lens is known, as well as the PSF. Unless stated otherwise, the following images were simulated with white Gaussian noise with standard deviation $\sigma$. We define the signal to noise ratio (SNR) of an  image $I$ with $N_{pix}$ pixels as:

\begin{equation}
SNR = \frac{1}{N_{pix}\sigma^2}\sum\limits_{N_{pix}} I^2 \label{ep:SNR}
\end{equation}

\subsubsection{Case 1: simulation with no lens light}

\begin{figure}
    \centering
    \includegraphics[scale = 0.25]{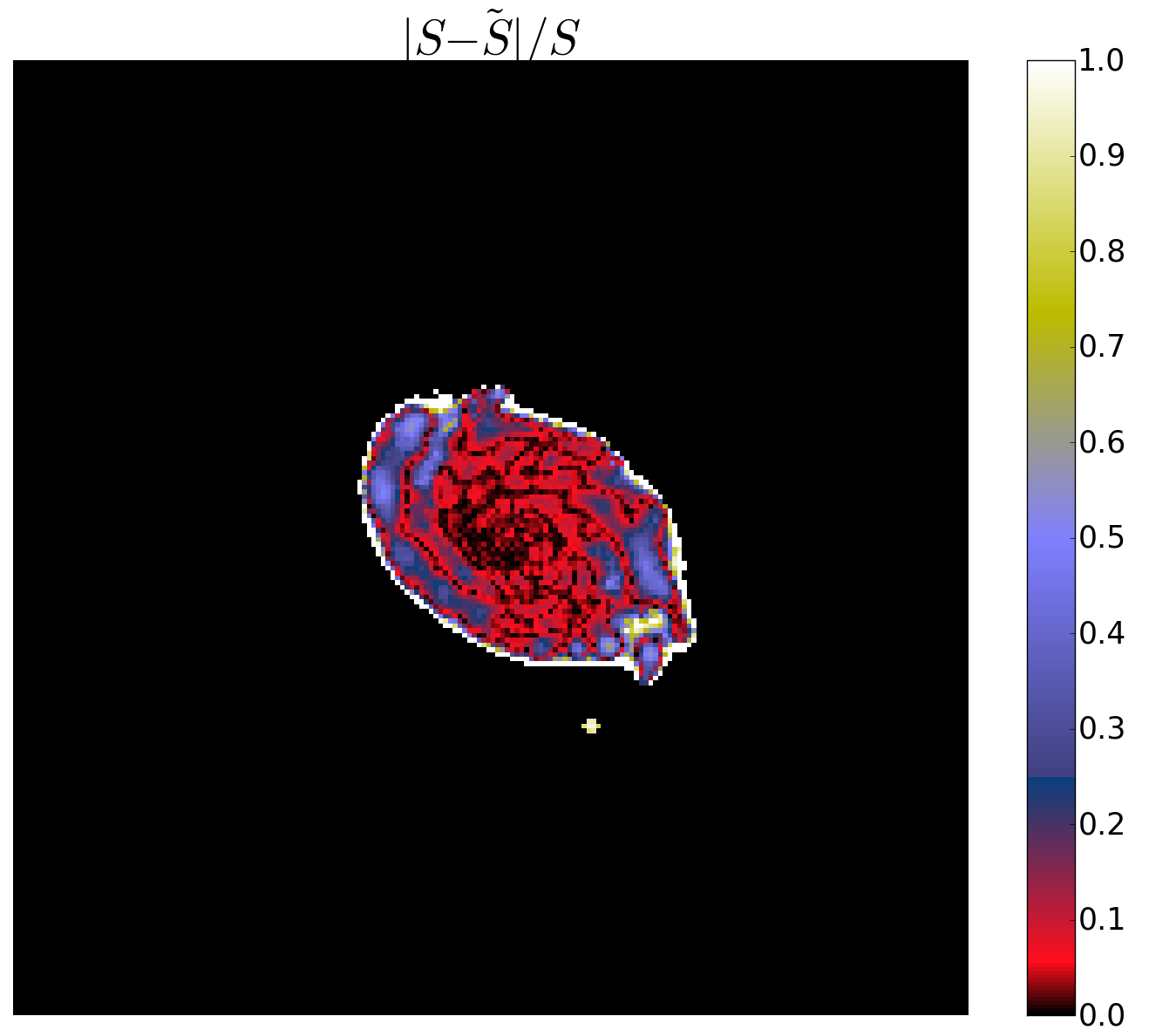}
    \caption{Residuals of the reconstruction from {\tt SLIT} relative to the true light profile of the source.}
    \label{fig:Rel_SLIT}
\end{figure}

We first reconstruct an image of a lensed galaxy with no foreground light. The simulation contains Gaussian white noise with $SNR=50$. We used $50$ iterations of Algo.~\ref{algo:SLIT}. The results are presented in Fig~\ref{fig:Results_SLIT} and Fig.~\ref{fig:Rel_SLIT}, illustrating the quality of the reconstruction. In Fig.~\ref{fig:Rel_SLIT} in particular, we show that the central region of the source galaxy, where the flux is larger, is reconstructed with less than $10\%$ error. The error increases in the outer parts of the galaxy where the flux is smaller. In this figure, the relative error is set to zero at locations where the flux in the source is zero, which does not account for false detection.

In this simulation, there are 4 times more pixels in the source plane image than in the lens plane, i.e. in image plane, the observable is an image of 100 by 100 pixels, while the source is reconstructed  on a 200 by 200 pixels. 

\subsubsection{Case 2: simulation with both lens and source light}

\begin{figure*}[t!]
\centering
    \includegraphics[trim = 5cm 1cm 3cm 1cm, clip = true,scale = 0.17]{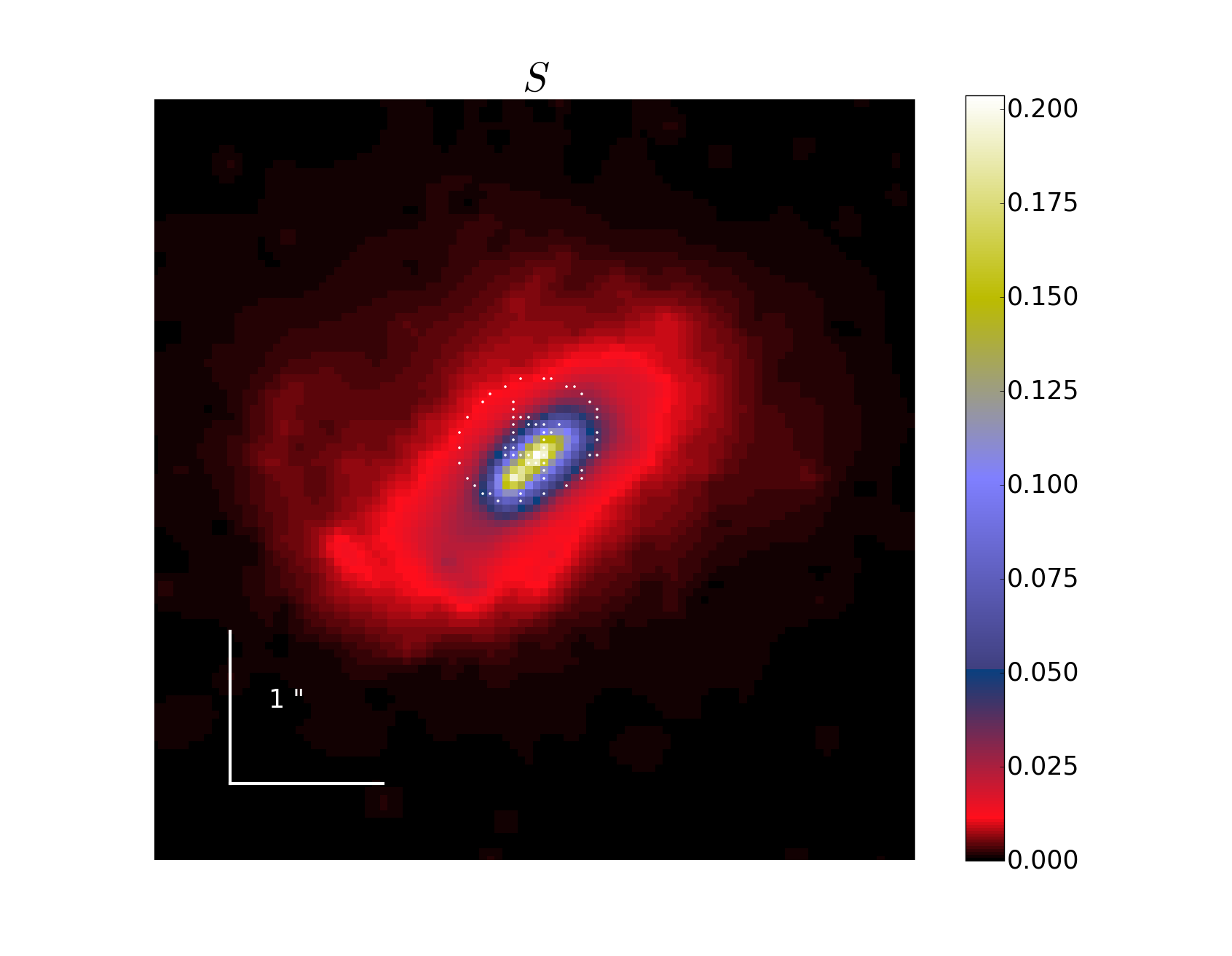}
    \includegraphics[trim = 5cm 1cm 3cm 1cm, clip = true,scale = 0.17]{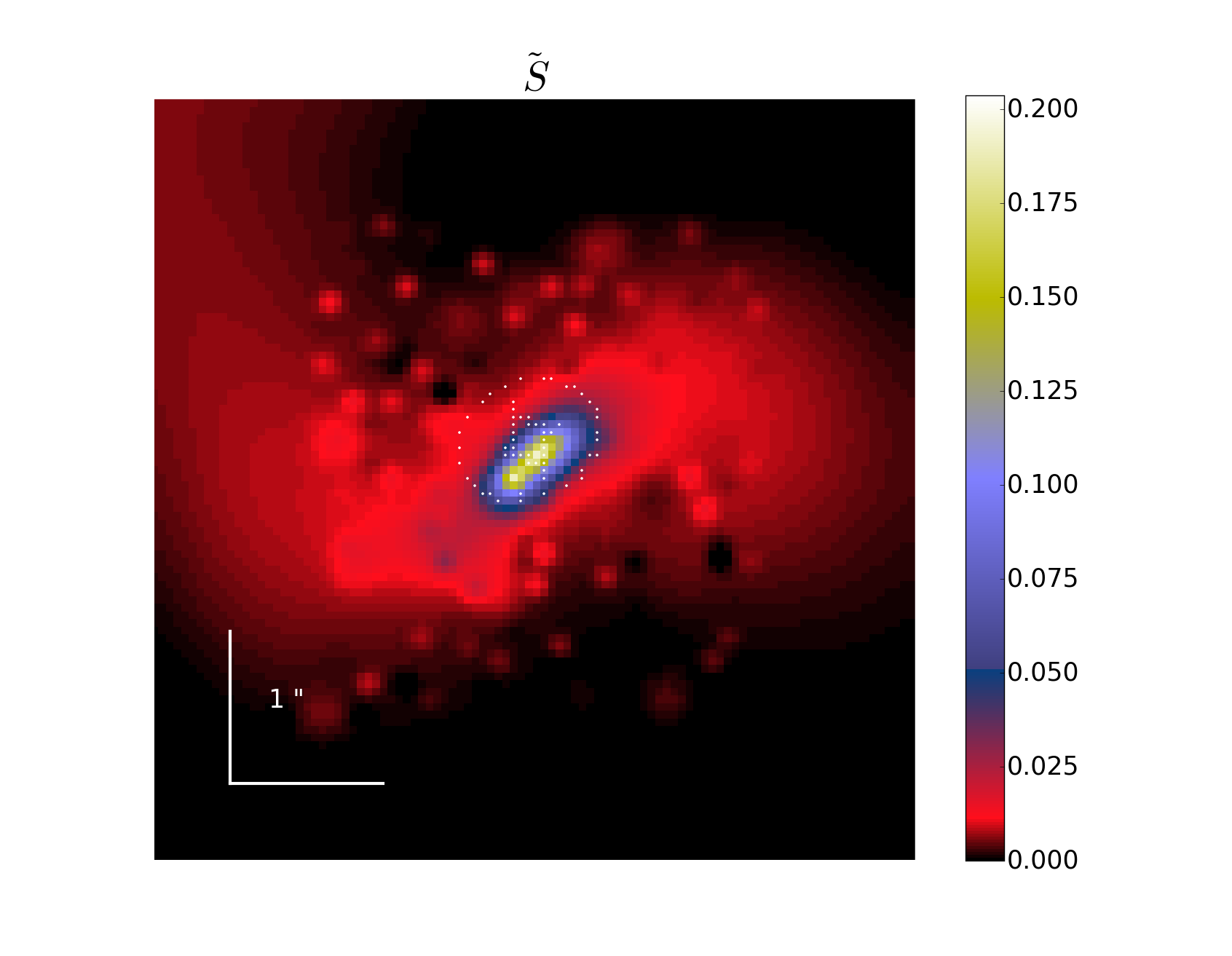}
    \includegraphics[trim = 5cm 1cm 3cm 1cm, clip = true,scale = 0.17]{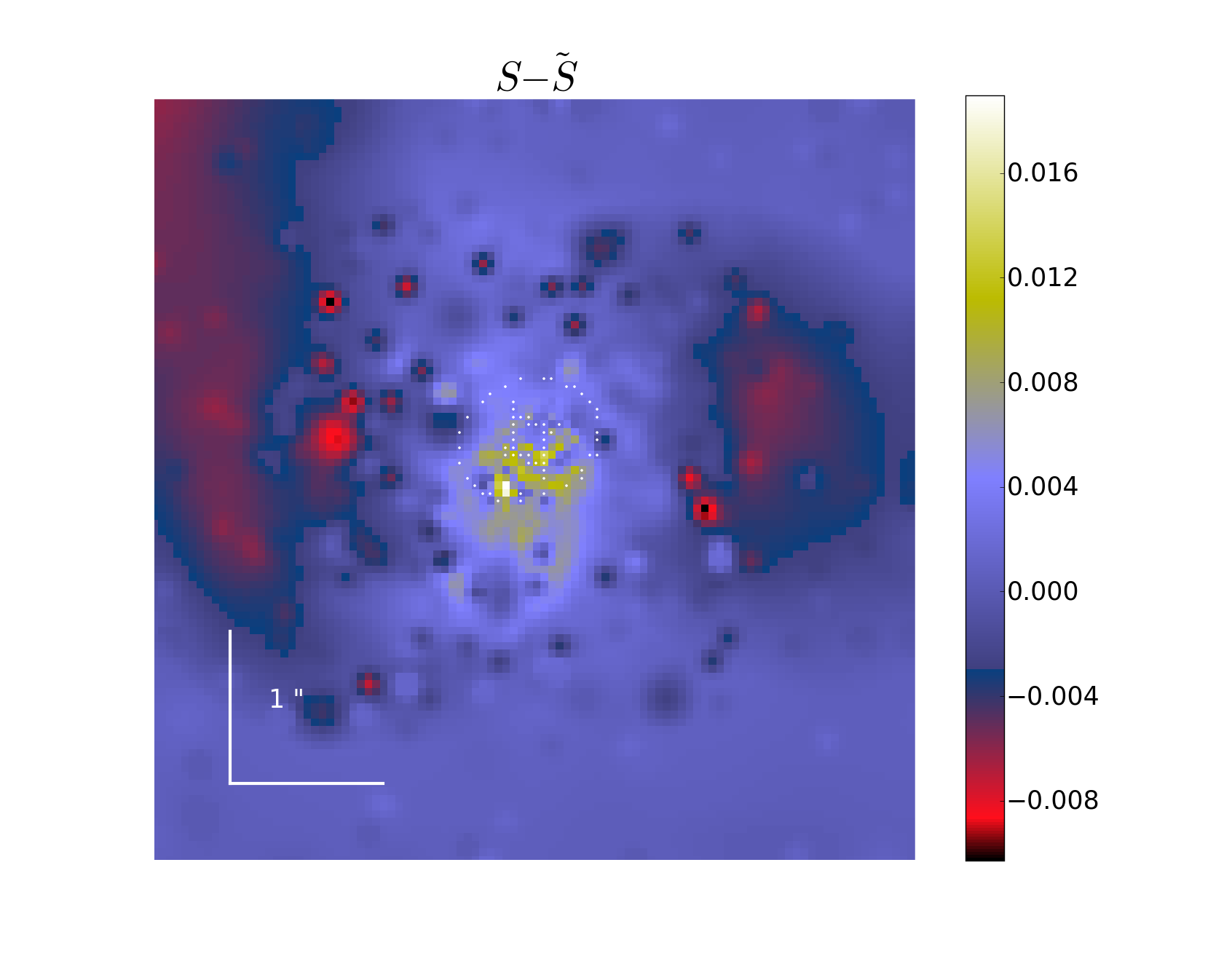} \\
    \includegraphics[trim = 5cm 1cm 3cm 1cm, clip = true,scale = 0.17]{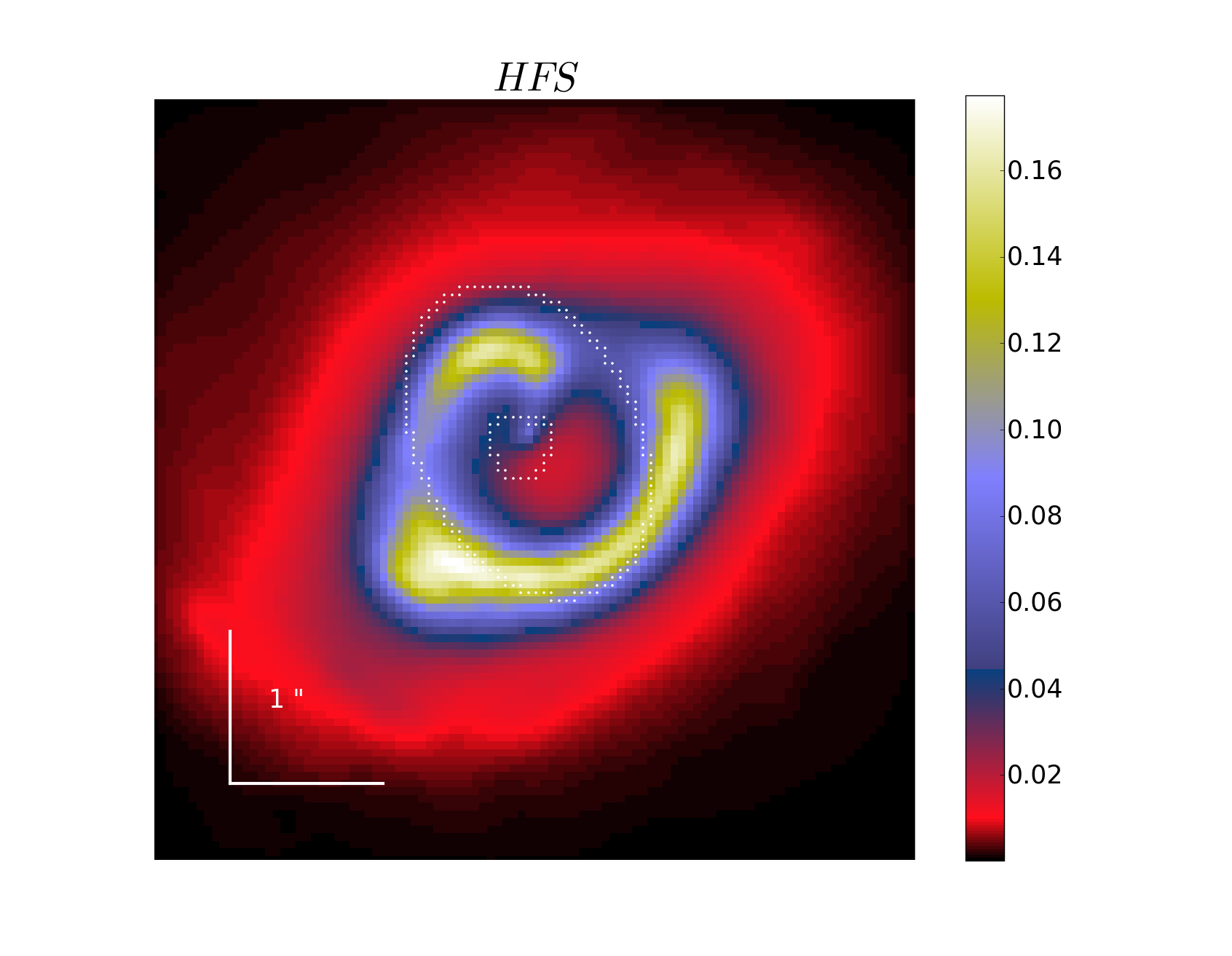}
    \includegraphics[trim = 5cm 1cm 3cm 1cm, clip = true,scale = 0.17]{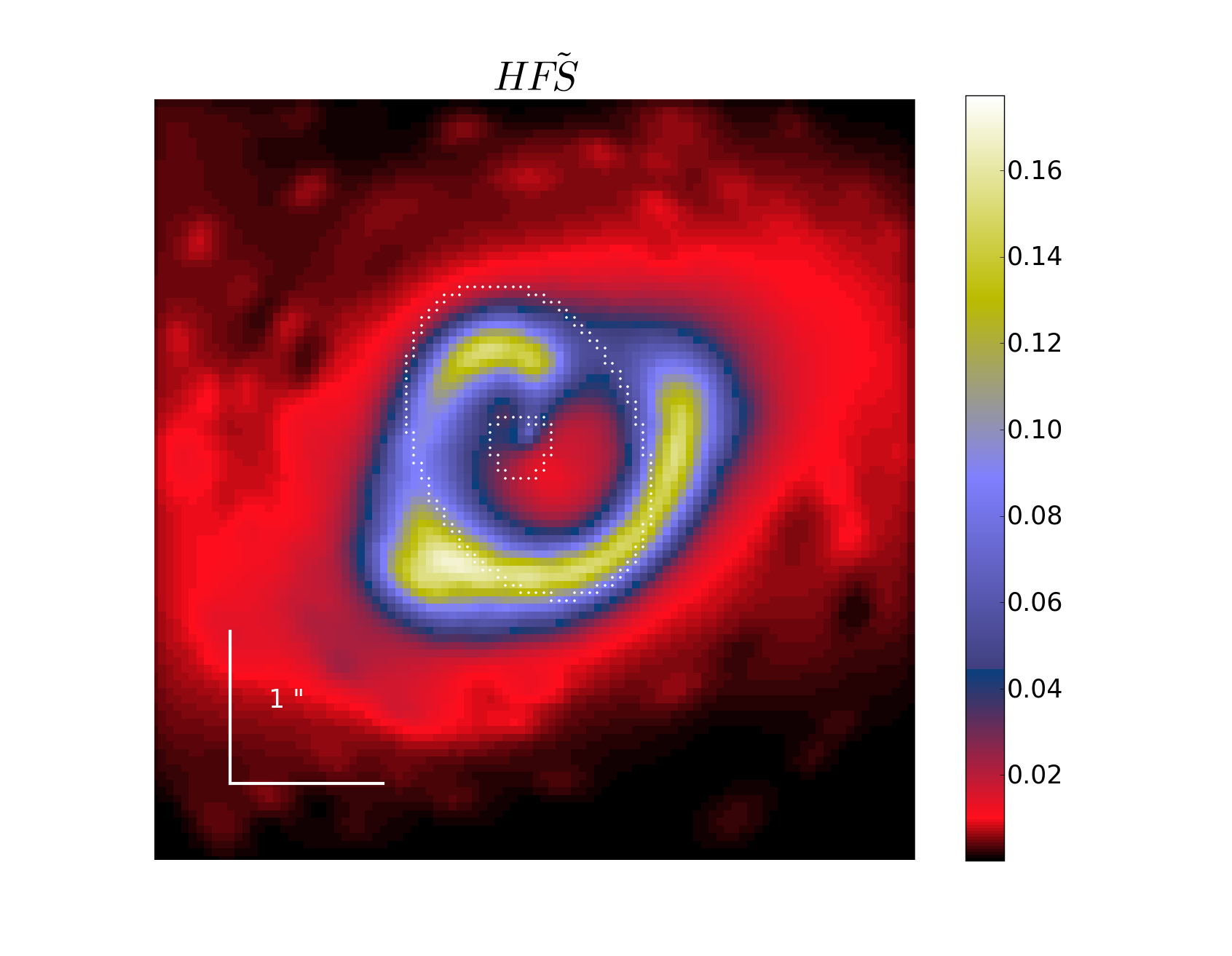}
    \includegraphics[trim = 5cm 1cm 3cm 1cm, clip = true,scale = 0.17]{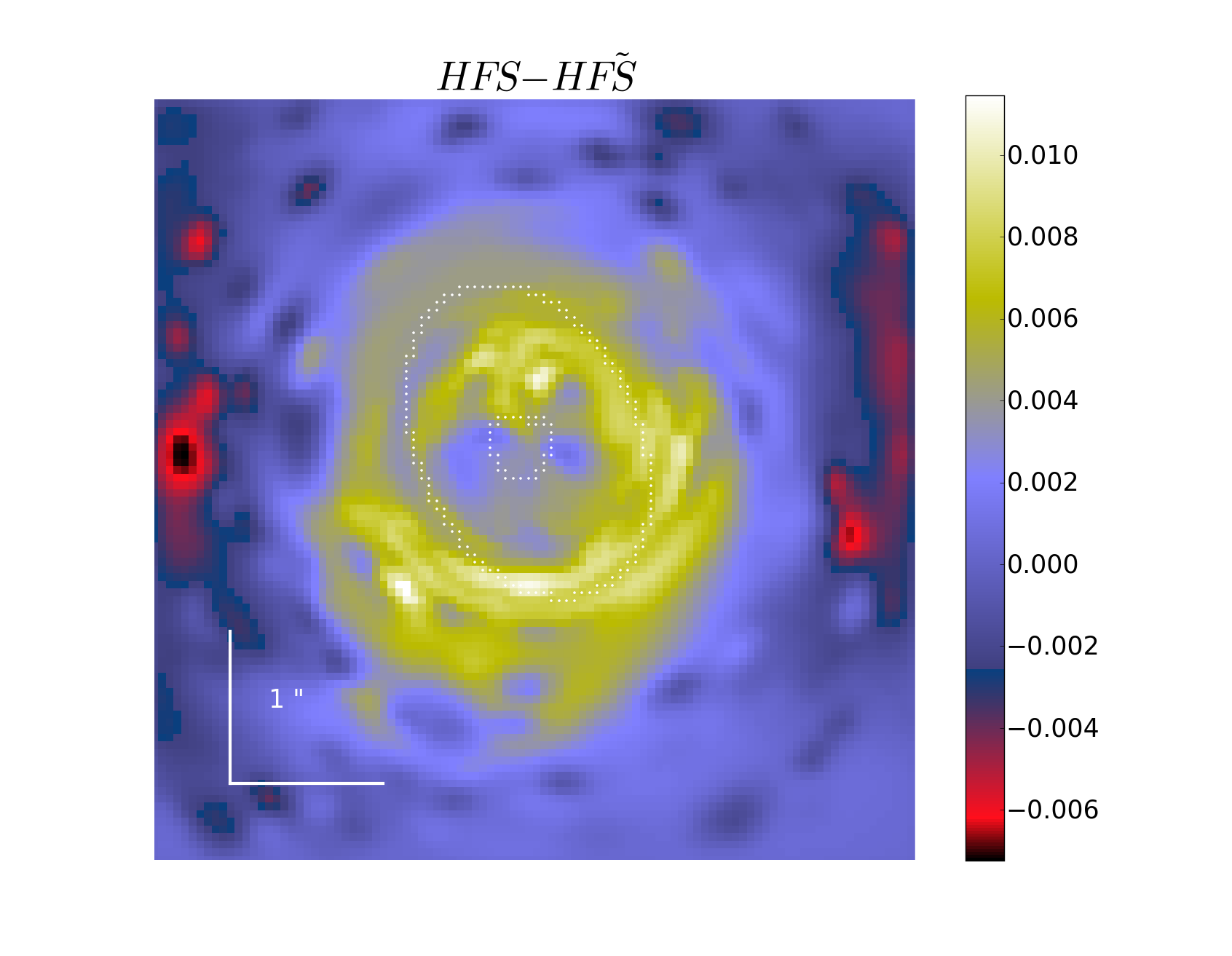}\\
    \includegraphics[trim = 5cm 1cm 3cm 1cm, clip = true,scale = 0.17]{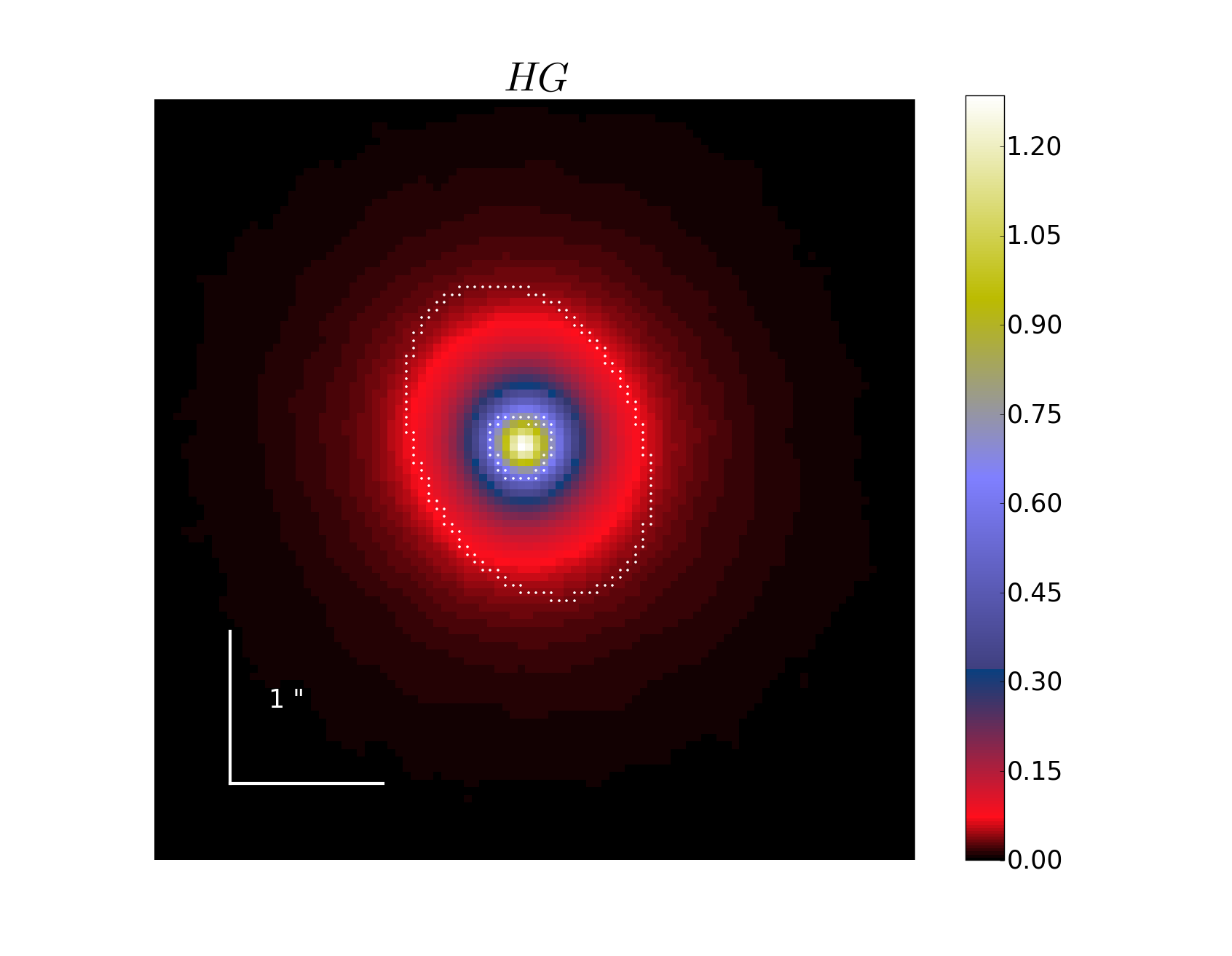}
    \includegraphics[trim = 5cm 1cm 3cm 1cm, clip = true,scale = 0.17]{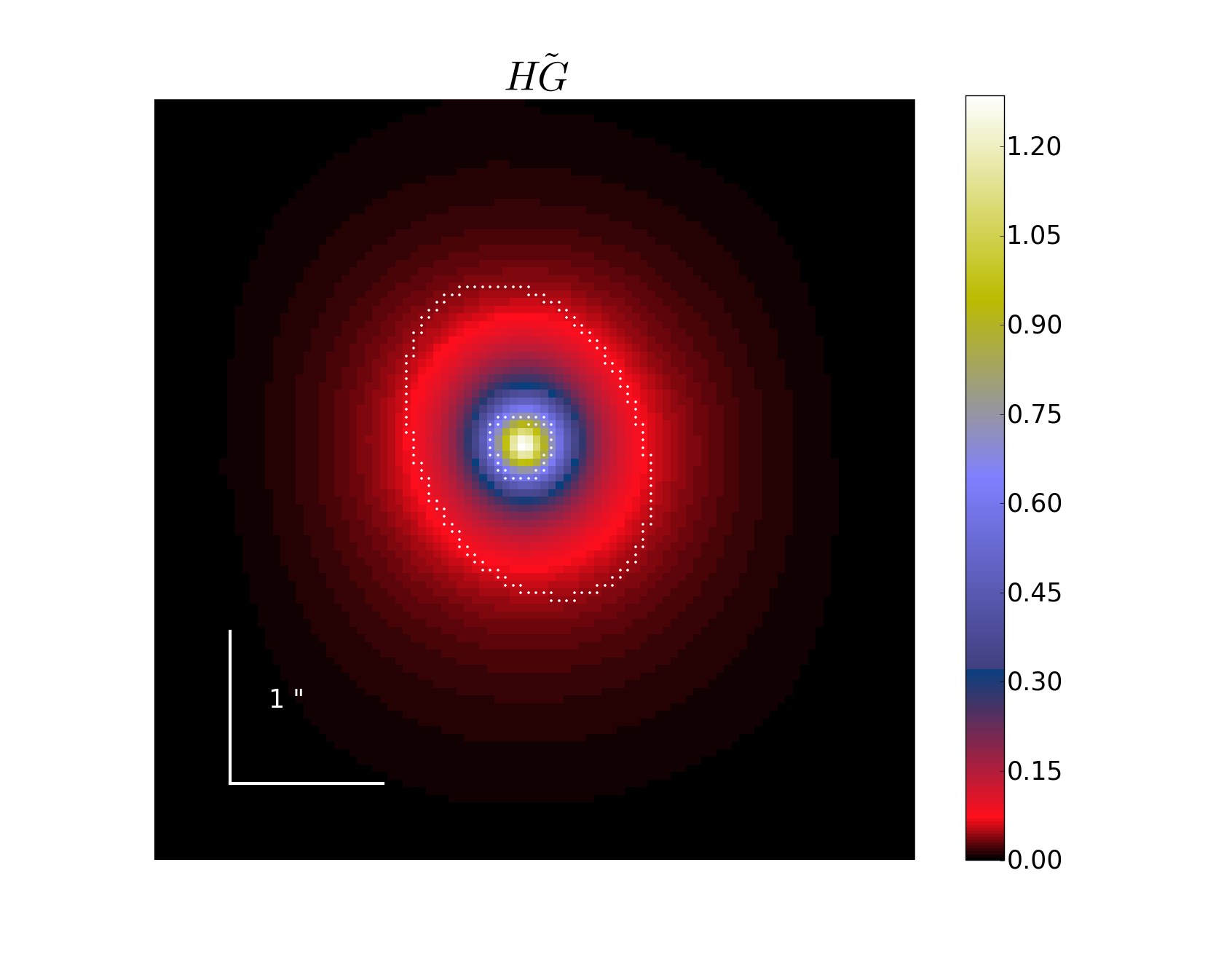}
    \includegraphics[trim = 5cm 1cm 3cm 1cm, clip = true,scale = 0.17]{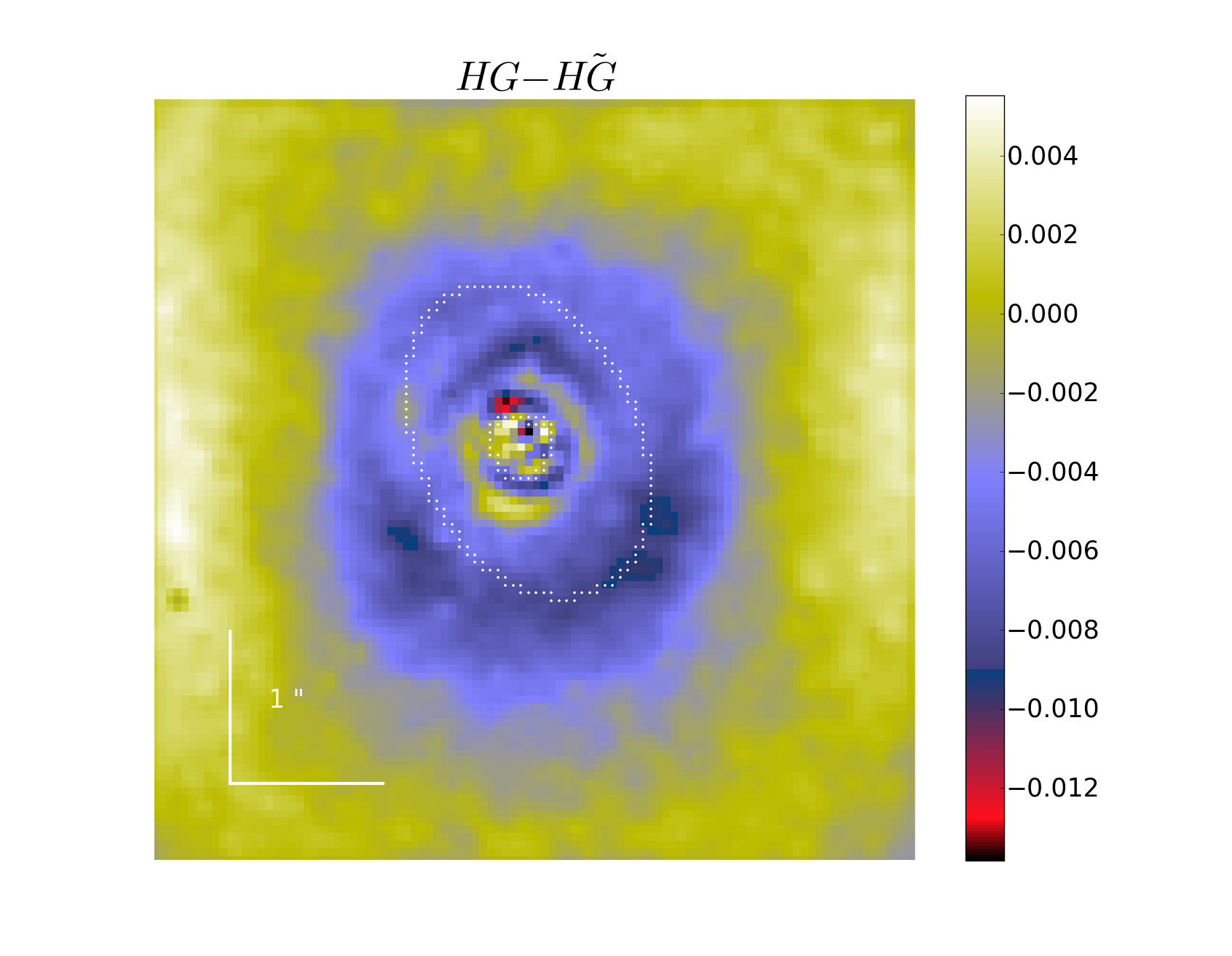}\\
    \includegraphics[trim = 5cm 1cm 3cm 1cm, clip = true,scale = 0.17]{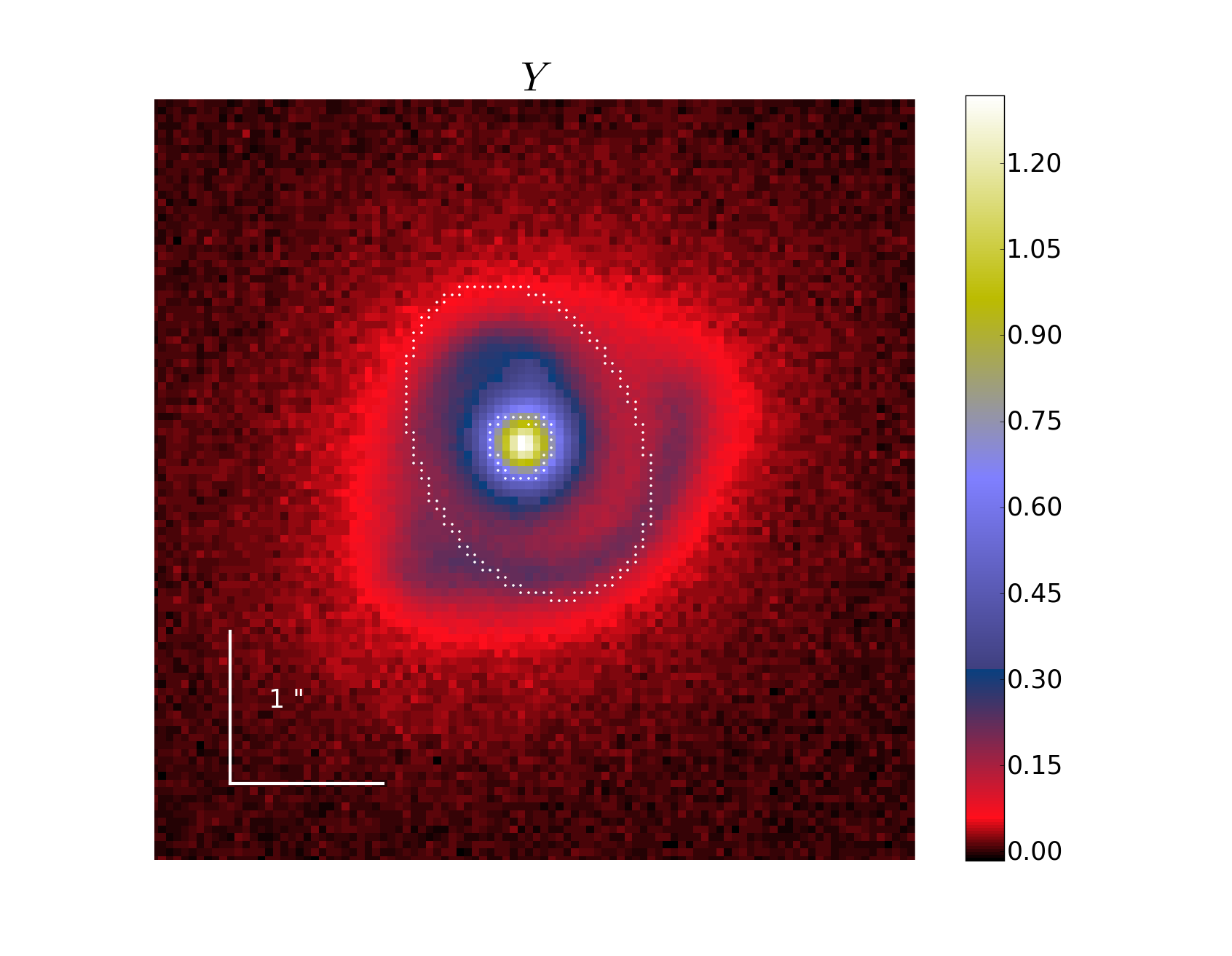}
    \includegraphics[trim = 5cm 1cm 3cm 1cm, clip = true,scale = 0.17]{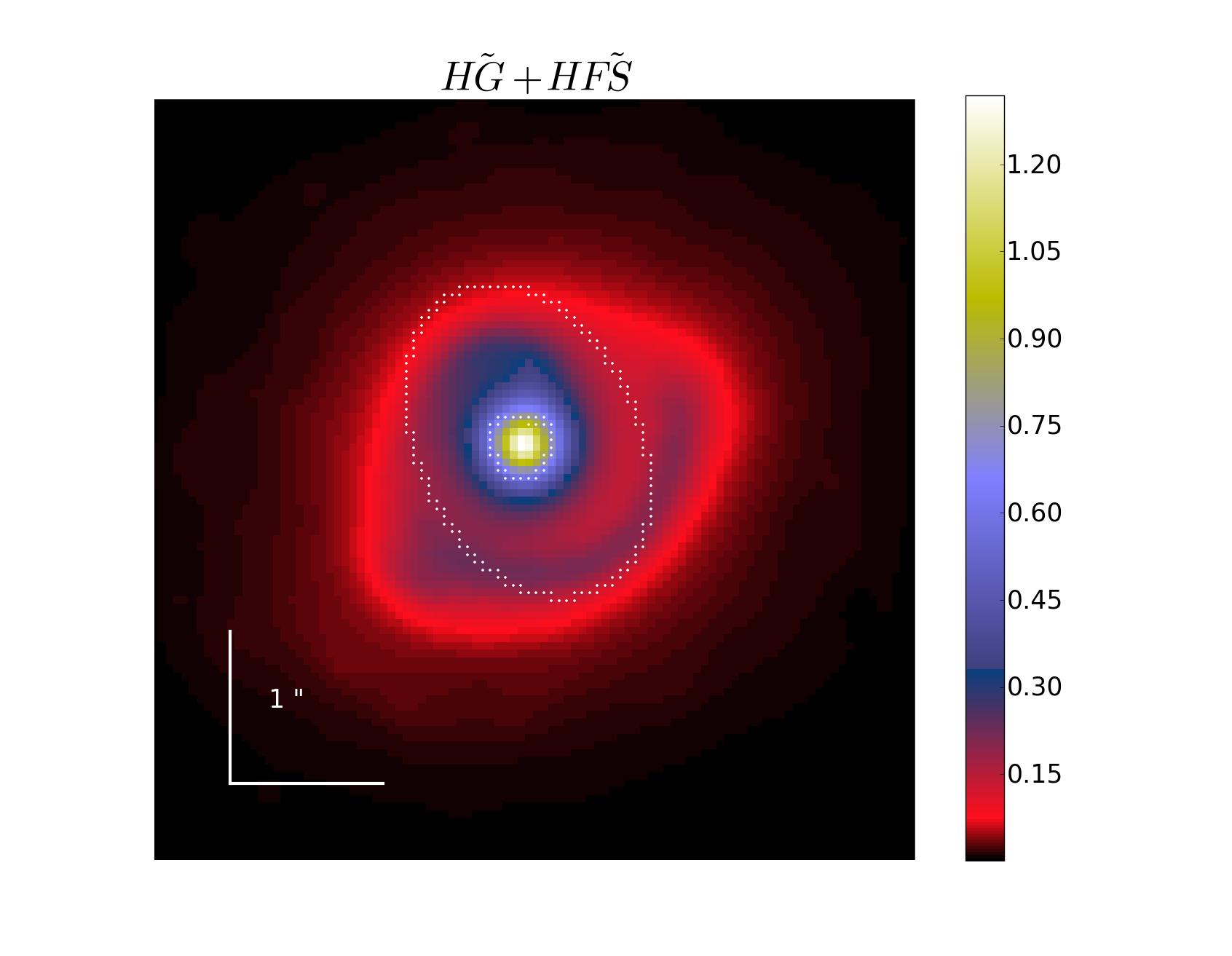}
    \includegraphics[trim = 5cm 1cm 3cm 1cm, clip = true,scale = 0.17]{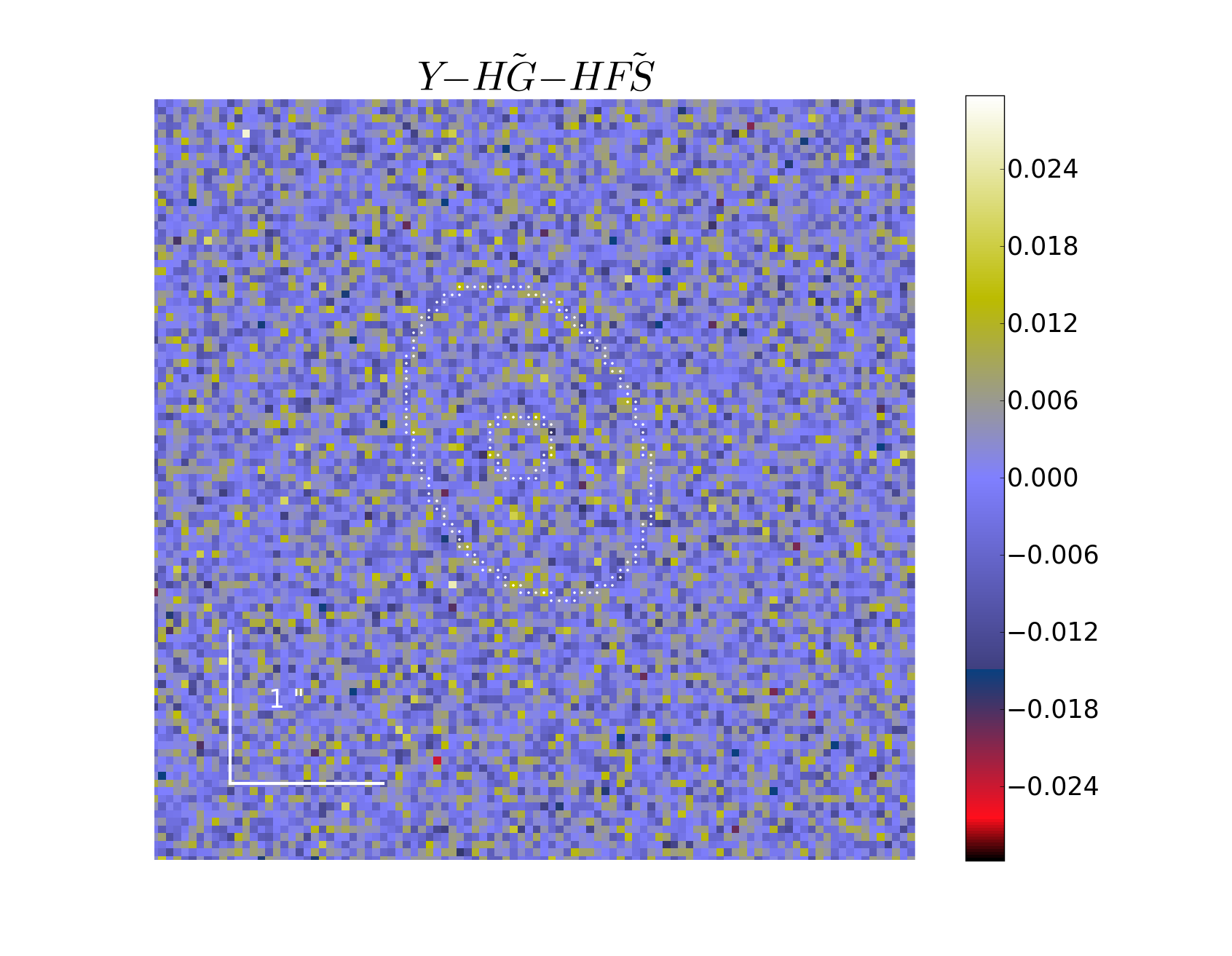}
    \caption{Illustration of the {\tt SLIT\_MCA} algorithm with simulated data. {\it Left:} simulated ground truths. From top to bottom are shown the original un-lensed source, its lensed version convolved with the PSF, the lensing galaxy (convolved with the PSF) and the full simulated system with noise.  {\it Middle:} the output of the {\tt SLIT\_MCA} algorithm. {\it Right:} the differences between the two previous panels. The original and reconstructed images are displayed with the same colour cuts. The residuals in bottom right panel are shown with cuts set to $\pm 5\sigma$. White dots show the positions of pixels crossed by critical lines in lens plane and by caustics in source plane.}\label{fig:Results_SLIT_MCA}
\end{figure*}

We then applied Algo.~\ref{algo:MCA_SLIT} to the simulated images of a full lens system. In this case we recover both the convolved lens light profile and the source light profile (Fig~\ref{fig:Results_SLIT_MCA}). We enforce the sparsity of each solution by using enough iterations of the algorithm to perform an efficient separation. A difficulty here is to choose the numbers of iterations and sub-iterations such that both components converge to a sparse solution. In our experiments, $N_{iter} = N_{subiter}$ ensures similar quality in the reconstructions of both components.

	The results show no structure in the residuals and visually good separation between the lens and the source as well as a good reconstruction of the source without significant leakage between the two. However, the residuals in the first three lines of the Figure show that the source flux was slightly overestimated at larger scales, while the lens galaxy was slightly underestimated. The amplitude of the phenomenon reaches no more than $10\sigma$ of the noise level given the amplitudes displayed in the last column.

\subsection{Comparison with {\tt lenstronomy}}

We tested our source reconstruction technique on three other simulations with various source morphologies and lens mass profiles, including one generated with the {\tt lenstronomy} package. We compared our reconstructions of three lensed sources with the ones computed by S. Birrer, the author of {\tt lenstronomy}. In order to avoid favouring one method over the other, we tried as much as possible to choose representations for true sources that do not correspond to either code's decomposition for reconstructing sources. Sources generated with {\tt SLIT} were extracted from HFF images and convolved with a Gaussian kernel with a full width at half maximum of five pixels. This produces a smooth version of the noisy HFF images to which we then subtract the median value of the image in order to set the sky background to zero. All remaining negative values in the image are set to zero. The image generated with {\tt lenstronomy} uses a source from a jpeg image of NGC 1300 from NASA, ESA. The image resolution is degraded by a factor 25 and decomposed over the shapelet dictionary \citep{Refregier2003} using enough coefficients (11476) to accurately recover the morphology of the image. Despite {\tt lenstronomy} relying on shapelets to solve the source inversion problem, the number of coefficients that it is possible to recover in the reconstruction is much smaller than the number of coefficients used in generating the true source. Therefore, the basis set of the reconstruction is different from the one used in generating the true source. The three systems tested here were made from sources with different morphologies and different lens profiles Tab.~\ref{tab:Simu}. 
In this exercise we test our methods on simulated images that were generated with different procedures. This comparison therefore allows to show how robust these techniques are to the underlying mapping between source and lens plane.

\begin{table*}
	\centering
	\begin{tabular}{c|c|c|c|c|c|c}
	Image number & Source Origin & Source processing & Factor & PSF & Noise & Lens model  \\
    \hline
    1 & NASA, ESA jpeg& Shapelets & 5 & Gaussian & P+G & SPEP \\
    2 & HFF fields& $\star G_-$ & 4 & Gaussian & SNR = 10 & SIE \\
    3 & HFF fields & $\star G_-$ & 2 & Gaussian & SNR = 100 & SIS \\
	\end{tabular}
	\caption{Description of the simulated images. Symbol $\star G_-$ stands for convolution by a Gaussian kernel and subtraction of the median (see text). Column factor stands for the resolution factor between source and lens plane. For instance, in image 1, the source has 5 times more pixels on the side than the image. Gaussian PSFs were used in all three images with a FWHM of 2 pixels, P+G stands for Gaussian poisson mixture. Gaussian nois with $\sigma = 2$ was used. The poisson noise value at pixel $i$ is drawn from a Gaussian distribution with $\sigma = \sqrt{f_i}$, $f_i$ being the flux in pixel $i$. }
  	\label{tab:Simu}
\end{table*}

\begin{figure*}[t!]
\includegraphics[trim = 1.cm 0cm 0cm 0cm, clip = true,scale = 0.42]{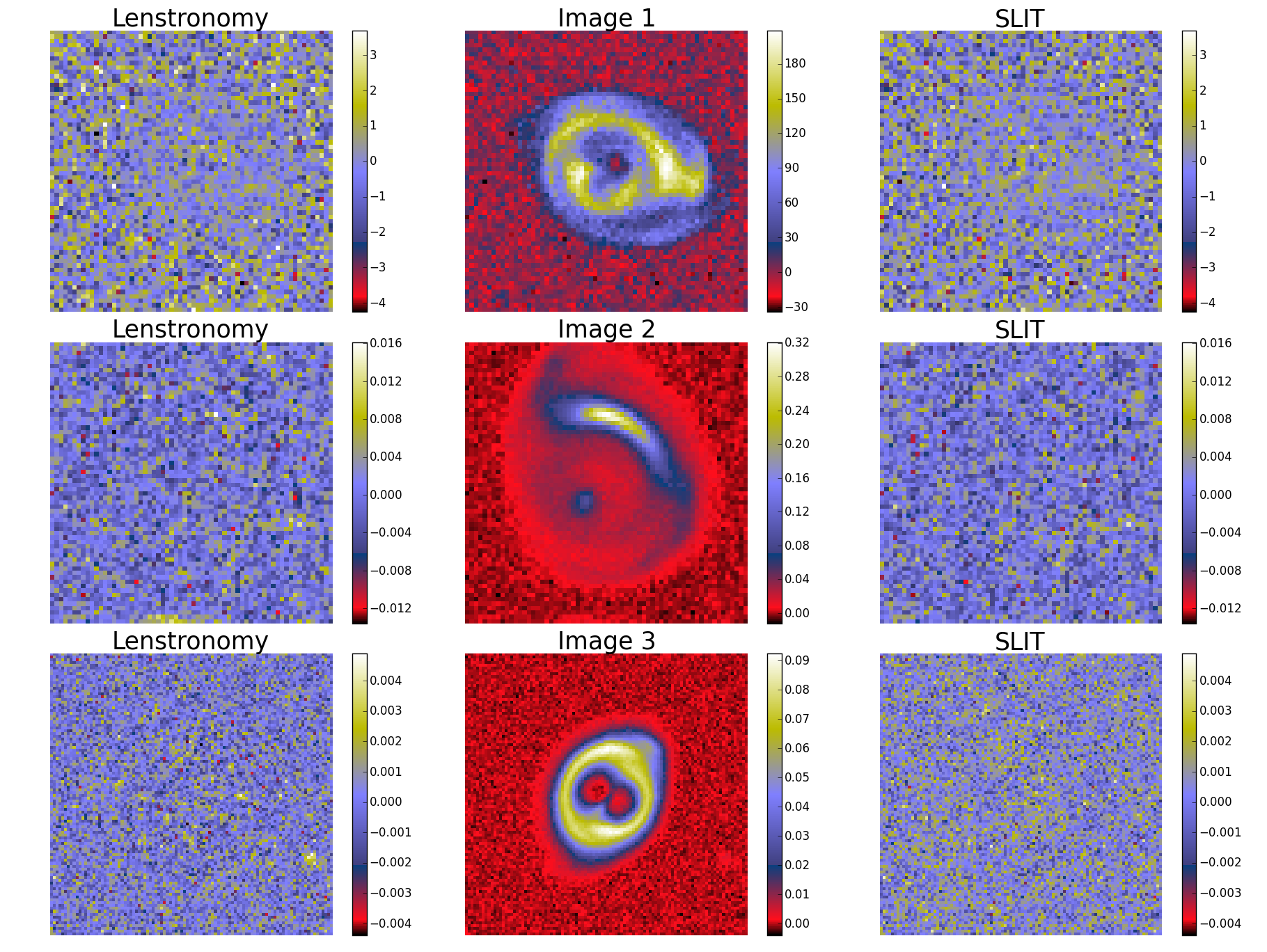}
\caption{Reconstructions with {\tt Lenstronomy} and {\tt SLIT} in image plane. The middle panels represent respectively from top to bottom, simulated images 1, 2 and 3. The left panels show the corresponding residuals after reconstruction with {\tt lenstronomy}, while the right panels show the residuals obtained with {\tt SLIT}. }\label{fig:Res_Images}
\end{figure*}

\begin{figure*}[h!]
\centering
\begin{tabular}{ll}
\rotatebox{90}{\LARGE \hspace{1.5cm} \bf SLIT \hspace{2.5cm} SLIT \hspace{1.8cm} Simulations \hspace{0.8cm} Lenstronomy \hspace{0.5cm} Lenstronomy}
&
\includegraphics[scale = 0.63]{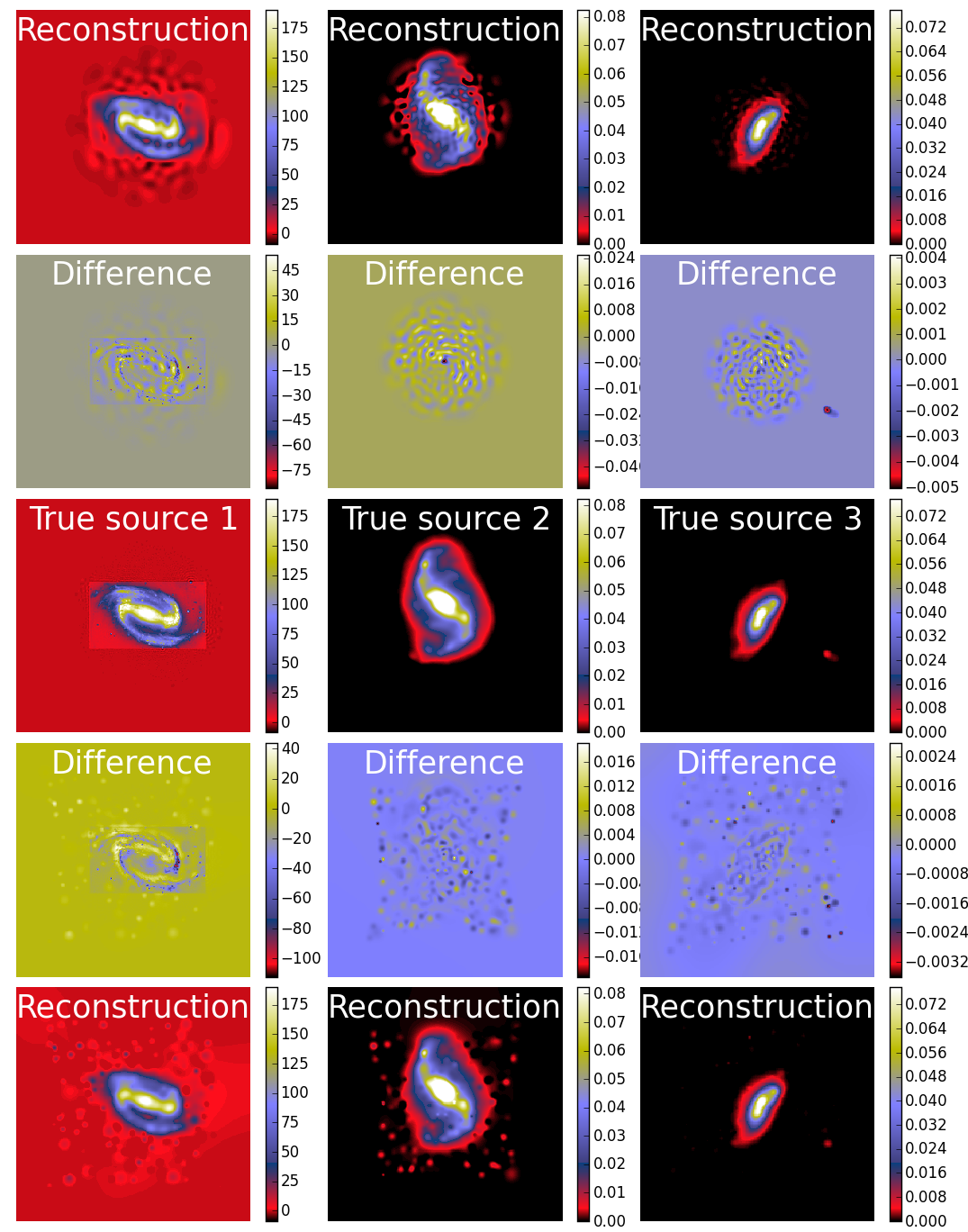}
\end{tabular}
\caption{Reconstructions with {\tt Lenstronomy} and {\tt SLIT} in source plane. Panels from the middle row show the true sources used to generate respectively simulated images 1,2 and 3. The first row show the source reconstruction from {\tt lenstronomy}. The second row show the difference between the true sources and the sources reconstructed by {\tt lenstronomy}. The last row shows the source reconstruction from {\tt SLIT}. The penultimate row shows the difference between true sources and sources reconstructed with {\tt SLIT}. Panels between reconstructed and true images, show the difference between the two for the corresponding technique.}\label{fig:Res_Sources}
\end{figure*}

In order to compare the results of both methods we show the resulting reconstructions of the runs in Figs. \ref{fig:Res_Images} and \ref{fig:Res_Sources} and we use three metrics:
\begin{itemize}
\item Quality of the residuals (QoR), given by the relative standard deviation of the residuals for a model of the source , $\tilde{S}$:
\begin{equation}
QoR(\tilde{S}) = std\Big{(}\frac{Y-HF\tilde{S}}{\sigma}\Big{)}. \label{eq:Res}
\end{equation}
In cases of Gaussian and Poisson mixture noise, $\sigma = \sqrt{\sigma_G^2+f}$, where $\sigma_G$ is the standard deviation of the Gaussian component and $f$ is a 2-D map of the flux in the true noiseless model for image $Y$. Given that definition, the closest $QoR$ is from one, the better the reconstruction.
\item Quality of the source reconstruction relative to the true source , $S_{true}$, estimated with the Source Distortion Ratio \citep[SDR, ][]{Vincent2006}. The SDR is the logarithm of the inverse of the error on the source light profile, weighted by true flux of the source. As a result, the higher the SDR, the better the reconstruction of the source. We compute the SDR as:
\begin{equation}
SDR(\tilde{S}) = 10 log_{10}\frac{||S_{true}||}{||S_{true}-\tilde{S}||}. \label{eq:SDR}
\end{equation}
\item Computation time
\end{itemize}

The two metrics of quality and source residuals, were chosen to provide a measure of the quality of both reconstruction in both source and lens planes. While the $SDR$ of the sources is the most informative metric with regards to the quality of the reconstruction of the source, it is also necessary to ensure that both methods are able to reconstruct similarly well the observables, hence the role of metric $QoR$. The evaluation of these metrics for both methods are given in Tab.~\ref{tab:SDR}.

The residuals in Fig.~\ref{fig:Res_Images}, as well as the results for $QoR(\tilde{S})$ in Tab.~\ref{tab:SDR} show that both codes achieve similar quality of reconstruction. While {\tt lenstronomy} leaves a little bit more signal in the residuals, in particular in cases of smooth sources generated with {\tt SLIT}, {\tt SLIT}, on the contrary tends to create false detections at noise level at locations where the actual signal is zero, resulting in a slight over-fitting. In the case of {\tt lenstronomy}, over-fitting of outer regions with no signal is prevented by the fact the method relies on shapelets, which are localised around the centre of the images provided that the number of coefficients used in the reconstruction is kept small. The down-side of that strategy is that shapelets hardly represent companion galaxies in the source such as the one on the right side of image 3 in Fig. \ref{fig:Res_Sources}. To circumvent this problem, one needs to use a second set of shapelets positioned at a different location to represent a second displaced light component. This has deliberately not been done in this comparison. From looking at the reconstructions of image 1 and the corresponding SDRs, it appears that {\tt lenstronomy} performs slightly better at reconstructing truth images generated with high resolution and detailed features. Despite both reconstructions showing the same levels of details, false detections in {\tt SLIT} at locations where the truth signal is 0 contribute diminishing its SDR. Also, the large number of pixels in the source compared to the number of pixels in the observations (25 times more pixels in the source than in the image) makes the problem highly under-constrained. In {\tt lenstronomy}, this is overcome by computing a small number of shapelet coefficients and displaying them on an arbitrarily fine grid. With {\tt SLIT}, we optimise for each coefficient in the starlet dictionary, which means, more pixels in source plane equals more unknowns as the number of pixels increases.

Regarding computational time, while {\tt lenstronomy} runs in less than a second, a typical {\tt SLIT} run for examples such as the ones provided above will last between $\sim 100$ and $\sim 1000$ seconds. For a full run of {\tt SLIT\_MCA}, this number is multiplied by a factor 5 to 10. While this is a current weakness of our algorithm, we are confident that optimised packages for starlet transforms and linear optimisation as well as parallel computing  will allow us to lower these numbers by at least a factor 10.

\begin{table*}
    \centering
    \begin{tabular}{c|c|c|c|c|c|c}
         & \multicolumn{2}{|c|}{$SDR$} & \multicolumn{2}{|c|}{$QoR$} & \multicolumn{2}{|c|}{T (s) } \\
         \hline
         & {\tt Lenstronomy}  & {\tt SLIT} & {\tt Lenstronomy}  & {\tt SLIT} & {\tt Lenstronomy}  & {\tt SLIT}\\
         Image $1$ & $8.31$ & $6.39$ & $1.10$ & $1.05$ & $\sim 0.1$ & $\sim 100$ \\
         Image $2$ & $9.30$ & $13.06$ & $1.26$ & $1.21$ & $\sim 0.1$ & $\sim 400$ \\
         Image $3$ & $13.69$ & $16.09$ & $1.24$ & $1.20$ & $\sim 0.1$ &  $\sim 2000$
    \end{tabular}
    \caption{Comparison of reconstructions of three lensed sources by {\tt SLIT} and  {\tt lenstronomy}.}
    \label{tab:SDR}
\end{table*}


\subsection{Lens parameter optimisation}

In order to assess whether our technique holds the potential to be used in a full lens modelling context, we test its sensitivity to the density slope $\gamma$. In \cite{Koopmans2006,Koopmans2009}, the authors showed that real lens galaxies have, on average, a total mass density profile with a density slope $\gamma \sim 2$, i.e. isothermal. In Fig. 1 of \cite{Koopmans2009}, the authors show that the posterior probability distributions of a sample of 58 strong lenses are maximised for a density slope between 1.5 and 2.5, which sets the limits for the values of $\gamma$ investigated here. 

\begin{figure*}[t!]
\centering
\includegraphics[trim = 0.cm 0cm 3cm 1cm, clip = true,scale = 0.33]{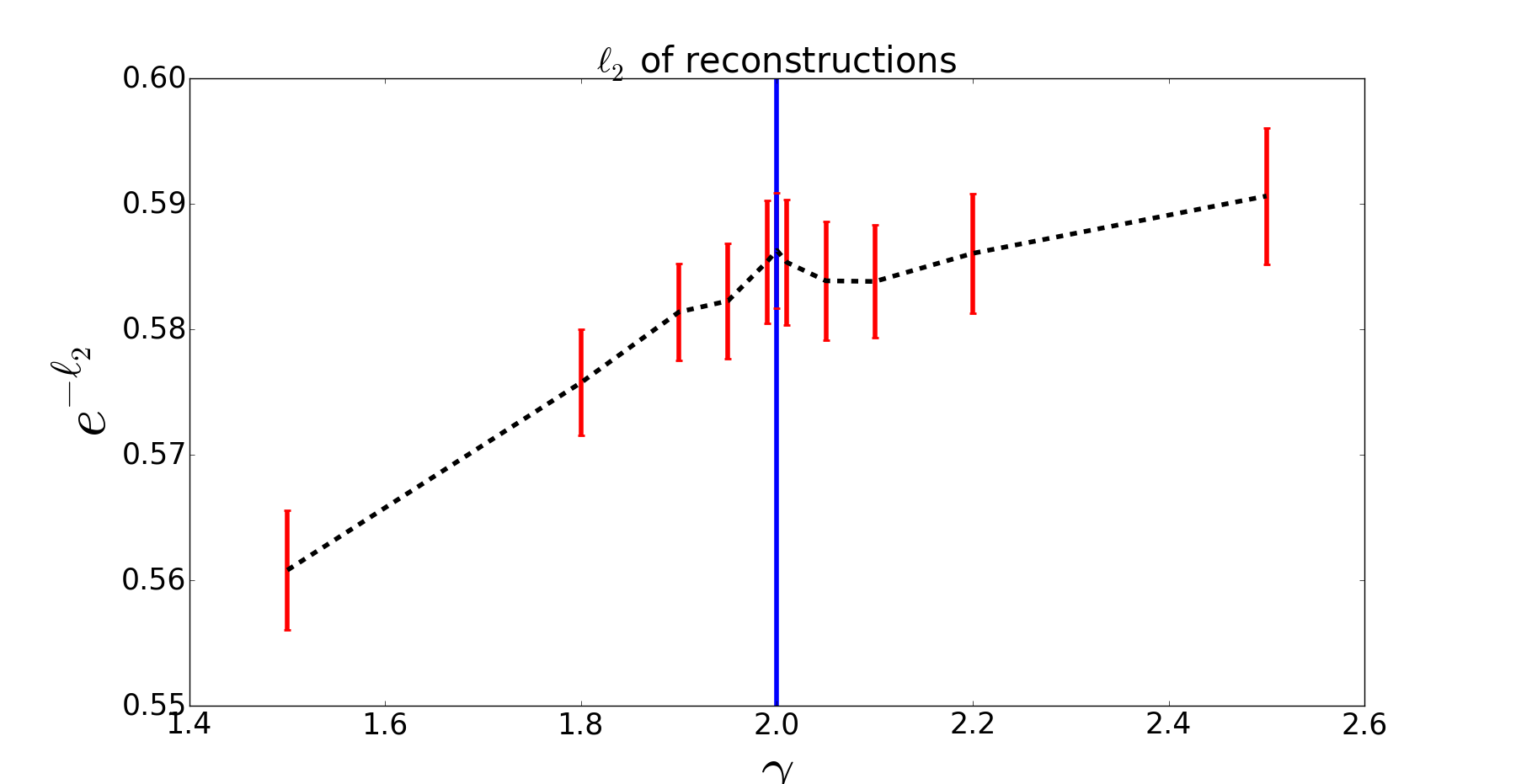}
\includegraphics[trim = 0.cm 0cm 3cm 1cm, clip = true,scale = 0.33]{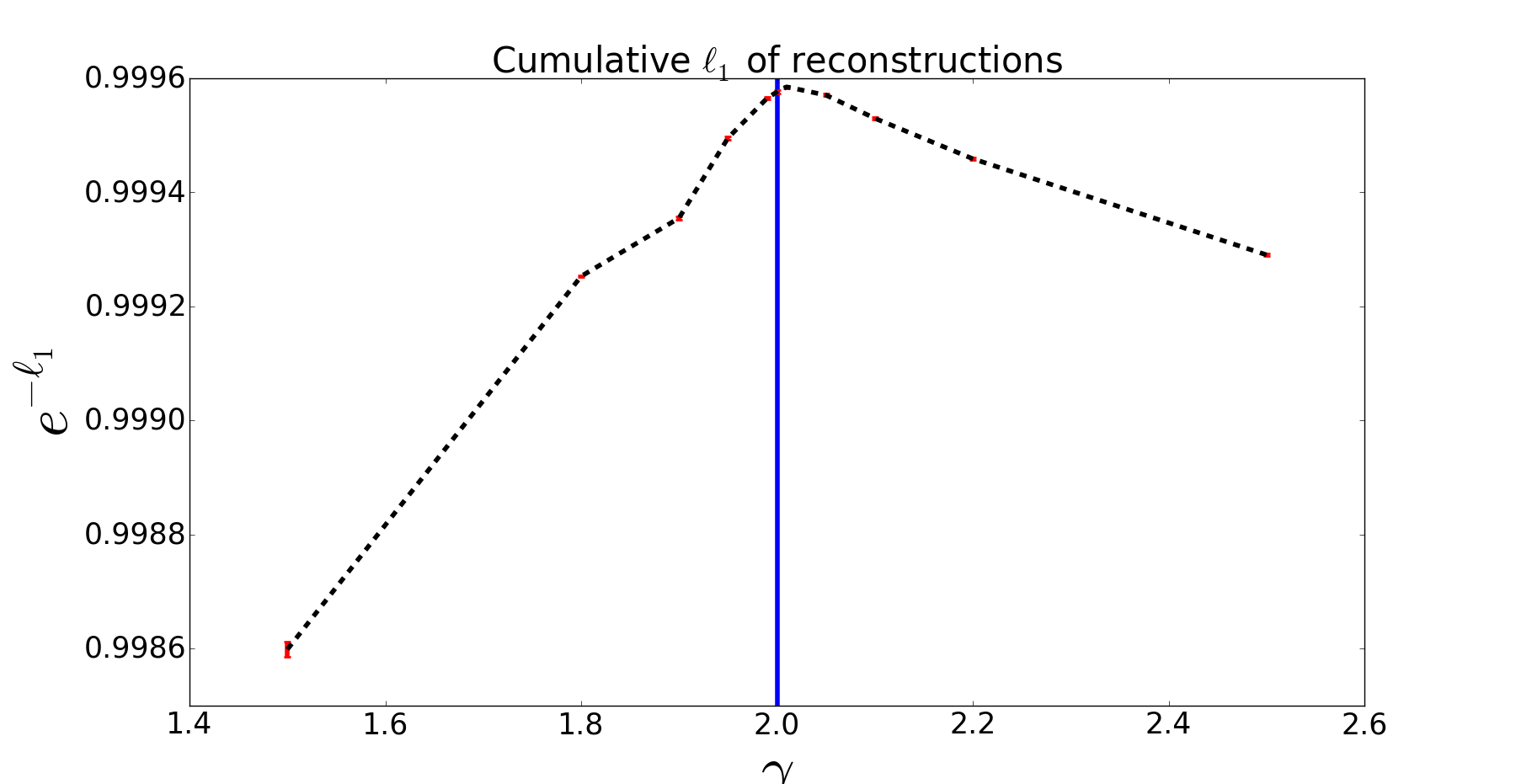}
\includegraphics[trim = 2.cm 0cm 3cm 1cm, clip = true,scale = 0.33]{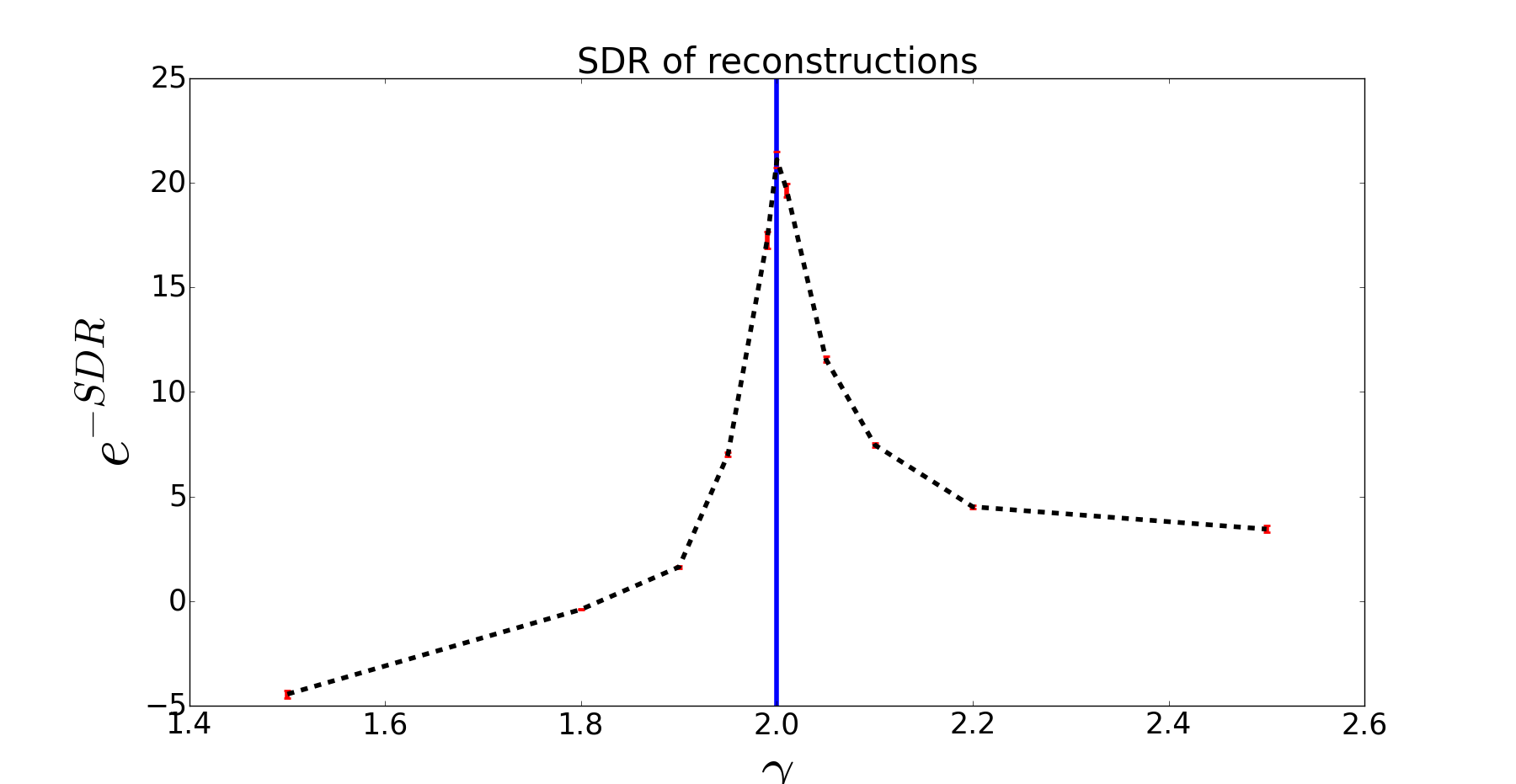}
\caption{Metrics of the reconstructions of system in Fig.~\ref{fig:System_Slope} as a function of mass density slope. The Top left panel shows the cumulative SDR of the source and galaxy light profile reconstruction. The top right panel shows the average of the residuals as $exp(-||Y-HF_{\kappa(\tilde{\gamma})}\Phi\alpha_S-\Phi\alpha_{G_H}||_2^2)$ over 100 noise realisations. The bottom panels displays the cumulative $\ell_1$-norm of $\alpha_S$ and $\alpha_{G_H}$ as $exp(-(\lambda_S||\alpha_S||_1+\lambda_G||\alpha_{G_H}||_1))$. The error bars show the standard deviation of these metrics over 100 noise realisations. The blue line shows the truth value $\gamma$.}
\label{fig:Slope}
\end{figure*}

\begin{figure}[t!]
\centering
\includegraphics[trim = 4.cm 1cm 3.5cm 1cm, clip = true,scale = 0.7]{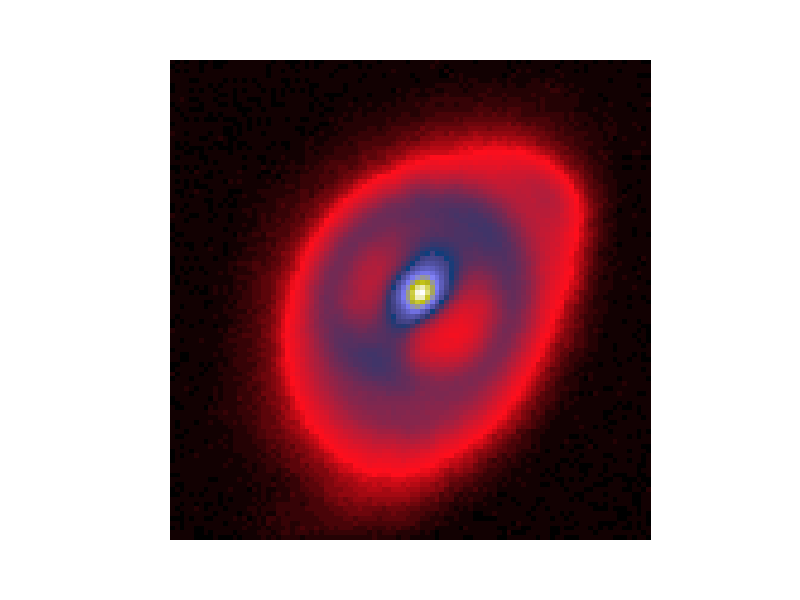}
\caption{Light profile of a simulated lens system (lens and lensed source light profiles) generated with a power law mass profile with $\gamma = 2$.}
\label{fig:System_Slope}
\end{figure}

In our analysis, we generate a lens system (shown in Fig.~\ref{fig:System_Slope}) with a power law mass density profile with $\gamma = 2$. The light profiles for the lens and source galaxies were drawn from HFF images. The PSF is a Gaussian profile with a FWHM of 2 pixels. We create 100 realisations of this system with additive Gaussian white noise at SNR 100. Each realisation is then reconstructed with {\tt SLIT\_MCA}, using mass models with density slopes $\tilde{\gamma}$ ranging from 1.5 to 2.5. The results are shown in Fig.~\ref{fig:Slope}. The true and the reconstructed profiles are shown in appendix \ref{an:2}. This appendix serves to give a visual impression of how sparsity allows to discriminate between lens mass models by discarding unphysical solutions that have a high $\ell_1$ norm in Starlets.

Fig.~\ref{fig:Slope} shows that the actual morphologies of $S$ and $G_H$  are recovered very accurately ($SDR\sim 20$) for $\gamma = 2$. In a real case study, the truth for $S$ and $G_H$ light distributions are not known, therefore, it is impossible to use the SDR to discriminate between lens mass model parameters. Instead, we have to rely on quantities derived from the observations or on properties of the reconstructed profiles. The top panel of Fig. \ref{fig:Slope} shows the likelihood defined as:
\begin{equation}
\mathcal{L}(\gamma)=exp(-||Y-HF_{\kappa(\tilde{\gamma})}\Phi\alpha_S-\Phi\alpha_{G_H}||_2^2), \label{eq:Like}
\end{equation}
 of a lens model with slope $\tilde{\gamma}$ to be the right model. Because of the strategy we choose for {\tt SLIT\_MCA}, which consists in optimising alternatively over the source and lens light profiles, the algorithm is very likely to estimate light profile models that minimise the residuals in the image, even in cases where the mass model does not correspond to the truth. Hence the relative flatness, compared to the error bars, of the likelihood profile from Fig.~\ref{fig:Slope}. In particular, we observe that, for $\tilde{\gamma}>2.2$, we achieve a likelihood as high or higher than the likelihood at $\gamma=2$, despite the lens model being wrong. This is caused by the extreme steepness of the mass density profile, which causes {\tt SLIT\_MCA} to model the source as its lensed version back in source plane. With these results, one would think {\tt SLIT\_MCA} unfit to be used in a full lens modelling framework. The strength of the algorithm lies in its potential to find the sparsest solution to a problem of lens light modelling. Since wrong mass models introduce artefacts in the reconstructions of light models, their $\ell_1$-norm is significantly higher than in the case of a reconstruction with a true model where light profiles are smooth. The middle panel of Fig.~\ref{fig:Slope} shows the cumulative $\ell_1$-norms of $\alpha_S$ and $\alpha_{G_H}$ as an argument of the likelihood. In this case, the metric is maximised around the true value for $\tilde{\gamma}$. Despite the sparsest solution being found for $\tilde{\gamma} = 2.01$, while the truth value for $\gamma$ sits at $2.00$, this result is still in the error bars estimated in \cite{Koopmans2009} over a sample of 58 lenses.

\section{Reproducible research}
\label{sec:Repro}

In order to allow for reproducibility of these results and also to provide a tool that can be used by the community, we provide the {\tt python} code that was used to implement the method we describe here. We also provide the routines used to create all plots shown in this paper as well as a readme that should allow users to run the method easily on their own data. The repository is available via the {\tt github} platform at \url{https://github.com/herjy/SLIT}. The only required libraries to run the code are the standard numpy, matplotlib, scipy and pyfits libraries. All functions and routines described in this paper were implemented by the authors. Tab.~\ref{tab:public_code} lists the products we make available in our public repository.


\begin{table*}[h!]
  \centering
  \begin{tabular}{@{} lcl @{}} 
  Product name  &Type & Description\\
  \hline
  Software products:\\
  {\tt SLIT}  & python package  & includes {\tt SLIT} implementation and visualisation tools  \\
  {\tt Lens.py}  & python code  & toolbox for lensing  \\
  {\tt Solve.py}  & python code  & python implementation of Algo.~\ref{algo:SLIT} and Algo.~\ref{algo:MCA_SLIT} and related tools  \\
  {\tt tools.py}  & python code  & code for various functions used in the minimisation process \\ & & such as the starlet transform or the FISTA iteration   \\
   & & Used to compute Fig.~\ref{fig:scales}\\
  Routines: \\
  {\tt Test\_SLIT.py}  & code (python)  & routines to reproduce Fig.~\ref{fig:Results_SLIT}.\\
  {\tt Test\_SLIT\_MCA.py}  & code (python)  & routines to reproduce Fig.~\ref{fig:Results_SLIT_MCA}.\\
  {\tt Quality.py}  & code (python)  & routines to reproduce Fig.~\ref{fig:Res_Images}, \ref{fig:Res_Sources} \& Tab.~\ref{tab:SDR}.\\
  {\tt Test\_sparsity.py}  & code (python)  & routines to reproduce Fig.~\ref{fig:NLE}.\\
  {\tt Results\_slope\_MCA.py}  & code (python)  & routines to reproduce Fig.~\ref{fig:System_Slope}, \ref{fig:Slope}, \ref{fig:Rec_gamma} \& \ref{fig:Rec_gamma2}.\\
  \hline

  \end{tabular}
  \caption{List of products made available in this paper in the spirit of reproducible research. All above material is available at the following url:  \url{https://github.com/herjy/SLIT}. \label{tab:public_code}}
\end{table*}

\section{Conclusion}
\label{conclusion}

We have developed a fully linear framework to separate the lens and source light profiles in strong lensing systems. We also to reconstruct the source shape as it was prior to the lensing effect at fixed lens mass model. As the problem is linear we were able to apply sparsity-based optimisation techniques to solve it, leading to two algorithms. 

The first algorithm, {\tt SLIT} (Sparse Lens Inversion Technique), decomposes the source plane on a basis of un-decimated wavelets (starlets). This allows us to represent the source in a non-analytical way, hence providing a sufficiently large number of degrees of freedom to capture small structures in the data. Using a sparse regularisation allows to reduce the effect of noise and artefacts on the reconstruction. {\tt SLIT} applies to lensed systems where the lens light has been removed and for a fixed mass profile.

The second algorithm, {\tt SLIT\_MCA} ({\tt SLIT} Morphological Component Analysis), also applies in the case of a fixed mass model but is able to deblend the lens light from the source as it reconstructs it in the source plane and deconvolves it from the instrumental PSF. The separation of lens and source light profiles in {\tt SLIT\_MCA} relies on the principle of morphological component analysis, but uses the distortion introduced by lensing itself as a way of discriminating between lens and source features. As is the case for {\tt SLIT}, {\tt SLIT\_MCA} does not use any analytical form for either the source or the lens representation. Both algorithms account for the instrumental PSF. 

We tested our algorithms with simulated images, showing that sources can be reconstructed while separating lens and source light profiles. We identify several advantages of our approach:

\begin{itemize}
\item the lens and source light profiles are pixelated numerical profiles, allowing a large number of degrees of freedom in their reconstruction;
\item the code implementing the algorithms is fully automated due to a careful automated computation of the regularisation parameter. It does not require any prior or assumption about the light profiles, of the source or the lens;
\item the performances of the algorithms are robust against pixel size in the sense that arbitrarily small pixel sizes can be used without leading to noise amplification or artefacts. The pixel size of both the source and lens can be an order of magnitude smaller than the PSF without negative impact on the results. On the contrary, adopting very small pixel sizes allows for detailed reconstruction of the source;
\item the {\tt python} code is public.
\end{itemize}

Furthermore, the linear framework of the algorithms presented here opens the possibilities for the developement of a lens modelling technique based on non-negative matrix factorisation \citep[NMF, ][]{Lee1999,Paatero1994} for the estimate of $F_{\kappa}$. This technique could allow us to solve Eq.~\ref{eq:Lens_G_PSF} in $F_{\kappa}$, $G$, $S$ simultaneously. The knowledge of $F_{\kappa}$ would in turn allow us to recover free form solutions for the density mass profile $\kappa$. We showed in Fig.~\ref{fig:Slope} how our strategy favoured the reconstruction of the sparsest solution, hinting at the possibility to be able to find a model for $F_{\kappa}$ that will favour the sparsest set of lens and source profiles. Because this problem introduces another degree of complexity due to the number of unknowns and the increasing degeneracies of the solutions, we delay this study to a later publication.

The main limitations of the algorithm, at this time, is its very high computational cost which scale to dozens minutes if not an hour when estimating very large source light profiles with {\tt SLIT\_MCA}. Although this makes the method unefficient as a possible minimiser in a Monte-Carlo Markov Chain sampling strategy, we are confident that upcoming optimised packages for linear optimisation will increase the speed of the algorithm. Also, the motivation behind the development of this technique is to be able to estimate free form lens mass models from NMF for instance, which requires less evaluations of {\tt SLIT\_MCA}.

To our knowledge it is the first time that an inverse problem is used to perform component separation inside a blind source reconstruction problem. In this sense the present work may extend beyond astrophysics and address a more general class of deblending and inversion problems. 

\begin{acknowledgements}
We would like to thank Danuta Paraficz, Markus Rexroth, Christoph Sch\"aefer, Sherry Suyu and Johan Richard for useful discussions on gravitational lensing and lens/source projection as well as Fran\c{c}ois Lanusse for his help on sparse regularisation and related algorithms.
This research is supported by the Swiss National Science Foundation (SNSF) and the European Community through the grant DEDALE (contract no. 665044). RJ also acknowledges support from the Swiss Society for Astronomy and Astrophysics (SSAA). S. Birrer acknowledges support by NASA through grants STSCI-GO-14630, STSCI-GO-15320, and STSCI-GO-15254 and by NSF through grant 1714953.
\end{acknowledgements}

\appendix

\section{Reconstructions for Lens mass model optimisation}
\label{an:2}

In this appendix, we show the average reconstructed lens and source light profiles as well as the corresponding truth profiles for reconstructions of a lens system with different slopes for a power law mass density profile. The profiles are shown in Fig.~\ref{fig:Rec_gamma} \& Fig.~\ref{fig:Rec_gamma2}. The light profile of lens and source galaxies reconstructed with wrong mass models show features and modes that are poorly represented by starlets and that visually do not correspond to the shape of galaxy light profile. This is particulaly visible in the most extreme cases: for $\gamma = 2.5$, the steepness of the slope leads to a mass model localised at the center of the image, such that the reconstructed source appears as a lensed galaxy; for $\gamma = 1.5$ and $\gamma = 1.8$, the light profile of the source is scattered accross the source plane and the lens light profile contains most of the signal from both source and galaxy light pofiles.

\begin{figure*}
\centering
\includegraphics[trim = 0cm 0cm 0cm 1cm, clip = true,scale = 0.2]{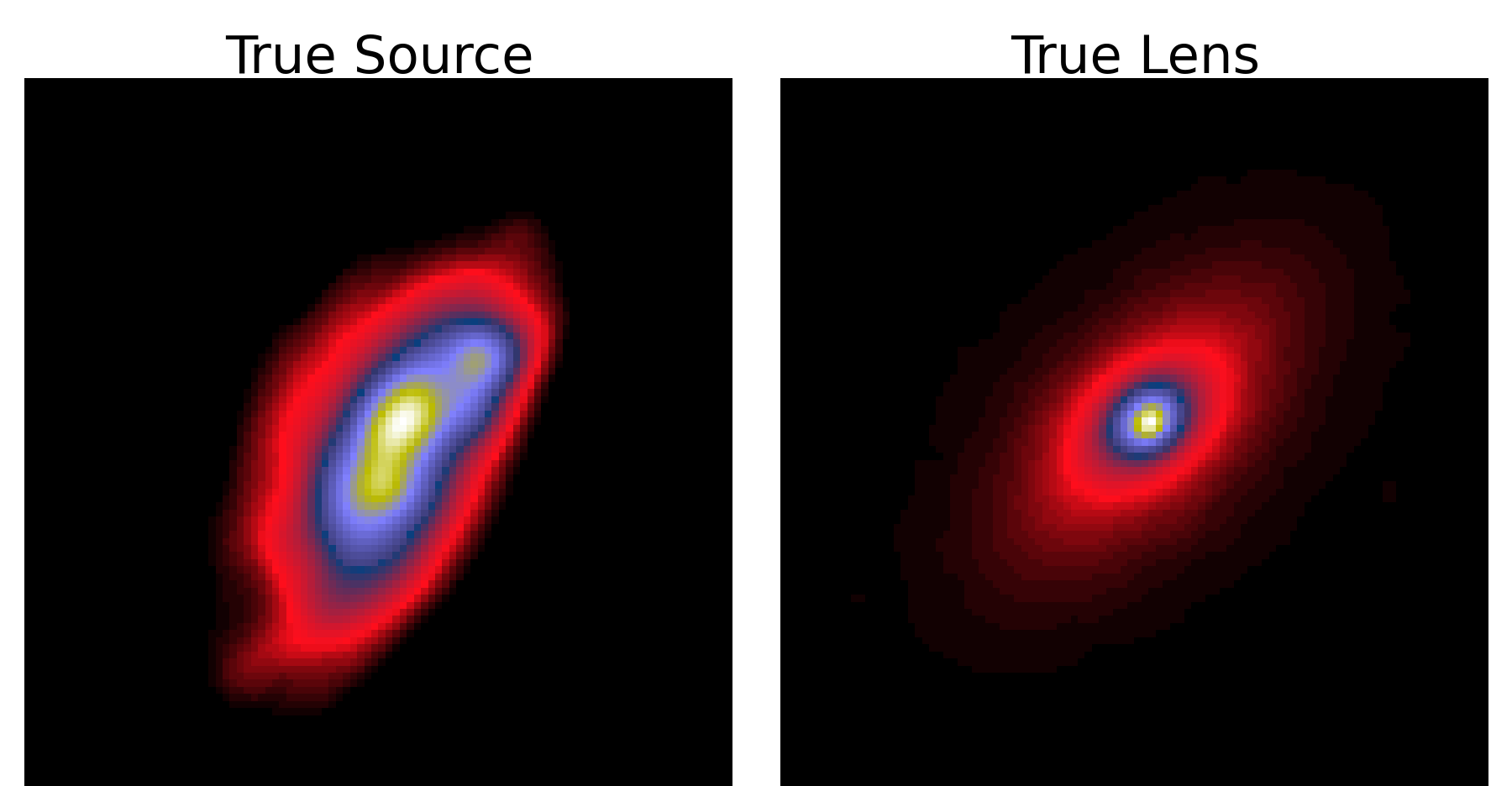}
\includegraphics[trim = 0cm 0cm 0cm 1cm, clip = true,scale = 0.2]
{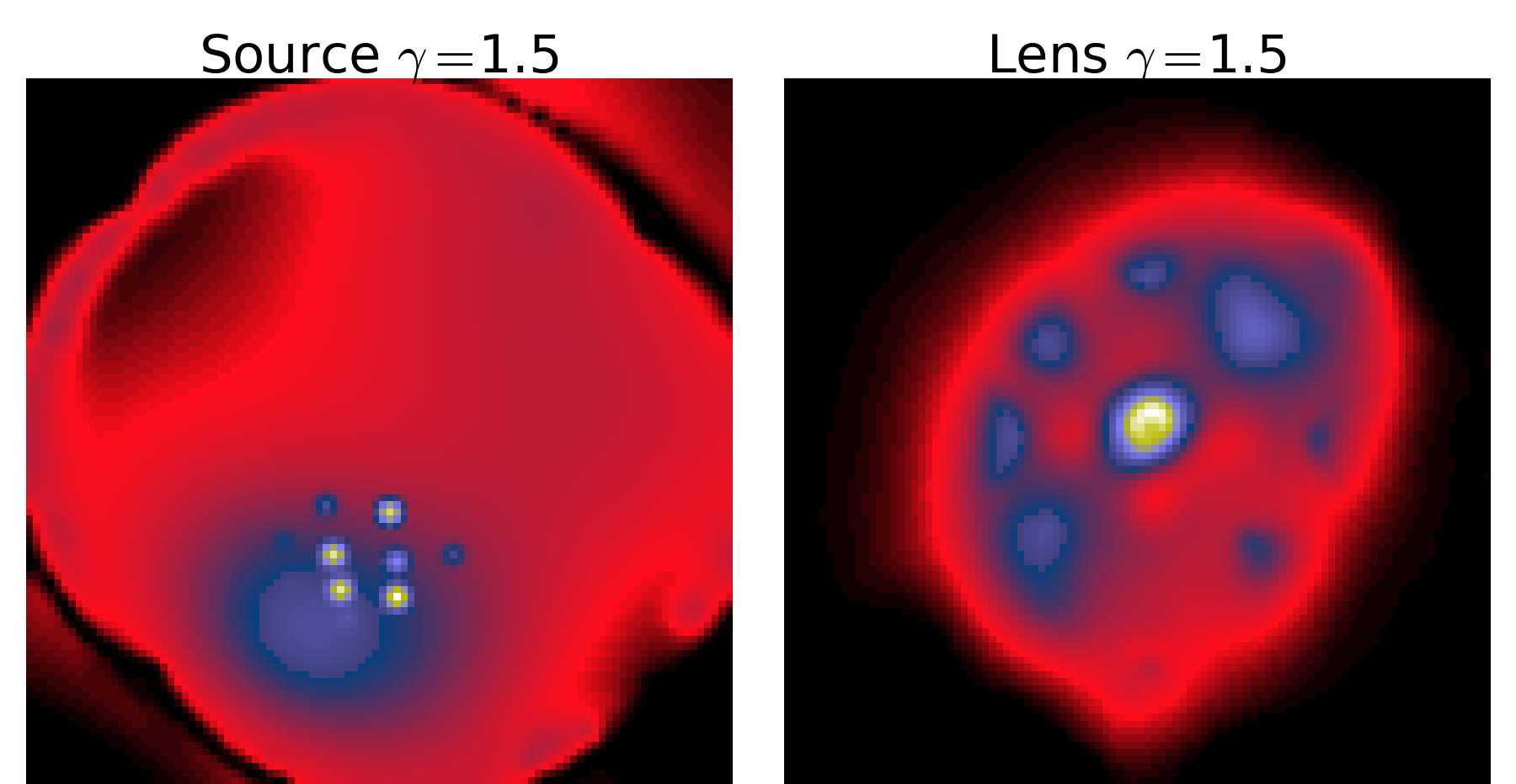}
\includegraphics[trim = 0cm 0cm 0cm 1cm, clip = true,scale = 0.2]
{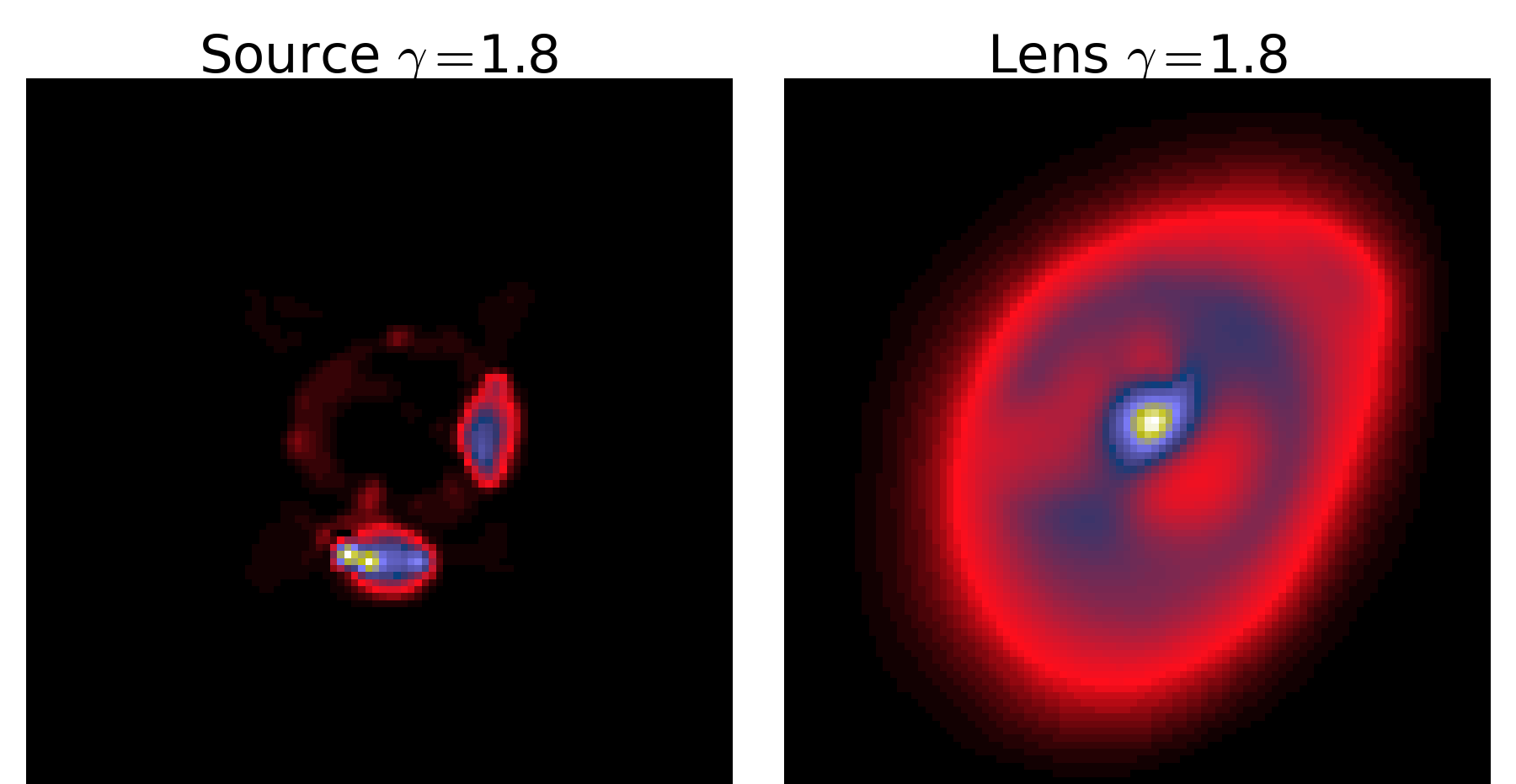}
\includegraphics[trim = 0cm 0cm 0cm 1cm, clip = true,scale = 0.2]
{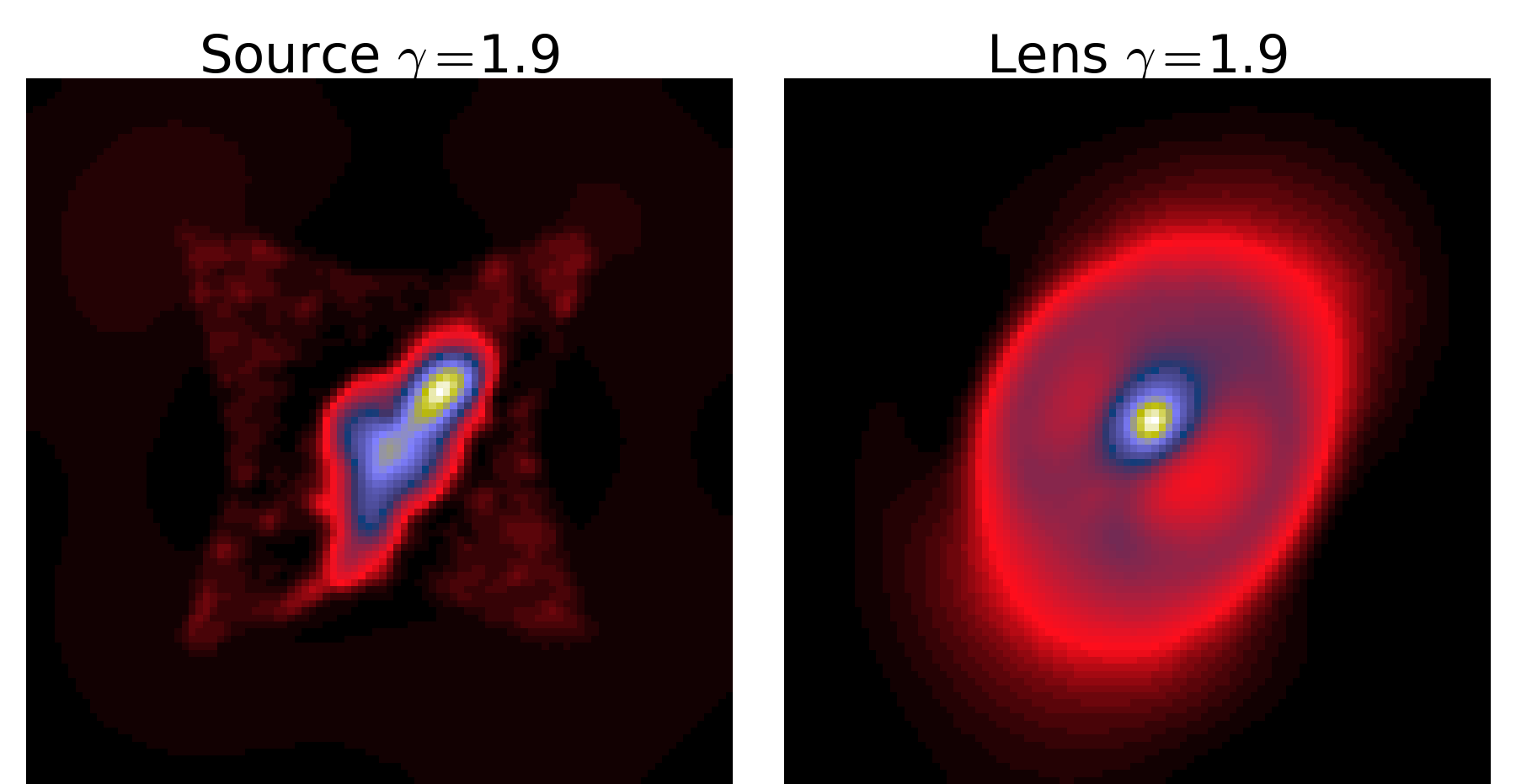}
\includegraphics[trim = 0cm 0cm 0cm 1cm, clip = true,scale = 0.2]
{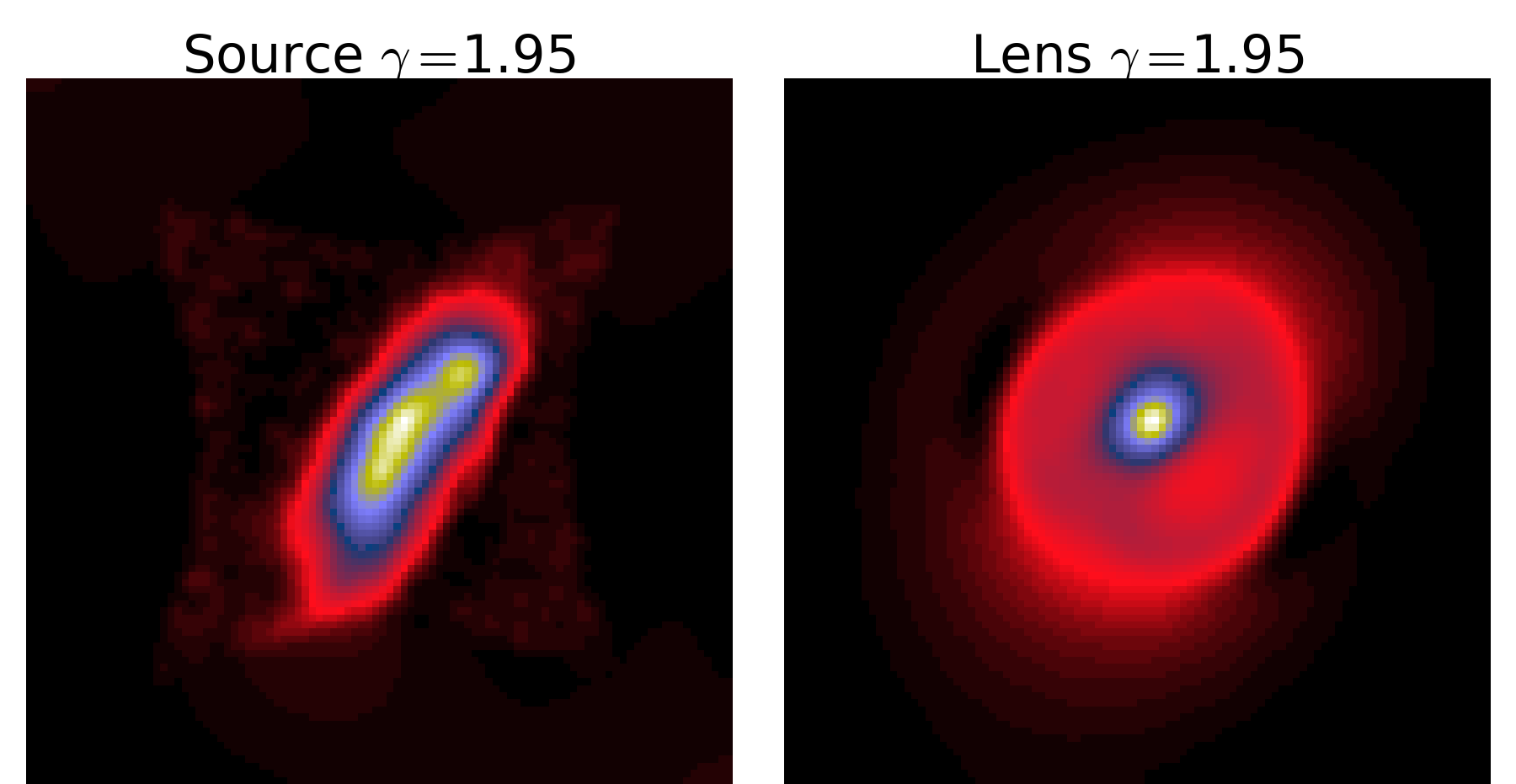}
\includegraphics[trim = 0cm 0cm 0cm 1cm, clip = true,scale = 0.2]
{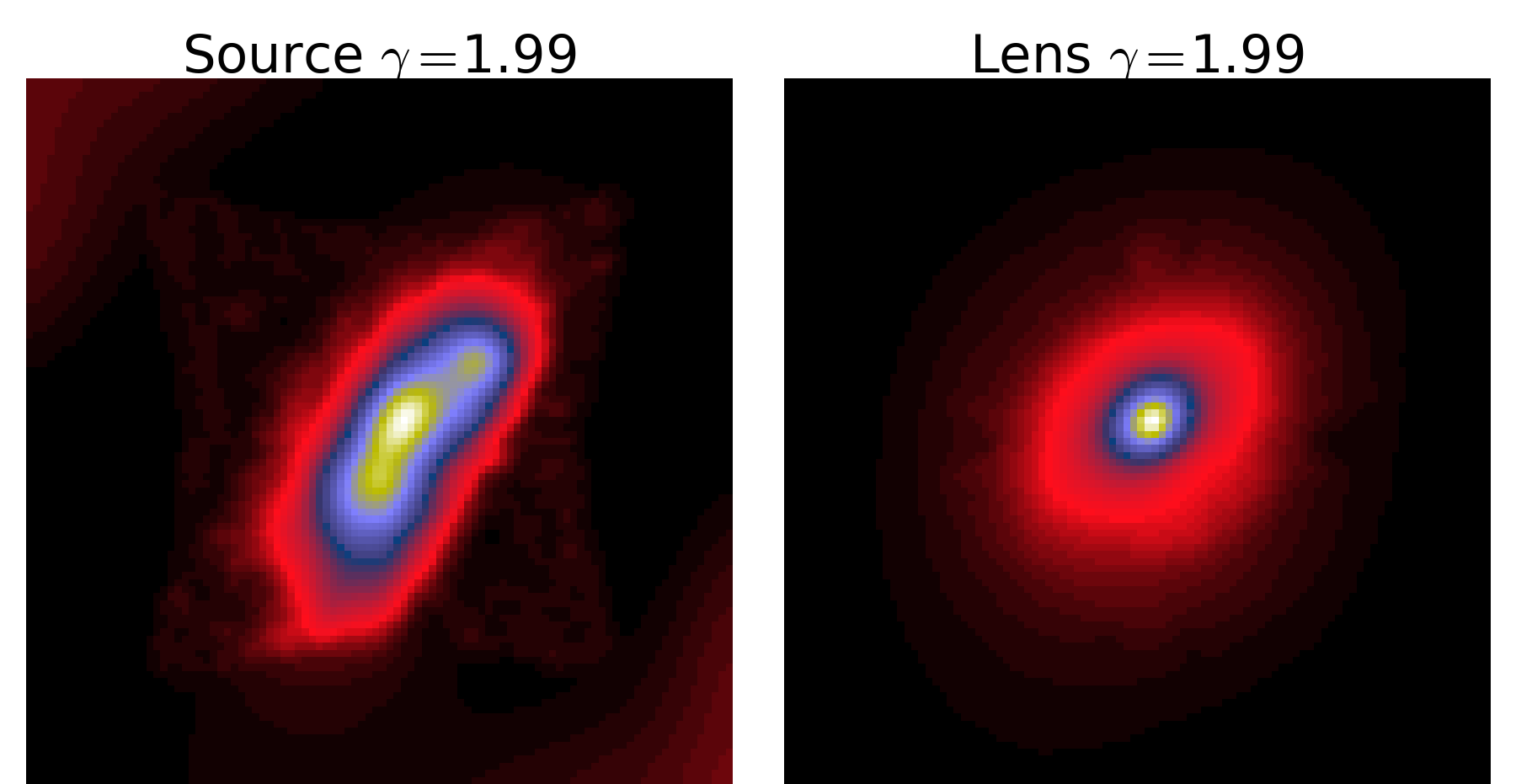}
\includegraphics[trim = 0cm 0cm 0cm 1cm, clip = true,scale = 0.2]
{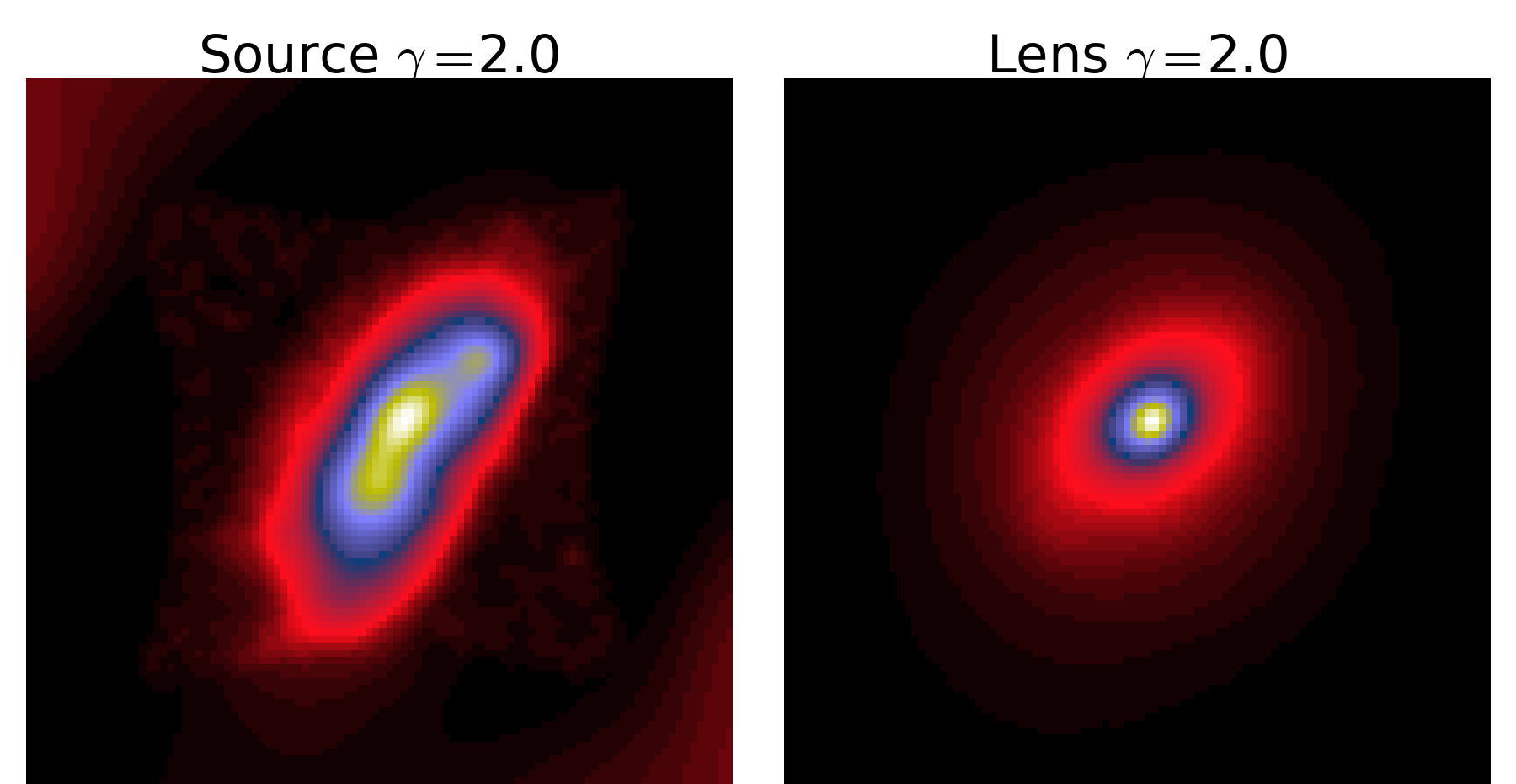}
\includegraphics[trim = 0cm 0cm 0cm 1cm, clip = true,scale = 0.2]
{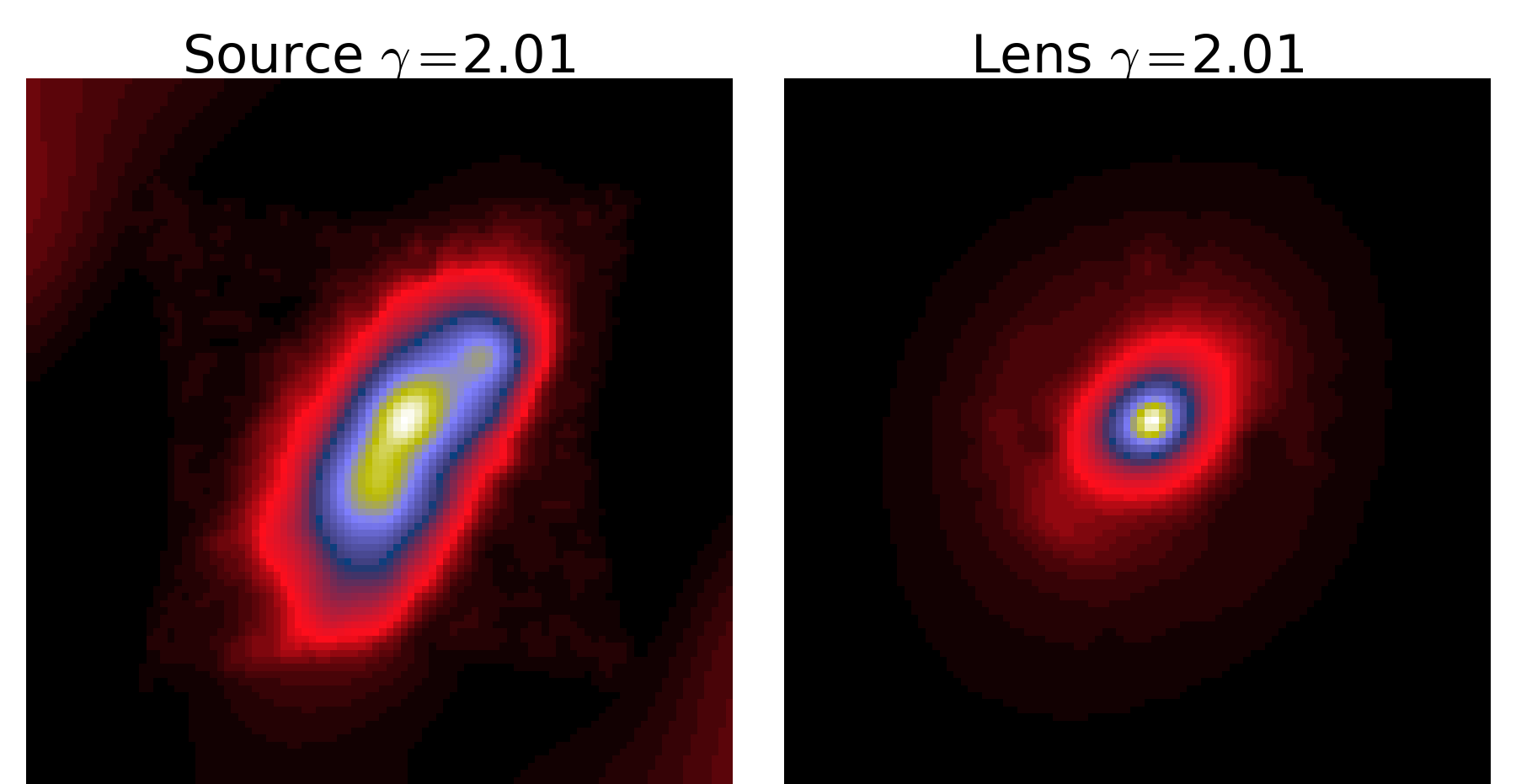}
\includegraphics[trim = 0cm 0cm 0cm 1cm, clip = true,scale = 0.2]
{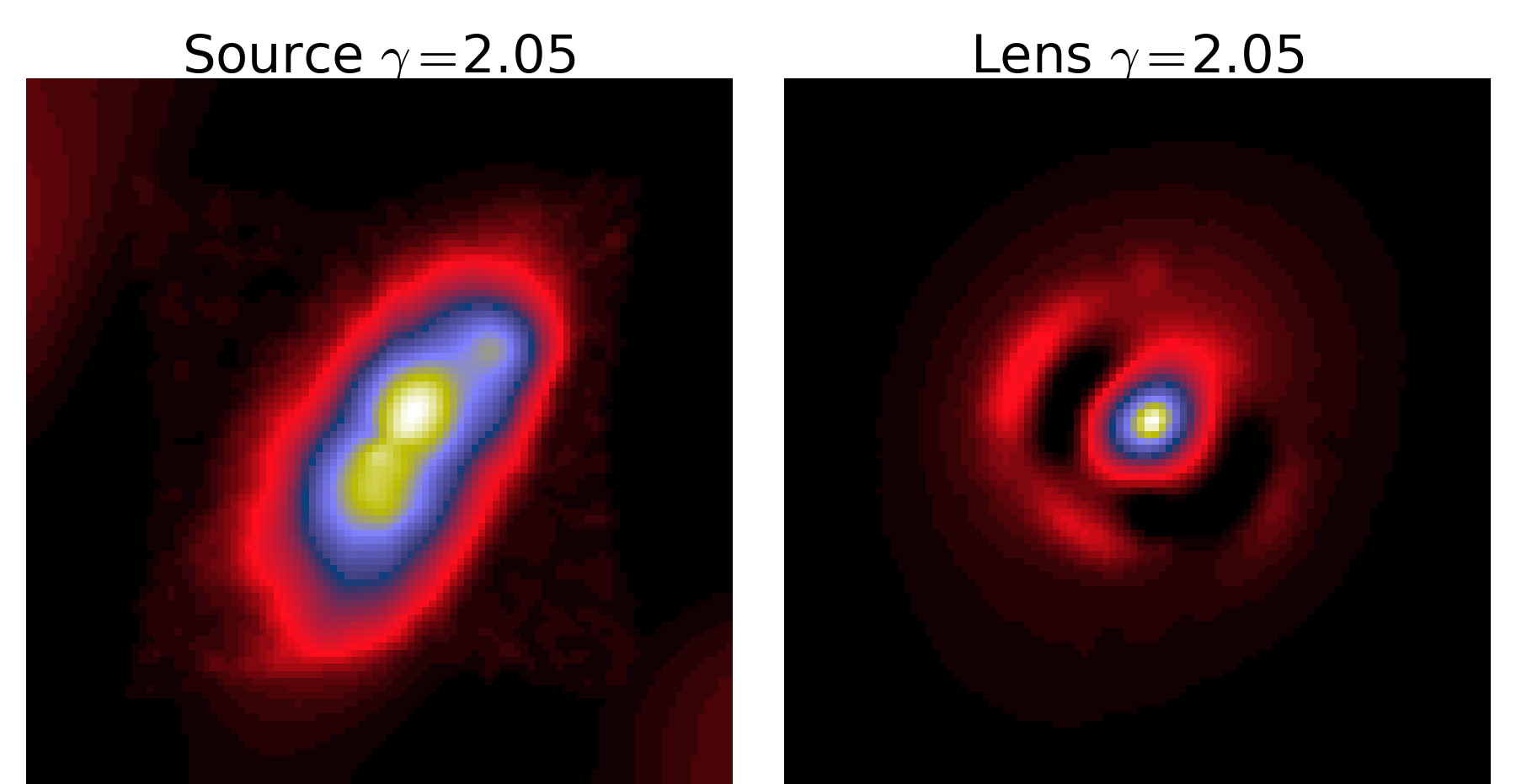}
\includegraphics[trim = 0cm 0cm 0cm 1cm, clip = true,scale = 0.2]
{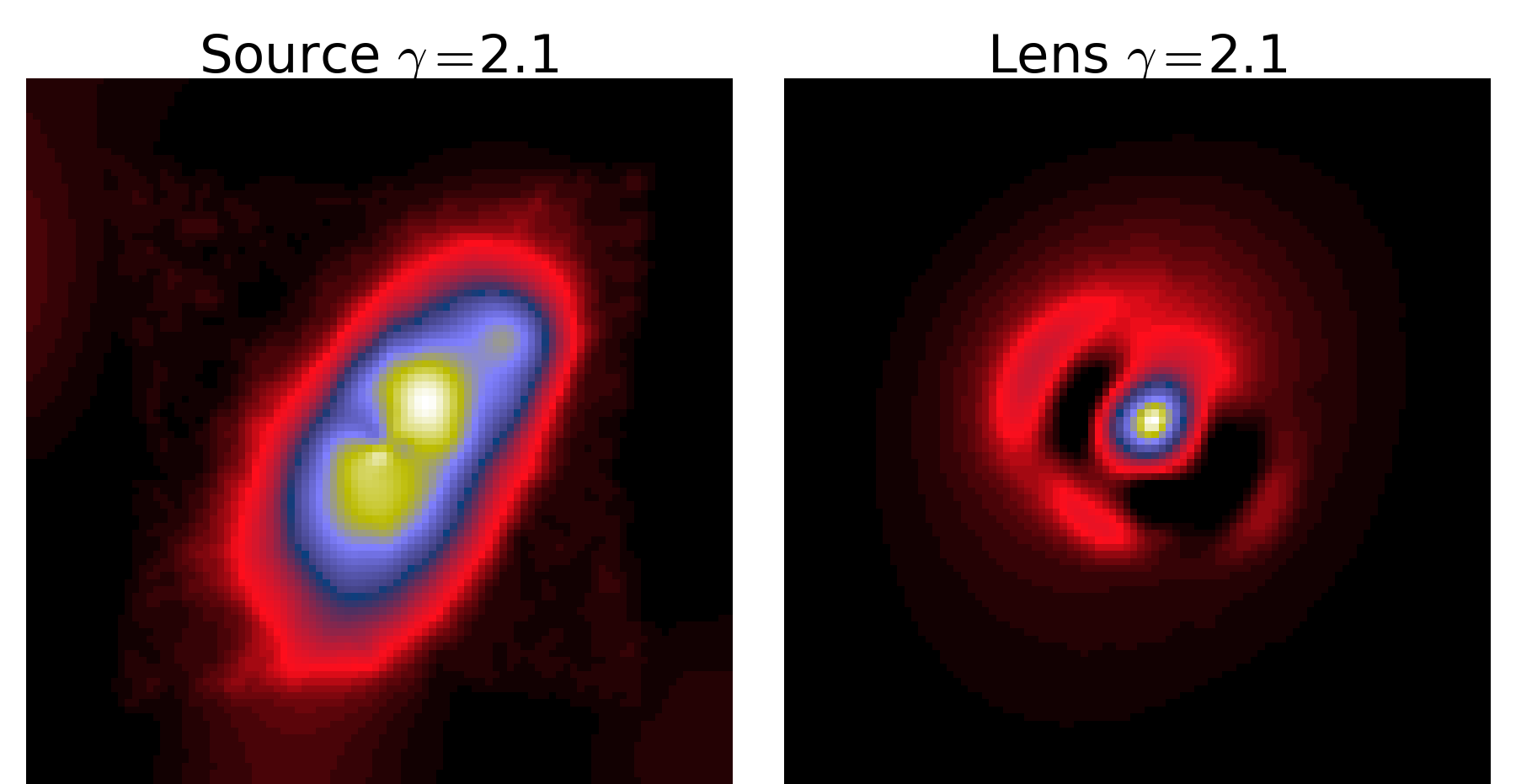}
\caption{Reconstructions of lens and source light profiles for various values of mass density slope of a lens generated with a mass density slope of 2. The first two panels show the true source (left hand-side) and lens (right hand side) light profiles used to generate the simulated images. The other couples of panels  from left to right and from top to bottom show the source and lens reconstructions for increasing values of $\tilde{\gamma}$.}
\label{fig:Rec_gamma}
\end{figure*}
\begin{figure*}
\centering
\includegraphics[trim = 0cm 0cm 0cm 1cm, clip = true,scale = 0.2]
{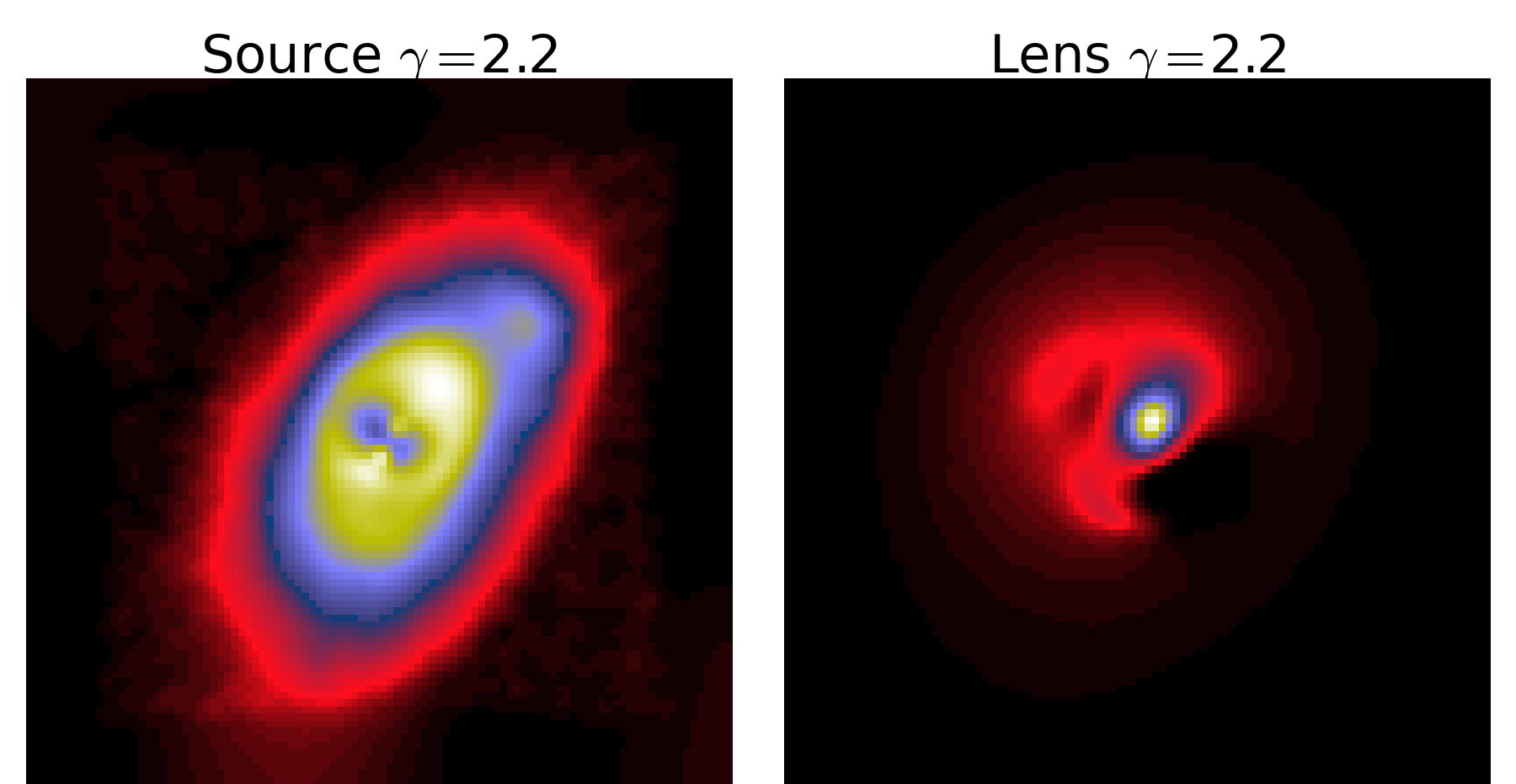}
\includegraphics[trim = 0cm 0cm 0cm 1cm, clip = true,scale = 0.2]
{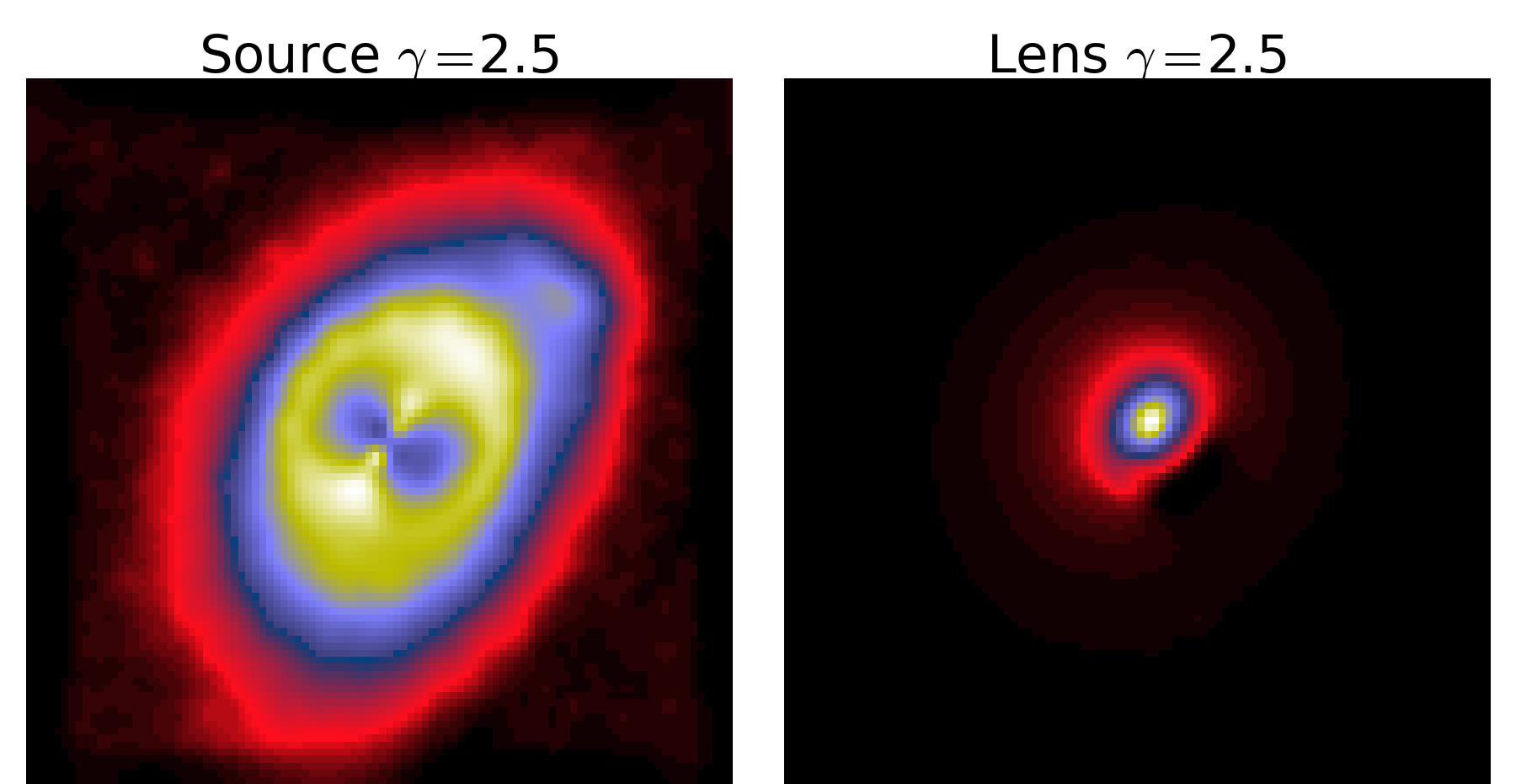}
\caption{Continuation of Fig.~\ref{fig:Rec_gamma}.}
\label{fig:Rec_gamma2}
\end{figure*}

\newpage

\bibliographystyle{aa} 
\bibliography{biblio} 
\end{document}